\documentclass[11pt,a4paper]{article}
\usepackage{amssymb}
\usepackage{amsmath}
\usepackage{graphicx}
\usepackage{subfig}
\usepackage[english]{babel}
\newcommand{\be}{\begin{equation}}
\newcommand{\ee}{\end{equation}}
\newcommand{\nn}{\nonumber}

\newcommand{\B}{{\cal B}}
\newcommand{\C}{{\cal C}}
\newcommand{\D}{{\cal D}}
\newcommand{\F}{{\cal F}}
\newcommand{\GH}{{\cal H}}
\newcommand{\I}{{\cal I}}

\newcommand{\cL}{{\cal L}}
\newcommand{\M}{{\cal M}}
\newcommand{\N}{{\cal N}}
\renewcommand{\P}{{\cal P}}
\newcommand{\Q}{{\cal Q}}
\renewcommand{\S}{{\cal S}}
\newcommand{\R}{{\cal R}}

\renewcommand{\O}{{\cal O}}
\newcommand{\G}{{\mathcal G}}
\renewcommand{\H}{{\mathcal H}}

\newcommand{\V}{{\mathcal V}}
\newcommand{\X}{{\cal X}}
\newcommand{\Y}{{\cal Y}}
\newcommand{\Z}{{\cal Z}}

\newcommand{\rmd}{{\rm d}}

\newcommand{\bt}[1]{{\bar t}}

\newcommand{\half}{{\tfrac{1}{2}}}

\newcommand{\pr}{\partial}
\newcommand{\vep}{\varepsilon}
\newcommand{\vphi}{\varphi}
\newcommand{\bvphi}{{\bar\varphi}}
\newcommand{\tphi}{\tilde \varphi}
\newcommand{\tbeta}{\tilde \beta}

\newcommand{\de}{{\delta}}
\newcommand{\si}{{\sigma}}
\newcommand{\bth}{{\bar \theta}}
\newcommand{\bphi}{{\bar \phi}}
\renewcommand{\d}{{\rm d}}
\newcommand{\bD}{\bar D}
\newcommand{\hG}{\hat {\mathcal G}}
\newcommand {\x}{{\rm x}}
\newcommand{\hS}{{\hat S}}
\newcommand{\hPsi}{{\hat \Psi}}

\newcommand {\tx}{{\tilde {\rm x}}}
\newcommand {\tF}{{\tilde F}}
\newcommand{\hJ}{{\hat J}}
\newcommand{\tr}{{\rm tr}}

\newcommand \ollr{{\raise 8pt\hbox{$\leftrightarrow  \! \! \! \! \! \!$}}}
\newcommand{\topcirc}{\mathaccent"7017}
\def\tosim#1{\mathrel{\mathop{\sim}\limits_{\scriptstyle{#1}}}}
\def\tolim#1{\mathrel{\mathop{\to}\limits_{\scriptstyle{#1}}}}

\csname @addtoreset\endcsname{equation}{section}
\begin{document}

\begin{titlepage}
\thispagestyle{empty}
\begin{flushright}
\small
DAMTP/11-65\\
arXiv:1108.5340[hep-th]\\
\today \\
\normalsize
\end{flushright}
\vskip 3cm
\centerline{\LARGE{\bf Remarks on Exact RG Equations}}
\bigskip
\vskip 3cm
\centerline{H. Osborn\footnote{ho@damtp.cam.ac.uk} and 
D.E. Twigg\footnote{det28@cam.ac.uk}}
\vskip 1cm
\centerline{Department of Applied Mathematics and Theoretical Physics,}
\centerline{Wilberforce Road, Cambridge, CB3 0WA, England}

\date{latest update: \today}
\begin{abstract}
Exact RG equations are discussed with emphasis on the role of the anomalous
dimension $\eta$. For the Polchinski equation this may be introduced as a free
parameter reflecting the freedom of such equations up to contributions which
vanish in the functional integral. The exact value of $\eta$ is only determined
by the requirement that there should exist a well defined non trivial limit at 
an IR fixed point. The determination of $\eta$ is related to the existence of
an exact marginal operator, for which an explicit form is given. The results 
are extended to the exact Wetterich RG equation for the one particle
irreducible action $\Gamma$ by a Legendre transformation. An alternative 
derivation of the derivative expansion is described. An application to
$\N=2$ supersymmetric theories in three dimensions is described where if an
IR fixed point exists then $\eta$ is not small.
\end{abstract}

\thispagestyle{empty}
\end{titlepage}

\newpage
\setcounter{footnote}{0}
\section{Introduction}

Historically the appreciation of  the role of the renormalisation group, which 
corresponds to varying the cut off scale, was essential to the modern 
understanding of quantum field theories. As a consequence it becomes clear 
that quantum field theories, which need a cut off in their formulation,
must be considered as belonging to an infinite dimensional 
space which may be parameterised in terms of couplings associated with 
all scalar operators consistent with the basic symmetries of the theory. 
The RG equations then describe flows in this space of quantum 
field theories under changes in the cut off scale $\Lambda$, the flow 
being determined by the requirement that physical observables are 
independent of the cut off, at least in the neighbourhood of fixed points, 
for energies much less than the cut off. This allows a continuum quantum 
field theory to be obtained which satisfies the required symmetries and is 
independent of the choice of cut off. A particular realisation may be 
obtained in terms of renormalisable theories where the space of quantum 
field theories is restricted to a submanifold, usually finite 
dimensional, which is invariant under RG flow. For renormalisable theories 
the RG flow equations are linear and the flow in the space of couplings is 
given in terms of the associated $\beta$-functions.

A particular realisation of the renormalisation group, for
quantum field theories where a cut off may be introduced through a modification
of the quadratic kinetic term, is obtained through 
exact functional non linear RG equations, due to Wilson \cite{Wilson} and 
developed in various alternative forms in \cite{Wegner,Phase,Polchinski}. 
Reviews covering different aspects are found in
\cite{MorrisRev,Bagnuls,Wett,Jan,Bertrand,Rosten,Fbook}. Such 
exact RG equations provide an in principle non perturbative definition of 
RG flow. Nevertheless such equations are restricted, at least in 
most applications, to theories containing just scalar fields, although
extensions to fermion fields are relatively straightforward. The exact RG
equations are also hard to 
approximate in a consistent fashion which allows calculable higher order 
corrections. Although it is possible to recover perturbative results, and 
rederive the perturbative results for $\beta$-functions, the methods 
involved tend to be distinct from those used in non perturbative 
approximations. A particular issue is connected with the anomalous 
dimension $\eta$ of the basic scalar fields in the theory. Although $\eta$ 
may be introduced as an additional parameter in the RG flow equations, 
where it is essentially arbitrary, the precise way in which it is to be 
determined is not always fully resolved.

The aim in this paper is to analyse such issues in more detail than hitherto. 
The RG equations determined how the theory varies with $t$, where 
$t \sim  - \ln \Lambda$.
We argue that $\eta$ is an arbitrary parameter until the we consider limit 
$t\to \infty$.  The requirement that there exists a well defined limit, with 
long range order, when $t\to \infty$ is the necessary and 
sufficient condition to determine $\eta$. In particular as was discussed 
by Wegner \cite{Phase}, and emphasised more recently by Rosten \cite{Rosten},
the existence of discrete values for $\eta$ is linked to the presence of
an exact marginal operators $\Z$. In this case $\Z$ generates a line of
equivalent fixed points which corresponding essentially to an overall 
rescaling of the fields. The limit $t\to \infty$ may also generate 
a trivial, so called high temperature, fixed point with no long range
order but then there is no marginal operator and $\eta$ is not determined.

In the next section we discuss in detail various aspects of the exact
RG equations as introduced by Wilson and its later alternative, with an
essentially similar form, developed by Polchinski \cite{Polchinski}. 
Although the analysis is not rigourous we attempt to make precise
the mathematical framework and avoid any particular choice of the cut off
function necessary in the derivation of the RG flow equation.
As usual for simplicity we consider a single scalar field
$\vphi$ which is rescaled by the cut off to be dimensionless and consider
the RG flow of the action functionals $\S_t[\vphi]$ as well as their
IR fixed points $\S_*[\vphi]$. The original Wilson/Polchinski equations
are non linear but there is a transformation such that 
$\S_t[\vphi]\to T[\vphi_t]$ that linearises the RG flow equation. 
The transformation is generated by the action of a functional operator
$e^{\frac{1}{2}\, \frac{\delta}{\delta \vphi} \, \cdot \, \G  \, \cdot \, 
\frac{\delta}{\delta \vphi}}$ for suitable $\G$. This is well defined
for $\G$ positive but such a transformation is not in general invertible.
At an IR fixed point $\S_*[\vphi]\to 
T_*[\vphi]$ with $T_*$ obeying a simple linear equation. 

The essential observables which appear in this framework are the critical
exponents $\lambda$\footnote{Here we choose $\lambda$ to have the
opposite sign to the standard conventions in discussions of critical phenomena.} 
which determine how the critical point is approached as $t\to \infty$
and which are associated  with scaling operators $\O$.
The number of negative $\lambda$ determine the extent to which $\S_0[\vphi]$
has to be tuned so as to lie on the critical surface where the RG flow
attains the IR fixed point. The RG equations determine the exponents
$\lambda$ as the eigenvalues of a functional differential operator
depending on $\S_*$. A subspace of the space of operators $\{ \O \}$ are
redundant operators corresponding to redefinitions $\vphi(x) \to \psi(x)$
where $\psi$ is a functional of $\vphi$. The marginal or zero mode operator
$\Z$ is a redundant operator determined by $\psi_\Z$ related to a constant
rescaling, or reparameterisation,  of $\vphi$.

In our treatment it is natural to consider also local scaling operators 
$\Phi_\Delta(x)$, functionals of $\vphi$, whose scaling dimensions $\Delta\ge 0$
are determined by a local functional eigenvalue equation.  For $\Delta = 0$ then
$\Phi_0 =1$, the identity. Any $\Phi_\Delta$ determines an associated $\O$ 
with $\lambda = \Delta -d$. A particular role in our discussion is played
by two exact scaling operators $\Phi_\delta$ and $\Phi_{d-\delta}$ 
where $\delta = \half (d-2+\eta)$. These both have the form
\be
\Phi_\Delta = X \cdot \vphi + Y \cdot \frac{\de}{\de \vphi} \S_*  \, ,
\label{PXY}
\ee
for appropriate linear operators $X,Y$. $\Phi_\delta$ gives directly $\psi_\Z$.
Requiring $\Phi_\delta=0$, for arbitrary $\eta$, clearly determines a
quadratic form for $\S_*$ and corresponds to the high temperature fixed point.
In this case $\Z=0$ reflecting the fact that $\eta$ is undetermined in this
case.
For $\eta=0$, when $\delta = \delta_0 = \half(d-2)$, imposing 
$\pr^2 \Phi_{\delta_0} \propto \Phi_{d-\delta_0}$ also determines $\S_*$ to be
quadratic in $\vphi$, corresponding to a free or Gaussian theory. 
Such a quadratic form depends on a parameter defining 
a line of equivalent IR fixed points which are generated by $\Z$.
An argument that $\eta=0$ implies a Gaussian theory in the context of
exact RG equations has also been given by Rosten \cite{Pohl}. 

We also discuss how solutions of the RG flow equations for $\S_t[\vphi]$
can be extended without essential modification to include a source $J$
coupled to $\vphi$. This allows the standard vacuum functional $W[J]$
to be directly related to $T[\vphi]$. The connection of the 
Wilson/Polchinski equation for $\S_t$ with the Wetterich equation 
for the RG flow of the one particle irreducible functional $\Gamma_t$
by a Legendre transform is also reviewed. For $\eta$ non zero it is still
feasible to relate the two equations although constructing a Legendre
transform with the desired properties involves solving some first order
differential equations. The precise form of the zero mode operator for
the Wetterich equation is constructed by considering the transform of $\Z$.
Various aspects of the discussion are illustrated by consideration of
the Gaussian case. This has a limit as $t \to \infty$ defining a massless
theory only when $\eta=0$, for $\eta\ne 0$ the Gaussian theory approaches
the high temperature fixed point. In either case the scaling operators
are explicitly constructed.

Although not directly related to our main discussion we also reconsider
more briefly the derivative expansion in section 3. This is based on a
modification of the Polchinski equation which is tantamount to expanding
$\S_t[\vphi]$ in terms of a normal ordered basis of monomials in $\vphi$.
This allows a derivation of equations, considered by us earlier \cite{DT},
to first order in the derivative expansion which maintains those exact
results for the full RG equations described earlier, namely the existence
of eigenfunctions corresponding to scaling dimensions $\half( d \mp 2 \pm \eta)$
and also a zero mode eigenfunction related to reparameterisation invariance.
In section 4 this approach is extended to a supersymmetric theory
in three dimensions with $\N=2$ supersymmetry. In this example $\eta$
is determined to be $\frac{1}{3}$ for there to be a non trivial IR fixed
point.

In appendix A we also outline extensions of part of our discussion to
more general RG flow equations without making the restrictions to 
obtain the Wilson/Polchinski form. In appendix B we also describe 
in a perturbative context how the zero mode operator is also present.

\section{Derivation and Analysis of Exact RG Equations}

There are many varieties of essentially equivalent RG flow equations
\cite{Wilson,Wegner,Phase,Polchinski}.
Assuming for simplicity just a single  scalar field $\phi(x)\in V_{\phi}$,
for $x\in {\mathbb R}^d$, 
they arise by considering regularised actions ${\hS}_\Lambda[\phi]$ 
depending on a cut off scale $\Lambda$ and understanding how these
should evolve as $\Lambda$ varies so as to ensure results independent
of the precise form of the cut off may be obtained. Requiring
\be
- \Lambda \frac{\pr}{\pr \Lambda}\,  e^{-\hS_\Lambda[\phi]}
= \frac{\delta}{\delta \phi} \cdot \bigg( \hPsi_\Lambda[\phi] \;
 e^{-\hS_\Lambda[\phi]} \bigg ) \, ,
\label{basic1}
\ee
for any $ \hPsi_\Lambda[\phi;x] \in V_\phi $, ensures that the basic 
functional integral defining the quantum field theory for the action
$\hS_\Lambda$ is invariant under changes in $\Lambda$. Although
the initial choice of $\hS_\Lambda$ for some $\Lambda = \Lambda_0$ is
largely arbitrary there are nevertheless features of the RG flow 
that are independent of the precise form of $\hS_{\Lambda_0}$,
at least for appropriate $\Psi_\Lambda$ and in suitable limits, which gives 
rise to the crucial notion of universality. \eqref{basic1} is of course 
equivalent to
\be
- \Lambda \frac{\pr}{\pr \Lambda}\,  \hS_\Lambda[\phi]
=  \hPsi_\Lambda[\phi] \cdot \frac{\delta}{\delta \phi}\, \hS_\Lambda[\phi]
- \frac{\delta}{\delta \phi} \cdot  \hPsi_\Lambda[\phi] \, .
\label{basic2}
\ee

In \eqref{basic1} and \eqref{basic2} 
\be
\phi \cdot \psi = \psi \cdot \phi = \int \d^d x \; \phi(x) \psi(x) \, , 
\label{XY}
\ee
for $\phi,\psi \in V_\phi$. Functional derivatives are here defined so that
$\delta F[\phi] = \delta\phi \cdot  \frac{\delta}{\delta \phi} F[\phi]$. 
We also define for $I(x,y)\in  V_\phi \times V_\phi$
\be
\phi \cdot I \cdot \psi = \psi \cdot I^T \cdot \phi
=  \int \d^d x\,\d^d y \; \phi(x)\, I(x,y) \psi(y) \, ,
\label{IG}
\ee
defining $I:\V_\phi \to \V_\phi$ with a functional trace
\be
\tr (I) = \int \d^d x \; I(x,x) \, .
\label{Itr}
\ee
As is commonplace it is natural to transform the fields to momentum
space, $\phi(x) \to  {\tilde \phi}(p)$, so that
\be
\phi \cdot \psi = 
\frac{1}{(2\pi)^d} \int \d^d p \; {\tilde \phi}(p) {\tilde \psi}(-p) \, .
\ee
For a translation invariant $I$ in \eqref{IG}, so that $I(x,y) \to G(x-y)$,
then in momentum space 
${\tilde I}(p,q) \to {\tilde G}(q) \, (2\pi)^d\de^d(p+q)$, 
\be
\phi \cdot G \cdot \psi = \frac{1}{(2\pi)^d} \int \d^d p \;
{\tilde \phi}(p) \, {\tilde G}(p) \,  {\tilde \psi}(-p) \, ,
\label{XYG}
\ee
and of course $\widetilde{G_1\cdot G_2}(p) = 
{\tilde G}_1(p){\tilde G}_2(p)$ and ${\tilde G}^T(p) = {\tilde G}(-p)$.
Furthermore in this case
\be
\tr (G) = V \,  \frac{1}{(2\pi)^d} \int \d^d p \; {\tilde G}(p) \,  ,
\label{Gtr}
\ee
where $V$ is the spatial volume, implicitly assuming a spatial cut off, such
as compactifying on a torus, but whose details are unspecified here. 

The cut off $\Lambda$ sets the fundamental scale and the equations
are further assumed to be reduced to dimensionless form by requiring
\be
\phi(x) = \Lambda^{\delta_0} \vphi(x \Lambda) \, , \quad
\frac{\delta}{\delta \phi(x)} =  \Lambda^{d- \delta_0}
\frac{\delta}{\delta \vphi(x\Lambda)}
\qquad \hPsi_\Lambda(x) = \Lambda^{\delta_0 } \Psi_t (x\Lambda) \, ,
\label{scale}
\ee
for some choice of $\delta_0$, with
\be
t = - \ln \Lambda/\Lambda_0  \, , \qquad 
\hS_\Lambda[\phi] = S_t [\vphi] \, .
\label{Sscale}
\ee

With the rescaling \eqref{scale}, \eqref{basic2} takes the form
\be
\bigg ( \frac{\pr}{\pr t} + D^{(\de_0)} \vphi \cdot \frac{\delta}{\delta \vphi} 
- d \, V\frac{\pr}{\pr V} \bigg ) \, S_t[\vphi]
=  \Psi_t[\vphi] \cdot \frac{\delta}{\delta \vphi} \, S_t[\vphi]
- \frac{\delta}{\delta \vphi} \cdot  \Psi_t[\vphi] \, ,
\label{basic3}
\ee
for $D^{(\de)}$ defined by
\be
D^{(\de)} \vphi(x) = \big ( x \cdot \pr_x + \delta\big )\, \vphi(x) \, , 
\qquad
D^{(\de)} \tphi(p) =  - \big ( p \cdot \pr_p + d - \delta\big )\, \tphi(p)  \, ,
\ee
and
\be
D^{(\de)} \vphi \cdot \frac{\delta}{\delta \vphi} = \int \d^d x \;
D^{(\de)} \vphi(x)  \, \frac{\delta}{\delta \vphi(x)} =
\frac{1}{(2\pi)^d} \int \d^d p \;
D^{(\de)} \tphi(p)  \, \frac{\delta}{\delta \tphi(p)}
\label{Deq}
\ee
Subsequently we frequently use
\be
D^{(\de)} \vphi \cdot \psi = - \vphi \cdot D^{(d-\de)} \psi  
= \vphi {\overleftarrow D}{}^{(\de)} \cdot \psi =
- \psi {\overleftarrow D}{}^{(d-\de)} \cdot \vphi \, .
\label{DDbar}
\ee
For a well defined RG equation $\Psi_t$, or ${\hat \Psi}_\Lambda$, should be 
such that $S_t[\vphi] \in \M$ under evolution in $t$ for all  finite $t>0$, so
that $S_t[\vphi]$ is always essentially local, as required for a Wilsonian 
action. In general the resulting equations should  also be in accord
with the expected irreversibility of RG flow towards IR fixed points
so that the detailed form for the initial $S_0$ becomes largely
irrelevant when $t \to \infty$.  Contributions to $S_t[\vphi]$ proportional
to the volume $V$  and independent of $\vphi$ are irrelevant for most purposes.
It is consistent to restrict to equivalence classes defined
by $S_t[\vphi] \sim S_t[\vphi] + c_t \, V$ which justifies the common neglect of
contributions involving $V$.
However such terms are generated by the RG flow in general so it is therefore 
necessary to include $V$ in the basis $\M$ if this is to be closed under RG flow. 
The $V$-dependent terms in \eqref{basic3}, which of course reflect the variation 
in the overall volume due to the rescaling of $x$ in \eqref{scale}, are also 
necessary to ensure a consistent derivative expansion when the leading term in 
$ S_t[\vphi]$ assumes constant $\vphi$.

Although the class of possible $\Psi_t$ satisfying the above conditions 
is not clear cut (a more general discussion is given in appendix A) we 
focus here on the choice due to Wilson which corresponds to taking
\be
\hPsi_\Lambda[\phi] = \frac{1}{2}\, 
{\hat G}_\Lambda \cdot \frac{\delta}{\delta \phi} 
\, \hS_\Lambda[\phi] - {\hat H}_\Lambda \cdot \phi \, ,
\ee
or, imposing agreement with \eqref{scale},\footnote{In the original
proposal the restriction ${\hat H}_\Lambda = - {\hat G}_\Lambda$ was 
also made, requiring by virtue
of \eqref{scGH1} $\delta_0 = \half d$, but this appears unnecessary.}
\be
{\hat G}_\Lambda(y) = \Lambda^{2\delta_0} G(y \Lambda) \, , \qquad 
{\hat H}_\Lambda (y) = \Lambda^d H(y \Lambda )\, ,
\label{scGH1}
\ee
then
\be
\Psi_t[\vphi] = \frac{1}{2}\; G \cdot \frac{\delta}{\delta \vphi}
\, S_t[\vphi] - H \cdot \vphi \, .
\label{scGH}
\ee
In this case \eqref{basic3} becomes
\begin{align}
\bigg ( \frac{\pr}{\pr t} & + \big (D^{(\de_0)} \vphi + H \cdot \vphi \big )
\cdot \frac{\delta}{\delta \vphi} - d \, V\frac{\pr}{\pr V} \bigg ) \, S_t[\vphi]\nn \\
&{} =  \frac{1}{2}\,  \frac{\delta}{\delta \vphi} \, S_t[\vphi]\cdot G
\cdot \frac{\delta}{\delta \vphi} \, S_t[\vphi]
-  \frac{1}{2}\, \frac{\delta}{\delta \vphi} \cdot G \cdot 
\frac{\delta}{\delta \vphi} \, S_t[\vphi] + \tr (H) \, .
\label{basic4}
\end{align}
Clearly it is sufficient to assume $G$ is symmetric, $G=G^T$.
Extending standard results for partial different equations
\eqref{basic4} is a  well defined parabolic functional
differential equation for $S_t$ so long as $G$ is negative definite
and is then soluble for $S_t$ in terms of  $S_{0}$ for $t>0$.
The existence of solutions for general initial $S_{0}$ only for
$t$ increasing reflects the irreversibility of RG flow. The locality
requirements necessitate that ${\tilde G}(p)$ and ${\tilde H}(p)$
should be analytic in $p$ in the neighbourhood of $p=0$.

Apart from the $\tr(H)$ term the equation \eqref{basic4} is invariant under
\be
\tphi(p) \to e^{h(p)} \tphi(p) \, , \qquad
{\tilde G}(p) \to e^{h(p)+h(-p)} {\tilde G}(p) \, , \qquad 
{\tilde H}(p)  \to {\tilde H}(p) + p \cdot \pr_p \, h(p) \, .
\label{freeh}
\ee
For consistency with locality $h(p)$ should also be required to be analytic
in $p$ for $p \approx 0$. The freedom in \eqref{freeh}
shows that ${\tilde H}(p)$ may be chosen at will save for ${\tilde H}(0)$
although in $D^{(\de_0)}\vphi + H \cdot \vphi$ only $\delta_0 + {\tilde H}(0)$ 
has significance.

A variant RG equation due to Polchinski \cite{Polchinski} is obtained by 
writing
\be
S_t [\vphi] = \frac{1}{2} \, \vphi \cdot \G^{-1} \cdot \vphi  
- \frac{1}{2d} \, \tr \big ( G \cdot \G^{-1} \big )   + \S_t [\vphi] \, ,
\label{pol1}
\ee
for $\G$ symmetric and with ${\tilde \G}(p)$ chosen in due course as the 
regularised propagator ensuring a finite functional integral.
Substituting in \eqref{basic4} gives
\begin{align}
\bigg ( \frac{\pr}{\pr t}&  + \big (D^{(\de_0)} \vphi - G \cdot \G^{-1} \cdot
\vphi + H \cdot \vphi \big ) \cdot \frac{\delta}{\delta \vphi} 
- d \, V\frac{\pr}{\pr V} \bigg ) \, \S_t[\vphi] \nn \\
= {}&  \frac{1}{2}\,  \frac{\delta}{\delta \vphi} \, \S_t[\vphi]\cdot G
\cdot \frac{\delta}{\delta \vphi} \, \S_t[\vphi]
-  \frac{1}{2}\, \frac{\delta}{\delta \vphi} \cdot G \cdot
\frac{\delta}{\delta \vphi} \, \S_t[\vphi] \nn \\ 
\noalign{\vskip 2pt}
&{} - \big (D^{(\de_0)} \vphi + H \cdot \vphi \big ) \cdot \G^{-1} \cdot \vphi 
+ \tfrac{1}{2} \, \vphi \cdot \G^{-1}\cdot G \cdot \G^{-1}  \cdot \vphi 
+ \tr \big ( H -  G \cdot \G^{-1} \big ) \, ,
\label{basic5}
\end{align}
Choosing now
\be
H =  G \cdot \G^{-1} + \half \eta \, 1 \, , 
\label{Heq0}
\ee
and also requiring
\be
D^{(d-\de_0)} \, \G^{-1} + \G^{-1} {\overleftarrow D}{}^{(d-\de_0)} = 
\G^{-1}\cdot G \cdot \G^{-1}  \, ,
\label{Geq}
\ee
\eqref{basic5} becomes the Polchinski RG equation \cite{Polchinski}, extended 
to include the parameter $\eta$ \cite{Ball},
\begin{align}
\bigg ( \frac{\pr}{\pr t}&  + D^{(\de)} \vphi
\cdot \frac{\delta}{\delta \vphi}- d \, V\frac{\pr}{\pr V} \bigg ) \, 
\S_t[\vphi] \nn \\
= {}&  \frac{1}{2}\,  \frac{\delta}{\delta \vphi} \, \S_t[\vphi]\cdot G
\cdot \frac{\delta}{\delta \vphi} \, \S_t[\vphi]
-  \frac{1}{2}\, \frac{\delta}{\delta \vphi} \cdot G \cdot
\frac{\delta}{\delta \vphi} \, \S_t[\vphi] 
- \frac{1}{2} \, \eta \, \big ( \vphi \cdot \G^{-1} \cdot \vphi - \tr (1) \big )\, ,
\label{basic6}
\end{align}
for a modified scaling dimension
\be
\de = \de_0 + \half \eta \, .
\label{dede}
\ee

Since $\G^{-1} \cdot (
D^{(d-\de_0)} \, \G^{-1} + \G^{-1} {\overleftarrow D}{}^{(d-\de_0)}) \cdot \G^{-1}
= - D^{(\de_0)} \, \G - \G  {\overleftarrow D}{}^{(\de_0)}$ from \eqref{Geq}
\be
D^{(\de_0)} \, \G + \G  {\overleftarrow D}{}^{(\de_0)} = - G \, ,
\label{Geq2}
\ee
which implies, for translation invariant $\G$,
\be
\big ( x \cdot \pr_x + 2\delta_0 ) \, \G(x) = - G(x) \, , \qquad
\big ( p \cdot \pr_p +  d - 2\delta_0 )\, {\tilde \G}(p) = {\tilde G}(p) \, .
\label{Geq3}
\ee
In order to ensure that IR behaviour is not modified by the introduction 
of a cut off the regularised propagator $\G$ must have the same
long distance behaviour as for a free field, so that $\delta$ has then
to be  identified with the canonical dimension of the scalar field,  giving
\be
\delta_0 = \half ( d-2) \, ,
\label{deq}
\ee
and with standard conventions this requires
\be
 {\tilde \G}(p)  = \frac{K(p^2)}{p^2} \, , \quad K(0)=1 \, , 
\quad K(p^2) \tolim{p^2 \to \infty} 0 \, ,
\label{GK}
\ee
for $K(p^2)$ a smooth cut off function, analytic in $p^2$ in the
neighbourhood of $p^2=0$. The expression given by \eqref{GK} for $\G$,
\eqref{Geq3} then determines $G$ in terms of the cut off $K$
\be
{\tilde G}(p) = 2 K'(p^2) \, .
\label{GKp}
\ee
Alternatively
\be
{\tilde \G}_\Lambda(p)  = \frac{K(p^2/\Lambda^2)}{p^2} \quad
\Rightarrow \quad {\dot \G} \equiv \frac{\pr}{\pr t} {\G}_\Lambda \, 
\Big |_{\Lambda=1} = G \, .
\ee
These results ensure that $ \tr \big ( G \cdot \G^{-1} \big ) /2d$ is equal,
neglecting divergent surface terms arising in integration by parts, to
$ \half \, \tr \, \ln ( \G^{-1} ) $ so this contribution in \eqref{pol1}
just cancels the one loop functional determinant arising from the
functional integration of $ \frac{1}{2} \, \vphi \cdot \G^{-1} \cdot \vphi$.

In terms of the cut off function $K$  the Polchinski RG equation 
\eqref{basic6} may then be written explicitly as
\begin{align}
\bigg (\frac{\pr}{\pr t} + D^{(\de)} \vphi
\cdot \frac{\de}{ \de \vphi} - d \, V\frac{\pr}{\pr V}\bigg ) \S_t = {} & 
 \frac{1}{(2\pi)^d}\int \d^dp \, K'(p^2) \bigg ( 
\frac{\delta \S_t}{\delta \tphi(p)}\, \frac{\delta \S_t}{\delta \tphi(-p)}
- \frac{\delta^2 \S_t} {\delta \tphi(p)\, \delta \tphi(-p)} 
\bigg ) \nn \\
&{}  -  \frac{\eta}{2(2\pi)^d} \int \d^d p \;  \bigg ( \tphi(p) \, 
\frac{p^2}{K(p^2)} \, \tphi(- p) - V \bigg )  \, .
\label{rg5}
\end{align}
For this parabolic equation to be well defined for increasing $t$
we must have
\be
K'(p^2) < 0 \, ,
\label{Kdash}
\ee
which is compatible with \eqref{GK}. By combining \eqref{dede} with \eqref{deq}
we have
\be
\de = \half (d-2+\eta) \, .
\label{deeq}
\ee

For $\eta=0$ the Polchinski equation \eqref{basic6} is of course
identical with the Wilson equation \eqref{basic5} if we set $H=0$
although there is now, by virtue of \eqref{pol1}, a Gaussian solution 
for $\S_t=0$. 

In each of the RG flow equations the coefficient functions,
$G,H$ in \eqref{basic4} or $G,\G^{-1}$ in \eqref{basic6}, have
no short distance singularities, as a consequence of the
analyticity assumptions for ${\tilde G}(p),{\tilde H}(p)$, so 
that $S_t$  or $\S_t$ should not contain any such singularities under $t$
evolution for finite $t$. In this sense the  RG flow preserves locality.

\subsection{Functional Space}

For non linear differential equations it is necessary to specify
the class of functions and associated boundary conditions which 
form the appropriate solution space.
The functional differential RG equations described here are assumed to act
on a infinite dimensional vector space of action functionals $\M$ 
spanned by monomials in $\vphi$ 
\be
\big \{ \, V , \P_n[\vphi] ,  n=1,2,\dots \big \} \, , \quad
\P_n[\vphi ] = \int {\textstyle \prod_{r=1}^n \d^d x_r \, \vphi(x_r)} \;
G_n(x_1,\dots , x_n)\, ,
\label{mono}
\ee
where $\{\P_n[\vphi]\}$ are defined in terms of symmetric functions 
$\{G_n(x_1, \dots,x_n)\}$  which are restricted 
so that each $G_n$ is translation invariant 
\be
G_n(x_1+a, \dots , x_n +a)  = G_n(x_1,\dots , x_n) \, , 
\ee
and also connected requiring
\be
G_n(x_{1,a}, \dots , x_{n,a}) \to 0 \quad \mbox{as} \quad a\to \infty \quad
x_{r,a} = \begin{cases}x_r + a \, , & r\in S\subset\{1, \dots ,n\}\\
x_r \, , & r\in \{1, \dots, n\} \setminus S\end{cases} \, ,
\ee
for all proper subsets $S$. Furthermore the cut off should ensure that
all $G_n$ are quasi-local in that they are
non singular at short distances, $G_n(x_1,\dots , x_n)$ is regular
for any $x_r-x_s\to 0$, $r\ne s$.  Equivalently 
\be
\int {\textstyle \prod_{r=1,n}} \d^d x_r \; e^{i \, p_r \cdot x_r} \;
G_n(x_1,\dots , x_n) = (2\pi)^d \delta^d \big (
{\textstyle \sum_{r=1,n}} p_r \big ) \, {\tilde G}_n(p_1,\dots , p_n) \, ,
\ee
where $ {\tilde G}_n(p_1,\dots , p_n) $, restricted to $\sum_{r=1,n}p_r=0$, 
is a symmetric function  analytic in 
each $p_r$ in the neighbourhood of $p_r=0$. By expanding 
$ {\tilde G}_n(p_1,\dots , p_n) $ about $p_r=0$ we may consider a basis of
local operators $\M_{\rm local}$ where $G_n$ in \eqref{mono} is restricted
so that $G_n(x_1,\dots , x_n) = 0$ if $x_{rs} \ne 0$, for all $r\ne s$, or
${\tilde G}_n(p_1,\dots , p_n)$ is just a polynomial in each $p_r$. The
local operators then have the form
\be
 \P_n[\vphi] = \int \d^d x \; P_{n,s}[\vphi;x] \, ,
\ee
for
\be
P_{n,s}[\vphi;x] = {\textstyle \prod} \, \pr^{s_i} \vphi(x)
= {\rm O}(\vphi^n, \pr^s) \, , \qquad s = {\textstyle \sum}_i s_i \, ,
\label{localphi}
\ee
depending only on $\vphi(x)$ and its derivatives, all indices contracted to 
form a  scalar. Of course such local operators \eqref{localphi} may be 
extended to include also operators with non zero spin.

For functionals of finite order in $\vphi$ as in  \eqref{mono}
then an alternative normal ordered form, relative to a two point
function $\G$, is defined by
\be
\N_{\G}(\P_n[\vphi]) = e^{- \frac{1}{2}  \,
\frac{\de}{\de \vphi} \cdot \G \cdot \frac{\de}{\de \vphi} }\P_n[\vphi] \, .
\label{normal}
\ee
In perturbative expansions, with a propagator $\G$, this removes
contractions between different $\vphi$ in $\P_n[\vphi]$.
In \eqref{basic6} we may write $\vphi \cdot \G^{-1} \cdot \vphi - \tr (1) = 
\N_\G ( \vphi \cdot \G^{-1} \cdot \vphi )$. Although the $\tr(1)$ term 
involves a divergent $p$-integration, as exhibited in \eqref{rg5}, its 
subtraction ensures contributions arising from 
$\vphi \cdot \G^{-1} \cdot \vphi$ are well defined in later manipulations.

\subsection{Linearisation of RG Equations}
\label{linear}

The exact RG equations are non linear but as shown by Rosten \cite{Rosten},
and also in  \cite{Morris}, they
may be linearised and the RG flow is then determined by standard
$\beta$-functions.  Defining
\be
\Y = \frac{1}{2}  \, 
\frac{\de}{\de \vphi} \cdot \G \cdot \frac{\de}{\de \vphi} \, ,
\label{GY}
\ee
which may be expressed in an analogous fashion to \eqref{IG}, then, 
using \eqref{Geq2} with $\delta_0$ given by \eqref{deq},
\be
\Big [ D^{(\de_0)} \vphi \cdot \frac{\delta}{\delta \vphi}  , \, \Y \, \Big ] = 
\frac{1}{2}\, \frac{\de}{\de \vphi} \cdot G \cdot \frac{\de}{\de \vphi}  \, ,
\label{DY}
\ee
so that
\be
D^{(\de_0)} \vphi \cdot \frac{\delta}{\delta \vphi} \, e^\Y
= e^\Y \bigg ( D^{(\de_0)} \vphi \cdot \frac{\delta}{\delta \vphi} 
+ \frac{1}{2}\, \frac{\de}{\de \vphi} \cdot G \cdot \frac{\de}{\de \vphi} \bigg ) \, .
\label{DYeq}
\ee
Hence from \eqref{basic6} for $\eta=0$, 
\be
\bigg ( \frac{\pr}{\pr t} \, + \, D^{(\de_0)} \vphi \cdot 
\frac{\delta}{\delta \vphi}  - d \, V \frac{\pr}{\pr V} \bigg ) 
\Big ( e^\Y \, e^{-S_t[\vphi]} \Big ) = 0 \, .
\label{St1}
\ee
So long as $\G$ is positive definite, requiring  $K(p^2)>0$,
$e^\Y$ has a well defined action but is not invertible for any general 
$\S_t[\vphi]$. For $t=0$ 
\be
 e^\Y \, e^{-\S_0[\vphi]} = e^{T[\vphi] + c \, V} \, ,
\label{FT0}
\ee
where $T[\vphi]+ c \, V $, $\pr_V T[\vphi]=0, \, \delta_\vphi c=0$,  may be 
evaluated by a perturbative expansion in terms  of the contributions  for all 
connected vacuum Feynman graphs with propagators given by ${\tilde \G}(p)$, as in 
\eqref{GK} and which are singular at $p^2=0$, and vertices 
determined by $\S_0 [\vphi]$ which may be assumed to be a conventional action
formed by a finite rotationally invariant polynomial in the scalar field 
$\vphi$ and its derivatives. As shown later $T[\vphi]$ is equivalent 
to the normal vacuum functional $W[J]$ and is inherently non local.  
In terms of $T[\vphi]$ the linear 
equation \eqref{St1} for the $t$ dependence is easily solved giving
\be
 e^\Y \, e^{-\S_t[\vphi]} = e^{T[\vphi_t] + c \, e^{dt} \, V}\, , 
\label{FT}
\ee
for
\be 
\vphi_t(x) = e^{-\frac{1}{2}(d-2)t} \; \vphi (e^{-t} x)  \quad \mbox{or} \quad
\tphi_t(p) = e^{\frac{1}{2}(d+2)t} \; \tphi (e^{t} p) \, .
\label{phit}
\ee
In terms of the original field $\phi$ before rescaling $\vphi_t(x) 
= \Lambda_0 {\!}^{-\frac{1}{2}(d-2)} \phi ( x /\Lambda_0)$.

\eqref{FT} and \eqref{FT0} may be combined to give
\be
e^\Y \, e^{-\S_t[\vphi]} =  e^{\Y_t} \, e^{-\S_0[\vphi_t]} 
\; e^{c( e^{dt} -1 ) V } \, ,
\label{YYS}
\ee
where 
\be
\Y_t  = \frac{1}{2}  \,
\frac{\de}{\de \vphi_t} \cdot \G \cdot \frac{\de}{\de \vphi_t} \, ,\qquad
\frac{\de}{\de \vphi_t(x)} 
= e^{-\frac{1}{2}(d+2)t} \frac{\de}{\de \vphi(e^{-t}x)} \, .
\ee
\eqref{YYS} may be solved for $\S_t$ in the form \cite{Keller,Ball2,Golner}
\be
e^{-\S_t[\vphi]}  = e^{\Y_t - \Y} \,  e^{-\S_0[\vphi_t]} \; e^{c( e^{dt} -1 ) V } \, ,
\label{Ssol1}
\ee
since
\be
\Y_t - \Y = \  \frac{1}{2}  \, \frac{\de}{\de \vphi_t} \cdot \X_t 
\cdot \frac{\de}{\de \vphi_t} \, , \qquad {\tilde \X}_t(p) =
\frac{1}{p^2}\big ( K(e^{-2t}p^2) - K(p^2) \big ) >0 \ \ \mbox{for} \ \ t>0 \, ,
\label{Ssol2}
\ee
as a consequence of \eqref{Kdash}. Hence $e^{\Y_t - \Y}$ is well defined
so long as $t\ge 0$. The solution verifies that $\S_t$ does not have
singularities for $p\approx 0$ for suitably local $\S_0$.

Such linearisation allows a direct connection with standard linear
RG flow equations involving $\beta$-functions.
In general a RG flow $\S_t[\vphi]$ determines a trajectory in an 
infinite dimensional space of all possible $\{\S[\vphi]\}$ actions 
consistent with the symmetries of the initial $\S_0[\vphi]$.
Coordinates on this space may be identified with an infinite set of couplings 
$\{g\}$ so that from \eqref{FT}
\be
\S_t[\vphi] = \S[\vphi;g_t]  \quad \Rightarrow \quad T[\vphi_t] = 
T[\vphi;g_t] \, .
\label{STt}
\ee
Clearly $T[\vphi;g] = T[\vphi_{-t};g_t]$. Defining the $\beta$-functions as 
usual as the tangent vectors to the RG flow of the couplings $g_t$,
\be
\frac{\d}{\d t}\, g_t =  \beta(g_t) \, ,
\label{betaf}
\ee
then \eqref{St1} and \eqref{STt} require
\be
\bigg ( D^{(\de_0)}  \vphi \cdot \frac{\delta}{\delta \vphi} -
 d \, V \frac{\pr}{\pr V} 
+ \beta(g) \cdot \frac{\pr}{\pr g} \bigg ) T[\vphi;g] = 0 \, ,
\label{CS1}
\ee
which is a standard linear RG equation for 
$\beta(g) \cdot \frac{\pr}{\pr g} = \beta^i(g) \frac{\pr}{\pr g^i}$.

If instead we assume the $\S_t$  evolves according to the Polchinski
equation  \eqref{basic6} with $\eta \ne 0$ then using
\be
e^\Y \Big (  \vphi \cdot \G^{-1} \cdot \vphi - \tr(1)  - 
\vphi \cdot \frac{\de}{ \de \vphi} \Big ) =
 \Big ( \vphi \cdot \G^{-1} \cdot \vphi + 
\vphi \cdot \frac{\de}{ \de \vphi} \Big ) \,  e^\Y \, ,
\label{Yeq}
\ee
which follows from
\begin{align}
\Big [ \Y , \, \vphi \cdot  \frac{\de}{ \de \vphi} \Big ] = {}& 2 \Y \, , 
\qquad \big [ \Y , \, \vphi \cdot \G^{-1} \cdot \vphi \big ] = 
2 \vphi \cdot \frac{\de}{ \de \vphi} + \tr(1) \, ,
\end{align}
\eqref{St1} is replaced by
\be
\bigg ( \frac{\pr}{\pr t} \, + \, 
D^{(\de-\eta)}   \vphi \cdot \frac{\de}{ \de \vphi}
- d \, V \frac{\pr}{\pr V} - \half \eta \, \vphi \cdot  \G^{-1} \cdot \vphi \bigg )
\Big ( e^\Y \, e^{-\S_t[\vphi]} \Big ) = 0 \, .
\label{St2}
\ee
The solution \eqref{FT} becomes instead
\be
 e^\Y \, e^{-\S_t[\vphi]} = e^{T[\vphi_t] + \frac{1}{2} \, \vphi_t \cdot h
\cdot \vphi_t -  \frac{1}{2} \, \vphi \cdot h \cdot \vphi + c \, e^{dt} \,V } \, ,
\label{FT2}
\ee
where now
\be
\vphi_t(x) = e^{-\frac{1}{2}(d-2-\eta)t} \; \vphi (e^{-t} x) \, ,
\label{scphi}
\ee
and we require
\be
D^{(\de-\eta)}  \vphi \cdot \frac{\de}{ \de \vphi} 
\, \big ( \vphi \cdot h \cdot \vphi \big)  = 
- \eta \, \vphi \cdot \G^{-1} \cdot \vphi \, ,
\label{heq}
\ee
which is equivalent to
\be
D^{(d-\de+\eta)} \, h + h {\overleftarrow D}{}^{(d-\de+\eta)} =
\eta\; \G^{-1}  \, .
\label{heq2}
\ee
In \eqref{FT2} we have imposed the initial condition that the
vacuum functional $T[\vphi]$ is given by $e^\Y e^{-\S_0[\vphi]}$ and so
is identical with $T$ in \eqref{FT0}. Writing $\vphi \cdot h \cdot \vphi$
in the form given by  \eqref{XYG} 
then \eqref{heq2} requires
\be
\Big ( 1 + \half \eta - p^2 \frac{\pr}{\pr p^2} \Big ) \tilde h(p) 
= \half \eta \, \frac{p^2}{K(p^2)} \, .
\ee
This has a solution, so long as $\eta <2$,
\be
{\tilde h}(p) = \frac{p^2}{K(p^2)} \big ( 1 - \sigma_*(p^2) \big ) \, , \qquad
\sigma_*(p^2) = K(p^2)   (p^2)^{\frac{1}{2}\eta} \int_0^{p^2}
\!\! \d x \; x^{-\frac{1}{2}\eta} \frac{\d}{\d x} \frac{1}{K(x)} \, ,
\label{hsol}
\ee
where we have imposed the condition that $h(p)$ should be analytic, for
general $\eta$,  when $p\approx 0$ to eliminate solutions of the 
homogeneous equation proportional to $(p^2)^{1+\frac{1}{2}\eta}$. The expression
\eqref{hsol} gives for any $\eta<2$
\be
\sigma_*(p^2) \sim - \frac{K'(0)}{1 - \half \eta} \, p^2  \, , \quad h(p) \sim p^2
\qquad \mbox{as} \quad p^2 \to 0 \, ,
\label{s0}
\ee
As a special case
\be
\sigma_*(p^2)\big |_{\eta=0} = 1-K(p^2) \, , \qquad
{\tilde h}(p) \big |_{\eta=0} =p^2 \, .
\label{hzero}
\ee
 If $\eta<0$ then it is possible to integrate
by parts so as to combine the two terms in the expression for ${\tilde h}(p)$ in
\eqref{hsol} into a single integral.

For the Polchinski RG equation with $\eta\ne 0$  there is still a flow in 
the space of couplings as determined by \eqref{STt} but the trajectory 
$g_t$ is
now modified as is therefore the beta function $\beta(g,\eta)$, 
given by \eqref{betaf}, which now depends on $\eta$. In this case
\be
T[\vphi_t;g] + \tfrac{1}{2} \, \vphi_t \cdot h
\cdot \vphi_t -  \tfrac{1}{2} \, \vphi \cdot h \cdot \vphi = T[\vphi;g_t] \, ,
\ee
and instead of \eqref{CS1} we now have
\be
\bigg ( D^{(\de-\eta)} \,  \vphi \cdot \frac{\de}{ \de \vphi} -
 d \, V \frac{\pr}{\pr V} +
 \beta(g,\eta) \cdot \frac{\pr}{\pr g} \bigg ) T[\vphi;g] = \half
 \eta \, \vphi \cdot \G^{-1} \cdot \vphi \ \, ,
\label{CS2}
\ee
although $T[\vphi;g]$ is unchanged. However the apparent dependence on $\eta$
is superfluous. For a general set of couplings $\{g\}$ rescaling of the
field $\vphi$ may be compensated by a corresponding change in $g$ giving
\be
\bigg ( \vphi \cdot \frac{\de}{ \de \vphi}+ \kappa(g)\cdot \frac{\pr}{\pr g} \bigg )
\S[\vphi;g] = - \vphi \cdot \G^{-1} \cdot \vphi \, ,
\ee
for some appropriate $\kappa(g)$. Equivalently $\S[\vphi;g] = \S[e^s \vphi;g_s]
+ \half (e^{2s}- 1) \,  \vphi \cdot \G^{-1} \cdot \vphi$ where $\frac{\d}{\d s}g_s
= \kappa(g_s)$.
Using \eqref{Yeq} this ensures
\be
\bigg ( \vphi \cdot \frac{\de}{ \de \vphi} -  
\kappa(g)\cdot \frac{\pr}{\pr g} \bigg )
T[\vphi;g] = - \vphi \cdot \G^{-1} \cdot \vphi \, ,
\label{Tdiff}
\ee
Applying \eqref{Tdiff} reduces \eqref{CS2} to  \eqref{CS1} so long as
\be
\beta(g,\eta) = \beta(g) - \half \eta \, \kappa(g) \, .
\ee

\subsection{IR Fixed Points}

In the above discussion $\eta$ is undetermined and may be chosen at will.
The crucial constraint, which provides a determination of $\eta$, is that 
there should exist a well defined non zero non trivial limit as $t\to \infty$,
\be
\S_t [ \vphi ] \tolim{t\to \infty} \S_*[\vphi] = \S[\vphi;g_*] \, .
\label{Slim}
\ee
where from \eqref{basic6} we must have
\begin{align}
E[\vphi] \equiv {}& D^{(\de)} \vphi \cdot \frac{\delta}{\delta \vphi}  \S_*[\vphi] 
- d \, V \frac{\pr}{\pr V} \S_*[\vphi] 
 - \frac{1}{2}\,  \frac{\delta}{\delta \vphi} \, \S_*[\vphi]\cdot G
\cdot \frac{\delta}{\delta \vphi} \, \S_*[\vphi]
+ \frac{1}{2}\, \frac{\delta}{\delta \vphi} \cdot G \cdot
\frac{\delta}{\delta \vphi} \, \S_*[\vphi] \nn \\
= {}& -  \frac{1}{2} \, \eta \; \big ( \vphi \cdot \G^{-1} \cdot \vphi - \tr(1) \big ) \, ,
\label{fxS}
\end{align}
with $\de,\eta$ related by \eqref{deeq}.
In general non trivial IR fixed points, with long range order, 
are only possibly if $\S_0 [ \vphi ]$
is restricted to belong to a critical surface of codimension equal to
$N_{\rm rel}>0$, the number of relevant operators at the particular fixed 
point. There may of course be more than one IR fixed point, with differing 
possible $N_{\rm rel}$ and $\eta$, which may be obtained
as $t\to \infty$ depending on the precise initial  $\S_0 [ \vphi ]$.

In \eqref{FT2} then, with $\vphi_t$ given by \eqref{scphi}, it
is easy to verify that
\be
\vphi_t \cdot h \cdot \vphi_t \tosim{t\to \infty}  
e^{\eta\, t} \, \vphi \cdot \G_0{\!}^{-1} \cdot \vphi \, \qquad
\vphi \cdot \G_0{\!}^{-1} \cdot \vphi  = 
\frac{1}{(2\pi)^d} \int \d^d p \; \vphi(p) \, p^2 \,\vphi(-p)\, ,
\ee
where
\be
\G_0(y)= \frac{k}{(y^2)^{\frac{1}{2}(d-2)}}\, , \qquad 
{\tilde \G}_0(p)=\frac{1}{p^2} \, ,
\label{Gz}
\ee
is just the Green function for $-\pr^2$, 
which is independent of the cut off and so satisfies, instead of \eqref{Geq2},
\be
D^{(\de_0)} \, \G_0 + \G_0  {\overleftarrow D}{}^{(\de_0)} = 0  \, .
\label{Geq0}
\ee 

In order to ensure that \eqref{FT2} is compatible with \eqref{Slim}
it is necessary to require
\be
 T[\vphi_t]  \tosim{t\to \infty} - e^{\eta\, t} \half \, 
 \vphi \cdot \G_0{\!}^{-1} \cdot \vphi   + T_*[\vphi] \, .
\label{Tlim}
\ee
Except for the local term proportional to $e^{\eta t}$, which diverges for
$\eta>0$,  \eqref{Tlim} shows that $T[\vphi_t]$ has well defined limit as 
$t \to \infty$ if there is a non trivial IR fixed point.
Such a limit can only be possible
for one precise value of $\eta$ which, as demonstrated later,
can be identified with the anomalous scale dimension of $\vphi$ at the
fixed point. For a non trivial limit in \eqref{Tlim}  $T_*$ must
scale invariant satisfying $T_ *[\vphi_t] = T_ *[\vphi] $ and from 
\eqref{Slim}, \eqref{Tlim} we then have
\be
e^{T_*[\vphi] -  \frac{1}{2} \, \vphi \cdot h \cdot \vphi} 
= e^\Y \; e^{-\S_*[\vphi]} \, ,
\label{Tstar}
\ee
and $T[\vphi;g_*]= T_*[\vphi] - \half \, \vphi \cdot h \cdot \vphi$.
Assuming
\be
\beta(g_*,\eta) = 0 \, ,
\ee
with $\eta$ determined by the existence of a non trivial limit, then using \eqref{CS2} with
\eqref{heq} ensures that
\be
\bigg ( D^{(\de-\eta)} \vphi \cdot \frac{\de}{ \de \vphi} -  d \, V \frac{\pr}{\pr V} \bigg )
T_*[\vphi] = 0 \, .
\label{Tlinear}
\ee

As the fixed point is approached it is standard to assume
\be
\S_t[\vphi] \sim \S_*[\vphi] - \sum_{n\ge 0} 
e^{-\lambda_n t} \, \O_n[\vphi] \quad \mbox{as} \quad t \to \infty \, ,
\label{asymS}
\ee
for $\{ \O_n \}$ the set of all scaling operators and $\lambda_n=\Delta_n -d$,
where $\Delta_n$ is the scale dimension of an associated local operator, determine
the critical exponents. These may be ordered so that  
$\lambda_{n+1} > \lambda_n$  with $\lambda_0 = -d$  and $\O_0 \propto V$ 
corresponding to the identity operator, which is  independent of $\vphi$.
In general $\lambda_1= -\half(d+2-\eta)$ when $\O_1$ is essentially just
$\vphi$ itself. For $\lambda_n< 0$, which defines relevant operators, then 
$\O_n$, $n=0,1, \dots N_{\rm rel}-1$,  in general corresponds to a 
composite operator constructed from $\vphi^n$.
If there is a ${\mathbb Z}_2$ symmetry under 
$\vphi \leftrightarrow -\vphi$ so that $\S_0[\vphi]= \S_0[-\vphi]$ then the
sum in \eqref{asymS} may be restricted to even $n$ corresponding to
operators $\O_n[\vphi] = \O_n[- \vphi]$.
The critical surface corresponds to no contributions in \eqref{asymS}
with $\lambda_n <0$ so that \eqref{Slim} is well defined. This provided
$N_{\rm rel}$ conditions on $\S_0[\vphi]$, or equivalently the initial $g$. 
Under RG flow the couplings $g_t$ are therefore restricted to a critical 
subspace of codimension $N_{\rm rel}$ in the total space of couplings.

In conjunction with \eqref{FT} \eqref{asymS} would require $T[\vphi_t]$ to 
have contributions $\propto e^{-\lambda_n t}$ as $t\to \infty$ where 
in general the differing $\{\lambda_n\}$ are not commensurate. Such terms
may correspond to the divergences present at coincident points reflecting
the singular coefficient functions in the operator product expansion.

\subsection{Eigenvalue Equations}
\label{eigen}

The determination of the exponents $\lambda_n$ in \eqref{asymS} is
associated with an eigenvalue problem
\be
\Delta_{\S_*}  \O = \lambda \, \O \, ,   \qquad \O \in \M \, ,
\label{eig}
\ee
where $\Delta_{S}$  is the functional differential operator depending
on action functionals $S[\vphi] \in \M$
\be
\Delta_{S} = 
D^{(\de)} \vphi \cdot \frac{\delta}{\delta \vphi}  - d \, V \frac{\pr}{\pr V}
- \frac{\delta}{\delta \vphi} \, S[\vphi]\cdot G
\cdot \frac{\delta}{\delta \vphi} 
+ \frac{1}{2}\, \frac{\delta}{\delta \vphi} \cdot G \cdot
\frac{\delta}{\delta \vphi} \, ,
\label{DelS}
\ee
and in \eqref{eig} $S\to \S_*$, the fixed point action determined by \eqref{fxS},
with $\de$ determined in terms of $\eta$ by \eqref{deeq}.

Associated with \eqref{eig} there is a corresponding eigenvalue problem for 
local operators $\M' = \{ \Phi[\vphi;x] \}$, which are functionals of 
$\vphi$ depending also on $x$. $\M'$  is assumed to have a similar 
basis to \eqref{mono} for $\M$ but $V\to 1$, the identity operator, and
$G_n(x_1,\dots , x_n)\to G_n(x;x_1,\dots , x_n)$ which are similarly 
translationally invariant under a translation in $x$ and each $x_r$, 
connected, quasi-local and symmetric functions of $(x_1,\dots,x_n)$.
Additionally $\M'_{\rm{local}}$ is such that $\Phi(x) \in \M'_{\rm{local}}$
is formed just from $\vphi(x)$ and its derivatives as in \eqref{localphi}.
The eigenvalue equation determining the scaling dimension $\Delta$ becomes
\be
\Delta_{\S_*, {\rm loc}} \Phi_\Delta(x) = D^{(\Delta)} \Phi_\Delta(x) =
\big (x \cdot \pr + \Delta \big )\, \Phi_\Delta(x) \, ,
\label{loceig}
\ee
for $\Delta_{\S_*,{\rm loc}} $, acting on $\M'$, given by
\be
\Delta_{\S_*,{\rm loc}} =
D^{(\de)}  \vphi \cdot \frac{\delta}{\delta \vphi} 
- \frac{\delta}{\delta \vphi} \, \S_*[\vphi]\cdot G
\cdot \frac{\delta}{\delta \vphi}
+ \frac{1}{2}\, \frac{\delta}{\delta \vphi} \cdot G \cdot
\frac{\delta}{\delta \vphi} \, .
\label{DelSloc}
\ee
This is  the same operator as in \eqref{eig} but omitting the 
$V$ dependent terms.  Manifestly $\Delta_{\S_*,{\rm loc}} \, 1 =0$.
Defining $ 1 \cdot \Phi = \int \rmd^d x \; \Phi(x) = {\tilde \Phi}(0)$, so that 
$1\cdot 1 = V$, then from \eqref{loceig}  there is an
associated $\O\in \M$ satisfying \eqref{eig} since
\be
\Delta_{\S_*} \, ( 1 \cdot \Phi_\Delta ) =  
1 \cdot \Delta_{\S_*, {\rm loc}} \Phi_\Delta - d \, V \frac{\pr}{\pr V}
 ( 1 \cdot \Phi_\Delta )  = (\Delta - d) \, 1 \cdot \Phi_\Delta \, .
\label{LO}
\ee

For two cases exact local eigenoperators can be constructed starting from
\be
\Delta_{\S_*,{\rm loc}}\, \vphi = D^{(\de)} \vphi - G \cdot 
 \frac{\de}{\de \vphi} \S_*  \, ,
\ee
 and, by taking the derivative of \eqref{fxS},
\be
\Delta_{\S_*,{\rm loc}}\,  \frac{\de}{\de \vphi} \S_* =
{D}^{(d-\de)}\,  \frac{\de}{\de \vphi} \S_* 
- \eta \; \G^{-1} \cdot \vphi = 0 \, .
\label{diffS}
\ee
Assuming the form \eqref{PXY} then \eqref{loceig} requires
\begin{align}
X {\overleftarrow D}{}^{(d-\de)} + 
D^{(\Delta)} X = {}& - \eta \; Y \cdot \G^{-1} \, , \nn \\
Y {\overleftarrow D}{}^{(\de)} + 
D^{(\Delta)} Y  ={}&  X \cdot G \, ,
\label{XYeq}
\end{align}
where for locality ${\tilde X}(p),{\tilde Y}(p)$ are analytic for $p \approx 0$.
Using \eqref{Geq2}, \eqref{heq2} and also \eqref{Geq0} the solutions of 
\eqref{XYeq} give
\begin{subequations}
\begin{align}
\Phi_\delta = {}& \G_0 \cdot \Big ( h \cdot \vphi - (1 - h \cdot \G) \cdot
\frac{\de}{\de \vphi} \S_* \Big ) 
= \G_0 \cdot \Big ( \G^{-1} \cdot \vphi - (1 -  h \cdot \G) \cdot
\frac{\de}{\de \vphi} S_* \Big ) \, , \label{philoca} \\
\Phi_{d-\delta} = {}& \G_0{\!}^{-1}  \cdot \Big ( \vphi + \G  \cdot
\frac{\de}{\de \vphi} \S_* \Big ) =  \G_0{\!}^{-1}  \cdot  \G  \cdot
\frac{\de}{\de \vphi} S_*  \, .
\label{philoc}
\end{align}
\end{subequations}
The quasi-locality of $\Phi_\delta, \Phi_{d-\delta}$ follows since
$\G_0{\!}^{-1}  \cdot \G, \, \G_0 \cdot \G^{-1}$ and $\G_0 \cdot (1 -  h \cdot \G)$
have no singularities at $p=0$ (from \eqref{hsol} and \eqref{s0}
$1 - {\tilde \G}(p)\, {\tilde h}(p) = \sigma_*(p^2) = {\rm O}(p^2)$ 
as $p^2 \to 0$).
Corresponding to \eqref{philoca} and \eqref{philoc}, according to \eqref{LO},
are scaling operators belonging to $\M$ satisfying \eqref{eig} 
\begin{subequations}
\begin{align}
1 \cdot \Phi_\delta = {}& 1\cdot \vphi + \frac{1}{2-\eta}\;  1 \cdot G \cdot  
\frac{\delta}{\delta \vphi} \, \S_*[\vphi]  \, , \qquad \lambda =
- \half ( d + 2 - \eta ) \, , \label{exacta} \\
1 \cdot \Phi_{d-\delta} = {}& 1  \cdot  
\frac{\delta}{\delta \vphi} \, \S_*[\vphi]  \, , \hskip 3.6cm
\lambda =- \half ( d - 2 + \eta ) \, .
\label{exact}
\end{align}
\end{subequations}
For $\eta=0$, $h\to \G_0{\!}^{-1}$, and \eqref{philoca} becomes
\be
\Phi_{\delta_0} = \vphi - (\G_0 - \G) \cdot
\frac{\de}{\de \vphi} \S_* \, .
\label{Phi0}
\ee

Imposing
\be
\G_0{\!}^{-1}\cdot \Phi_{\delta_0} = Z \, \Phi_{d- \delta_0} \, ,
\label{eom}
\ee
restricts $\S_*$ to a Gaussian form
\be
\S_*[\vphi] = \half \, \vphi \cdot F_* \cdot \vphi  + c_* \, V \, ,
\qquad F_* = (1-Z) \, \big ( \G_0 - (1-Z) \, \G \big )^{-1} \, ,
\label{gauss0}
\ee
where $c_* V = \tr( G \cdot F_* ) /2d $  is determined by requiring $E[\vphi]=0$, 
according to \eqref{fxS}.
However when $d - \delta_0 = n \delta_0$ \eqref{eom} may be relaxed.

The space of operators $\{\O\}\subset \M$ includes a subspace $\{\O_\psi\}$ of 
redundant operators of the form
\be
\O_\psi = \psi \cdot  \frac{\de}{\de \vphi} \S_* 
 + \psi \cdot \G^{-1} \cdot \vphi  -  \frac{\de}{\de \vphi} \cdot \psi 
= \psi \cdot  \frac{\de}{\de \vphi} S_* -  \frac{\de}{\de \vphi} \cdot \psi \, ,
\label{red}
\ee
for $\psi[\phi;x] \in V_\phi$ and $\S_*$ related to $S_*$ as in \eqref{pol1}.
Redundant operators as in 
\eqref{red} may be regarded as associated with the freedom of choice for 
$\Psi_t$ in the RG flow equations as in \eqref{basic3}.  Clearly
$1 \cdot \Phi_{d-\delta}$ in \eqref{exact} is a redundant operator.

To show that $\{\O_\psi\}$ form a closed subspace we consider
\begin{align}
& \Delta_{\S_*} \bigg ( \psi \cdot  \frac{\de}{\de \vphi} \S_* -  
\frac{\de}{\de \vphi} \cdot \psi \bigg )\nn \\
& = \big ( \Delta_{\S_*,{\rm loc}} \psi  - D^{(\de)} \psi \big ) 
\cdot  \frac{\de}{\de \vphi} \S_* - \frac{\de}{\de \vphi} \cdot
 \big ( \Delta_{\S_*,{\rm loc}} \psi - D^{(\de)}  \psi \big ) 
-  \eta \, \psi \cdot \G^{-1} \cdot \vphi  \, ,
\label{aux1}
\end{align}
with $\Delta_{\S_*,{\rm loc}} $ as in \eqref{DelSloc}.
Furthermore, using \eqref{Geq},
\begin{align}
\Delta_{\S_*} \big ( \psi \cdot \G^{-1} \cdot \vphi  \big )
= {}& \big ( \Delta_{\S_*,{\rm loc}} \psi  - D^{(\de)}  \psi 
- G \cdot \G^{-1} \cdot \psi \big ) \cdot \G^{-1} \cdot \vphi \nn \\
&{} - \big ( G \cdot \G^{-1} \cdot \psi \big ) \cdot \frac{\de}{\de \vphi} \S_* 
+ \frac{\de}{\de \vphi} \cdot \big ( G \cdot \G^{-1} \cdot \psi \big )
+ \eta \, \psi \cdot \G^{-1} \cdot \vphi  \, .
\label{aux2}
\end{align}
With the definition \eqref{red} combining \eqref{aux1} and \eqref{aux2} then 
gives
\be
 \Delta_{\S_*} \, \O_\psi = 
\O_{{\tilde \Delta}_{\S_*}\, \psi} \, ,
\label{delpsi}
\ee
for
\be
{\tilde \Delta}_{\S_*} \, \psi = 
\Delta_{\S_*,{\rm loc}} \psi - G \cdot \G^{-1} \cdot \psi -
D^{(\de)}  \psi \, .
\label{tDel}
\ee
It is  then clear from 
\eqref{delpsi} that for redundant operators \eqref{eig} is equivalent to
\be
{\tilde \Delta}_{\S_*} \, \psi = \lambda \, \psi \, . 
\label{peig}
\ee

Using \eqref{Geq3} and \eqref{Geq0} then from \eqref{tDel}
\begin{align}
\G_0 \cdot \G^{-1} \cdot {\tilde \Delta}_{\S_*} \, \psi = {}& 
\Delta_{\S_*,{\rm loc}} \; \G_0\cdot \G^{-1} \cdot \psi -
D^{(\de)} \, 
\big  ( \G_0 \cdot \G^{-1} \cdot \psi \big )  \, .
\label{psieq}
\end{align}
Hence, for any local operator  $\Phi_\Delta$ satisfying \eqref{loceig}, 
\eqref{psieq} ensures that
\be
\Delta_{\S_*} \O_{\G \cdot \G_0{\!}^{-1} \cdot \Phi_\Delta}
= ( \Delta - \delta )\, \O_{\G \cdot \G_0{\!}^{-1} \cdot \Phi_\Delta} \, ,
\label{redop}
\ee
giving for each $\Phi_\Delta$ an associated redundant operator.

For non redundant operators
$\O$, which are defined so as to be linearly independent of all
$\{\O_\psi\}$, \eqref{eig} may be relaxed to require only
\be
\Delta_{\S_*} \O = \lambda \, \O + \O_\chi \, ,
\label{eigO2}
\ee
for some redundant operator $\O_\chi$ of the form \eqref{red},
as  \eqref{eigO2} may be reduced to \eqref{eig} since
\be
\Delta_{\S_*} ( \O + \O_\psi)  = \lambda \, (\O + \O_\psi)  \quad \mbox{for}
\quad \big ({\tilde \Delta}_{\S_*} - \lambda \big ) \psi = \chi \, .
\label{eigO3}
\ee
Assuming no accidental degeneracies of the eigenvalues of
non redundant and redundant operators ${\tilde \Delta}_{\S_*} - \lambda$ is
then invertible, so long as  $\O$ is not redundant, and $\psi$ may be found
in terms of $\chi$.

Using the operator $\Delta_{\S_*}$ equations for the variation in the fixed
point action $\S_*$ consequent on variations in the initial $\Psi$ 
which determine the RG equation may be found. Here $\Psi$ is given by
\eqref{scGH} and \eqref{Heq0}. For a variation of $\eta$ in \eqref{fxS}
\be
\Delta_{\S_*} \de \S_* = - \half \, \de \eta \, \O_\vphi \, ,
\label{var1}
\ee
and for variations in $G$ and hence $\G$, which are related by
\eqref{Geq}, then
\be
\Delta_{\S_*} \big (  \de \S_* + \half \, \vphi \cdot \de \, \G^{-1} \cdot
\vphi - \de\, \tr ( G \cdot \G^{-1} ) / 2d \big ) = 
\Delta_{\S_*} \de S_* = \O_\psi \, ,
\label{var2}
\ee
for
\be
\psi  = \frac{1}{2} \,  \de G \cdot \frac{\de} {\de \vphi} \S_* + 
\frac{1}{2} \, \de G \cdot \G^{-1} \cdot \vphi - 
 G \cdot \de \G^{-1} \cdot \vphi \, ,
\ee
and $\Delta_{\S_*}$ given by \eqref{DelS}.
Solutions of \eqref{var1} and \eqref{var2} lie within the space of 
redundant operators. However as discussed subsequently $\Delta_{\S_*}$ 
is expected to have  in general a non
trivial cokernel at non trivial fixed points. The quantisation
of $\eta$ at the fixed point requires that \eqref{var1} should have 
no solution for $ \de \S_*$.

Corresponding to the transformation $\S_* \to T_*$ in \eqref{Tstar}, which
 gives rise to a linear fixed point equation as in \eqref{Tlinear}, there
is an associated transformation for any $\Phi \in \M'$
with $\Phi(x) \to P_\Phi(x)$ given by
\be
P_\Phi(x) \, e^{T_*[\vphi]} = e^{\frac{1}{2} \, \vphi \cdot h \cdot \vphi} 
\; e^\Y \, \big (  e^{-\S_*[\vphi]}\,  \Phi(x) \big ) \, .
\label{defPP}
\ee
This transformation may also be extended to $\O \to P_\O$ for $\O\in \M$
but as shown later this need not always be well defined.
Directly from the definition \eqref{defPP} the identity is invariant, $P_1 = 1$.  
Using
\begin{align}
e^{-S_*[\vphi]} \Delta_{\S_*,{\rm loc}} \Phi  = {}&
\bigg ( D^{(\de)} \vphi \cdot \frac{\delta}{\delta \vphi}  - 
d \, V \frac{\pr}{\pr V} + \frac{1}{2}\, \frac{\delta}{\delta \vphi} \cdot G 
\cdot 
\frac{\delta}{\delta \vphi} \bigg ) \big ( e^{-S_*[\vphi]}\,\Phi \big )\nn \\ 
&{} + E[\vphi]  \,  e^{-S_*[\vphi]} \Phi \, ,
\end{align}
for $E[\vphi]$ given by \eqref{fxS}, with \eqref{DYeq} and \eqref{Yeq}
then
\be
e^{\frac{1}{2} \, \vphi \cdot h \cdot \vphi} 
\, e^\Y \, \big (  e^{-\S_*[\vphi]} \, \Delta_{\S_*,{\rm loc}} \Phi(x) \big ) 
= \bigg ( D^{(\de-\eta)} \vphi \cdot \frac{\delta}{\delta \vphi}  - 
d \, V \frac{\pr}{\pr V} \bigg ) \, e^{\frac{1}{2} \, \vphi \cdot h \cdot \vphi} 
\,  e^\Y \, \big (  e^{-\S_*[\vphi]} \,  \Phi(x) \big ) \, .
\ee
Hence \eqref{defPP} gives
\be
D^{(\de-\eta)} \vphi \cdot \frac{\delta}{\delta \vphi} \, P_\Phi =
P_{  \Delta_{\S_*,{\rm loc}} \Phi}
\ee
For an eigenoperator satisfying \eqref{loceig} then
\be
D^{(\de-\eta)} \vphi \cdot \frac{\delta}{\delta \vphi} \, P_{\Phi_\Delta} =
D^{(\Delta)} P_{ \Phi_\Delta} \, .
\label{DPP}
\ee
This equation is independent of $\S_*$ and does not therefore determine 
$\Delta$ in any non trivial fashion,
which is consistent with the transformation $\Phi \to P_\Phi$ not being
invertible as $e^{-\Y}$ acting on arbitrary functionals is ill defined.   

To consider the transformation of the exact eigenoperators
$\Phi_\delta$ and $\Phi_{d-\delta}$, defined in \eqref{philoca}
and \eqref{philoc}, we first obtain from \eqref{defPP} and \eqref{Tstar}
\begin{align}
P_\vphi ={}&  ( 1 - \G \cdot h ) \cdot \vphi + \G \cdot 
\frac{\delta}{\delta \vphi} T_*[\vphi] \, , \nn \\
P_{\frac{\delta}{\delta \vphi} \S_*[\vphi] } ={}&  h \cdot \vphi -
\frac{\delta}{\delta \vphi} T_*[\vphi]  \, .
\end{align}
Hence
\begin{subequations}
\begin{align}
P_{\Phi_\delta} = {}& \G_0 \cdot \frac{\delta}{\delta \vphi} T_*[\vphi] \, ,  
\label{P1} \\
P_{\Phi_{d-\delta}} = {}& \G_0{\!}^{-1} \cdot \vphi \, . \label{P2}
\end{align}
\end{subequations} 
Using \eqref{Tlinear} and \eqref{Geq0} it is straightforward to check that
\be
D^{(\de-\eta)} \vphi \cdot \frac{\delta}{\delta \vphi} \, P_{\Phi_\delta}
= D^{(\delta)} P_{\Phi_\delta} \, , \qquad
D^{(\de-\eta)} \vphi \cdot \frac{\delta}{\delta \vphi}  P_{\Phi_{d-\delta}}  = 
D^{(d-\delta)} P_{\Phi_{d-\delta}}  \, .
\ee

\eqref{P1}, \eqref{P2}  demonstrate that the transformation
$\O \to P_\O$, given by \eqref{defPP},
does not extend to arbitrary $\O \in \M$, since from \eqref{P1}
$1 \cdot P_{\Phi_\delta} $ is singular while from  \eqref{P2} 
$1 \cdot P_{\Phi_{d-\delta}}$ is zero while there are no such problems
for $1\cdot \Phi_\delta$ and $1\cdot \Phi_{d-\delta}$.  However for
a redundant operator, as in \eqref{red}, \eqref{defPP} gives
\be
P_{\O_\psi} = \vphi \cdot \G^{-1} \cdot P_\psi \, .
\label{redPP}
\ee
Using \eqref{Geq}
\begin{align}
D^{(\de-\eta)} \vphi \cdot \frac{\delta}{\delta \vphi} 
\big (  \vphi \cdot \G^{-1} \cdot P_\psi \big ) = {}&
 \vphi \cdot \G^{-1} \cdot \bigg ( 
 D^{(\de-\eta)} \vphi \cdot \frac{\delta}{\delta \vphi}
 - D^{(\delta)} - G \cdot \G^{-1} \cdot  \! \bigg )  P_\psi  \nn \\
 ={}&   \vphi \cdot \G^{-1} \cdot   P_{{\tilde \Delta}_{\S_*}\psi} \, ,
 \end{align} 
 as expected according to \eqref{delpsi} with ${\tilde \Delta}_{\S_*}$
 defined in \eqref{tDel}.

\subsection{Zero Modes}
\label{zerom}

Amongst the scaling operators appearing in the asymptotic expansion
\eqref{asymS} the zero modes, or marginal operators, $\Z$ for which
$\lambda_\Z=0$ so that
\be
\Delta_{\S_*} \Z = 0 \, ,
\label{Zeq}
\ee
are of especial interest. The fixed point action $\S_*[\vphi]$
is then arbitrary to the extent 
$\S_*[\vphi] \sim \S_*[\vphi] - \vep \, \Z[\vphi]$ for infinitesimal $\vep$. 
If this is integrable there is an associated line of fixed points. 
For simple scalar theories a  marginal operator $\Z$ can be obtained
by considering particular reparameterisations of the scalar field $\vphi$ but  are
redundant since they are then removable by such a redefinition, or 
equivalently are zero subject to the dynamical field equations. If the line
of fixed points generated by this redundant $\Z$ is parameterised by a 
variable $a$ then all $\S_*[\vphi;a]$ are equivalent. 

For a redundant operator then
\be
\Z = \O_{\psi_\Z} \, ,
\label{ZO}
\ee
for appropriate $\psi_\Z$.
Using \eqref{redop} and \eqref{philoca} the zero mode operator is determined
in terms of $\Phi_\delta$ by taking
\begin{align}
\psi_\Z= {}& \G \cdot \G_0{\!}^{-1} \cdot \Phi_\delta 
= \G \cdot h \cdot \vphi + 
( \G \cdot h\cdot \G - \G ) \cdot \frac{\delta}{\delta \vphi} \, \S_*[\vphi] \nn \\
= {}& \vphi + ( \G \cdot h\cdot \G - \G ) \cdot \frac{\delta}{\delta \vphi} \, 
S_*[\vphi] \, .
\label{zpsi}
\end{align}
This result for $\psi_\Z$ gives, using 
\eqref{red}, an  expression for $\Z[\vphi]$ identical with 
the operator constructed by  O'Dwyer and Osborn \cite{DO}. 

Although $\psi_\Z$ and hence $\Z$ is relatively complicated under
the transformation $\Z \to P_\Z$ given by \eqref{defPP} there are
significant simplifications. Using \eqref{zpsi} to express $\psi_\Z$
in terms of $P_\delta$ and then \eqref{P1} with \eqref{redPP} gives
\be
P_\Z[\vphi] =  \vphi \cdot \frac{\delta}{\delta \vphi} \, T_*[\vphi]  \, .
\label{PZ}
\ee
This demonstrates that 
letting $\S_*[\vphi] \to \S_*[\vphi] - \vep \,  \Z[\vphi]$ 
induces the associated transformation $T_*[\vphi] \to 
T_*[\vphi] +\vep \,  \vphi \cdot  \frac{\de}{\de \vphi} T_*[\vphi]$ and hence
that the line IR fixed points generated by $\Z$ corresponds
to a simple rescaling of $\vphi$ in $T_*$,
$T_*[\vphi] \sim T_*[\lambda \vphi]$ for all $\lambda >0$.

The existence of a zero mode $\Z$ satisfying \eqref{Zeq} is potentially
crucial in ensuring that the anomalous dimension $\eta$ is determined
at an IR fixed point. This requires that \eqref{var1}, corresponding to
varying $\eta$, has no solution. The presence of a zero mode demonstrated 
that $\Delta_{\S_*}$ has a non trivial kernel while the lack of solutions 
to \eqref{var1} in general would be a consequence of a non zero cokernel.
These would be identical if it were possible to construct a scalar product
with respect to which $\Delta_{\S_*}$ was self adjoint.

Associated with the marginal operator $\Z$ there is a corresponding
local operator $\Phi_\Z(x)$ satisfying \eqref{loceig} with $\Delta=d$.
To determine this explicitly we first construct a bilocal
functional $\F[\vphi;x,y]$ satisfying
\be
\Delta_{\S_*, {\rm loc}} \, \F(x,y)  = D_x{\!}^{(d-\delta)} \F(x,y)
+ \F(x,y) {\overleftarrow D}_y{\!}^{(\delta)} \, .
\label{Feig}
\ee
Suppressing the $x,y$ arguments this is satisfied by taking
\begin{align}
\F = \Phi_{d-\de} \; \Phi_\de + \G_0{\!}^{-1} \cdot \G \cdot S_*{\!}^{(2)}
\cdot ( 1 - \G \cdot h ) \cdot \G_0 + c \, \I \nn \\
= \Big ( \Phi_{d-\de} -  \G_0{\!}^{-1} \cdot \G \cdot \frac{\de}{\de \vphi}
\Big ) \, \Phi_\de +  (c+1) \, \I \, ,
\label{FFeq}
\end{align}
where $\Phi_{d-\de},  \Phi_\de$ are given by \eqref{philoc},  \eqref{philoca}
albeit the latter in \eqref{FFeq} is rewritten in terms of its transpose
\be
\Phi_\delta =  \vphi \cdot\G^{-1}  \cdot \G_0 - \frac{\de}{\de \vphi} S_*
\cdot  (1 -  \G \cdot h)  \cdot \G_0 \, ,
\ee
and  $S_*{\!}^{(2)}$ is also the symmetric bilocal functional defined by
\be
S_*{\!}^{(2)}[\vphi;x,y]  =
\frac{\delta^2}{\delta \vphi(x) \, \delta \vphi(y)} \, S_*[\vphi]  \, .
\ee
In \eqref{FFeq} $\I \to \de^d(x-y)$ which satisfies \eqref{Feig} trivially
since $\Delta_{\S_*, {\rm loc}} \, \I = 0$,
so that $c$ is, for the moment, an unconstrained constant. To verify that 
\eqref{FFeq} satisfies \eqref{Feig} it is sufficient to use
\be
\Delta_{\S_*, {\rm loc}} \, \Phi_{d-\de} \; \Phi_\de  - 
D^{(d-\delta)} \Phi_{d-\de} \; \Phi_\de -
\Phi_{d-\de} \; \Phi_\de {\overleftarrow D}{}^{(\delta)} =
\Phi_{d-\de} \, \overleftarrow{\frac{\de}{\de\vphi}} \cdot G \cdot 
\frac{\de}{\de\vphi} \, \Phi_\de \, , 
\ee
and, from  \eqref{fxS},
\begin{align}
\Delta_{\S_*, {\rm loc}}& \, S_*{\!}^{(2)} - 
D^{(d-\delta)}  S_*{\!}^{(2)}  -  S_*{\!}^{(2)}  {\overleftarrow D}{}^{(d-\delta)}
\nn \\
= {}& 
- \G^{-1} \cdot G \cdot  S_*{\!}^{(2)} -  S_*{\!}^{(2)} \cdot G \cdot \G^{-1}
+  S_*{\!}^{(2)} \cdot G \cdot  S_*{\!}^{(2)} \, ,
\end{align}
together with identities such as \eqref{Geq2}, \eqref{heq2} and \eqref{Geq0}.
In the same fashion as for $\Phi_{d-\de}$ and $\Phi_\de$, $\F$ is also
quasi-local. It is then sufficient from \eqref{Feig} to take
\be
\Phi_\Z (x) = \F(x,x) \, ,
\ee
so long as the coincident limit is non singular. This determines $c$, if
the leading term in $S_*{\!}^{(2)}$ is $\G^{-1}$ then $c=-1$. Note
that also
\be
\Z = \tr ( \F ) \, .
\ee

If we consider $\Phi_\Z \to P_{\Phi_\Z}$, as determined by \eqref{defPP}, then
\be
P_{\Phi_\Z} =  P_{\Phi_{d-\de}} \; P_{\Phi_\de} \, ,
\ee
where $ P_{\Phi_{d-\de}} \; P_{\Phi_\de}$ are given by \eqref{P1},\eqref{P2}.
In this case taking the coincident limit of $P_{\Phi_{d-\de}}(x) \; 
P_{\Phi_\de}(y)$ causes no problems and it is also trivial that
$D^{(\de-\eta)} \vphi \cdot \frac{\delta}{\delta \vphi} \, P_{\Phi_\Z} =
D^{(d)} P_{ \Phi_\Z}$ from \eqref{DPP} for $P_{\Phi_{d-\de}}$ and $P_{\Phi_\de}$.
However it is non trivial that $P_{\Phi_\Z}$ corresponds to a quasi-local
$\Phi_\Z$.

\subsection{Solution of RG Equations with a Source}

In quantum field theory it is natural to introduce a source term
$e^{\hJ \cdot \phi}$,  $\hJ \cdot \phi = \int \d^d x \; \hJ(x) \phi(x)$,  
into the functional integral so that all correlation functions are obtained
in terms of functional derivatives with respect to $\hJ$. In this
context we therefore consider RG flow equations starting from
${\hat S}_{\Lambda_0}[\phi,\hJ] = {\hat S}_{\Lambda_0}[\phi]- \hJ \cdot \phi$,
or following \eqref{scale}, \eqref{Sscale} and \eqref{deq},
\be
\S_0[\vphi,J] = \S_0[\vphi] - J \cdot \vphi \, , \qquad 
\hJ(x) = \Lambda^{\frac{1}{2}(d+2)} J(x\Lambda) \, ,
\label{init}
\ee
so that $\hJ \cdot \phi = J \cdot \vphi$. 
The RG flow equation \eqref{basic6} may be extended to the form
\begin{align}
\bigg ( \frac{\pr}{\pr t}&  + D^{(\de)} \vphi
\cdot \frac{\delta}{\delta \vphi} + D^{(d- \de)}  J
\cdot \frac{\delta}{\delta J} - d \, V\frac{\pr}{\pr V} 
\bigg ) \, \S_t[\vphi,J] \nn \\
= {}&  \frac{1}{2}\,  \frac{\delta}{\delta \vphi} \, \S_t[\vphi,J]\cdot G
\cdot \frac{\delta}{\delta \vphi} \, \S_t[\vphi,J]
-  \frac{1}{2}\, \frac{\delta}{\delta \vphi} \cdot G \cdot
\frac{\delta}{\delta \vphi} \, \S_t[\vphi,J] 
- \frac{1}{2} \, \eta \; \big ( \vphi \cdot \G^{-1} \cdot \vphi - \tr(1) \big )\, .
\label{SJeq}
\end{align}

We here show how a solution for $\S_t[\vphi,J]$, with the initial
condition \eqref{init}, may be found in terms of $\S_t[\vphi]$ 
by taking
\be
\S_t[\vphi,J] = \S_t[ \vphi + \B_t \cdot J ] 
- \half \, J \cdot \C_t \cdot J - J \cdot \D_t \cdot \vphi \, .
\label{SPJ}
\ee
Substituting \eqref{SPJ} into \eqref{SJeq} and using \eqref{basic6}
gives a solution so long as
\begin{align}
\frac{\pr}{\pr t} \, \B_t - D^{(\de)} \, \B_t - \B_t  \, 
{\overleftarrow D}{}^{(\de)} +  G \cdot \D_t {\!}^T= 0 \, , \nn \\
\frac{\pr}{\pr t} \, \D_t  - D^{(\de)} \,  \D_t - \D_t \,  
{\overleftarrow D}{}^{(d-\de)} + \eta \; \B_t{\!}^T \cdot \G^{-1}  = 0 \, , \nn \\
\frac{\pr}{\pr t} \, \C_t - \D^{(\de)} \, \C_t - \C_t \, 
{\overleftarrow D}{}^{(\de)} + \eta \; \B_t {\!}^T \cdot \G^{-1} \cdot \B_t 
+ \D_t \cdot G \cdot \D_t{\!}^T = 0 \, ,
\label{KCD}
\end{align}
with similar definitions to \eqref{DDbar}, and to ensure \eqref{init}
\be
\B_0 = 0 \, , \qquad \C_0 = 0 \, , \qquad \D_0 = 1 \, .
\ee
It is easy to see that it is sufficient to take
\be
\C_t = \D_t \cdot \B_t \, ,
\ee
so that  \eqref{KCD} may be reduced to just the coupled equations
\begin{align}
\frac{\pr}{\pr t} \, {\tilde \B}_t(p)
+ ( p \cdot \pr_p +2 - \eta ) \,  {\tilde \B}_t(p) = {}& - 2 K'(p^2) \,
{\tilde \D}_t (-p)\, , \nn \\
\frac{\pr}{\pr t} \, {\tilde \D}_t (p) + p \cdot \pr_p \, {\tilde \D}_t (p)
= {}& - \eta \, \frac{p^2}{K(p^2)} \, {\tilde \B}_t(-p) \, .
\end{align}
From this we may obtain
\be
\frac{\d}{\d t} \Big ( k(t) \, {\tilde \D}_t (e^t p) +
e^{2t}\, {\tilde \B}_t(-e^t p) \Big ) = 0 \, ,
\label{DKeq}
\ee 
for
\be
\quad k(t) = \frac{K(e^{2t}p^2)}{p^2} \, .
\label{keq}
\ee
Using the solution of \eqref{DKeq} we find
\be
\frac{\d}{\d t}\, {\tilde \D}_t (e^t p) - \eta \, {\tilde \D}_t (e^t p)
= - \eta\, \frac{k(0)}{k(t)} \, .
\ee
It is then easy to obtain
\begin{align}
{\tilde \D}_t ( p) = \frac{K( e^{-2t} p^2)}{K(p^2)} 
\big ( 1 - \sigma_t(p^2) \big ) \, , \qquad
{\tilde \B}_t( p) = \frac{ K(e^{-2t} p^2)} {p^2} \; \sigma_t(p^2)   \, ,
\end{align}
for 
\be
\sigma_t(p^2) = - K(p^2)  \int_0^t \d s \;
e^{\eta s} \, \frac{\d}{\d s}\frac{1}{K(e^{-2s}p^2)} \, .
\label{sigeq}
\ee
When $\eta=0$ this simplifies to
\be
{\tilde \D}_t ( p) = 1 \, , \qquad {\tilde \B}_t( p) = \frac{1}{p^2}
\Big ( K(e^{-2t} p^2) - K(p^2) \Big ) \, .
\ee

In the limit $t\to \infty$ 
\be
{\tilde \D}_* ( p) = \frac{1}{K(p^2)} 
\big ( 1 - \sigma_*(p^2) \big ) \, , \qquad
{\tilde \B}_*( p) = \frac{ 1} {p^2} \; \sigma_*(p^2)   \, , 
\ee
where $\sigma_*(p^2)$ is defined in \eqref{hsol}.
Hence
\be
\S_t[\vphi,J] \big |_{t\to \infty} \to \S_*[\vphi,J] = \S_*[\vphi + \B_* \cdot J]
- \half \, J \cdot \D_*\cdot \B_* \cdot J - J \cdot \D_* \cdot \vphi \, .
\label{Ssol}
\ee

This existence of a fixed point action $\S_*[\vphi,J]$ for arbitrary $J$
is a reflection that $J$ only couples to $\vphi$ and the  result essentially
corresponds just to a shift in $\vphi$.

\subsection{Relation of $T[\vphi]$ to Vacuum Functional $W[J]$}

In standard quantum field theory it is conventional to introduce the
generating functional $W$ for all $n$-point connected correlation functions
$W$ is given by the contributions for all connected vacuum
Feynman graphs in the presence of the source $J$ so that
\be
G^{(n)}(x_1, \dots , x_n) = \frac{\de^n}{\de J(x_1) \dots \de J(x_n)} \, 
W[J] \, \bigg |_{J=0} \, .
\label{corr}
\ee
Standard functional manipulations \cite{Coleman} show that, with $\Y$ given by 
\eqref{GY},
\be
e^{W[J]} = e^{\Y} \, e^{-\S_0[\vphi] + J \cdot \vphi}\; \Big |_{\vphi=0} \, .
\label{Wdef}
\ee

Following Rosten \cite{Rosten} we show how this is linked with $T[\vphi]$ 
defined by \eqref{FT}. It is straightforward to extend \eqref{St2}, starting 
from \eqref{SJeq}, to
\be
\bigg ( \frac{\pr}{\pr t} + 
D^{(\de-\eta)}  \vphi \cdot \frac{\de}{ \de \vphi}
+ D^{(d-\de)} J \cdot \frac{\de}{ \de J}
- \half \eta \, \vphi \cdot  \G^{-1} \cdot \vphi \bigg )
\Big ( e^\Y \, e^{-\S_t[\vphi,J]} \Big ) = 0 \, ,
\label{SJ2}
\ee
which has a similar solution to \eqref{FT2}
\be
 e^\Y \, e^{-\S_t[\vphi,J]} = e^{T[\vphi_t,J_t] + \frac{1}{2} \, \vphi_t \cdot h
\cdot \vphi_t -  \frac{1}{2} \, \vphi \cdot h \cdot \vphi +  c \, e^{dt} \,V } \, ,
\label{TJ2}
\ee
for $\vphi_t$ as in \eqref{scphi} and
\be
J_t (x) = e^{-\frac{1}{2}(d+2 -\eta)t} \, J(e^{-t}x) \, .
\ee
However using the Baker-Campbell-Hausdorff formula in the form
\be
e^{\Y} \, e^{J \cdot \vphi} =  
e^{J \cdot \, \left ( \vphi +  \G\cdot \frac{\de}{\de \vphi} \right )}  \; e^{\Y} 
= e^{\frac{1}{2}  J\cdot \G \cdot J + J \cdot  \vphi}  \;
e^{J \cdot \G\cdot \frac{\de}{\de \vphi} } \; e^\Y  \, ,
\label{BCH}
\ee
it follows from \eqref{init} and \eqref{FT2} that, for $t=0$, $T[\vphi,J]$ is
expressible just in terms of $T[\vphi]$
\be
T[\vphi,J] =  \tfrac{1}{2} \, J\cdot \G \cdot J + J \cdot \vphi + 
T[\, \vphi +  \G \cdot J \,] \, .
\label{TTe}
\ee 
so that in \eqref{TJ2} for any $t\ge 0$
\be
T[\vphi_t,J_t] = \tfrac{1}{2} \, J_t \cdot \G \cdot J_t + J_t \cdot \vphi_t + 
T[\, \vphi_t +  \G \cdot J_t \,] \, .
\label{TTeq}
\ee

The previous discussion of the behaviour at an IR fixed point can now
be extended to  $T[\vphi,J]$ since \eqref{Tlim} becomes
\be
 T[\vphi_t,J_t]  \tosim{t\to \infty} - e^{\eta\, t} \half \, 
 \vphi \cdot \G_0{\!}^{-1} \cdot \vphi   + T_*[\vphi,J] \, .
\label{Tlim2}
\ee
Using \eqref{Tlim} with \eqref{TTeq} with
\be
J_t \cdot \vphi_t = e^{\eta t} \, J \cdot \vphi \, , \qquad
J_t \cdot \G \cdot J_t  \tosim{t\to \infty} e^{\eta\, t} \, 
J \cdot \G_0 \cdot J \, 
\ee
since
\be
\G(e^t y )  \tosim{t\to \infty} e^{-(d-2)t} \, \G_0(y) \, ,
\label{Gas}
\ee
for $\G_0$ as in \eqref{Gz}, then
\be
T_*[\vphi , J] = T_*[\vphi + \G_0 \cdot J ] \, .
\label{Tsol}
\ee

In a similar fashion to \eqref{Tstar}
\be
e^{T_*[\vphi,J] -  \frac{1}{2} \, \vphi \cdot h \cdot \vphi} = 
e^\Y \; e^{-\S_*[\vphi,J]} \, .
\label{Tstar2}
\ee
Compatibility of this result with  \eqref{Tsol} and the solution for
$-\S_*[\vphi,J]$ in \eqref{Ssol} may be checked by
applying the Baker-Campbell-Hausdorff formula just as in \eqref{BCH} to give
\begin{align}
e^\Y \; e^{-\S_*[\vphi,J]} ={}& e^{\frac{1}{2}  J\cdot 
(\D_* \cdot \B_*+ \D_* \cdot \G  \cdot \D_*{\!}^T) \cdot J + 
J \cdot \D_* \cdot  \vphi}  \;
e^{J \cdot \D_* \cdot \G\cdot \frac{\de}{\de \vphi} } \; e^\Y \;
e^{-\S[\vphi+ \B_* \cdot J] } \nn \\
={}& e^{\frac{1}{2}  J\cdot (\D_*  \cdot \B_* +  \D_* \cdot \G  \cdot 
\D_*{\!}^T) \cdot J + J \cdot \D_* \cdot  \vphi}  \;
e^{J \cdot \D_* \cdot \G\cdot \frac{\de}{\de \vphi} } \; 
e^{T_*[\vphi + \B_* \cdot J] -  \frac{1}{2} \, 
(\vphi + J \cdot \B_*{\!}^T)  \cdot h \cdot (\vphi + \B_* \cdot J ) } \, ,
\end{align}
as a consequence of $\B_* + \G \cdot \D_*{\!}^T = \G_0$, 
$\D_* = \G_0\cdot h$
and $\D_* \cdot \B_* + \D_* \cdot \G \cdot \D_*{\!}^T = 
\G_0 \cdot h \cdot \G_0 $,
noting that, from \eqref{hsol}, ${\tilde h}(p) = 
(1- \sigma_*(p^2))/{\tilde \G}(p)$.

For the generating functional for connected correlation functions $W[J]$, 
\eqref{Wdef} and \eqref{TJ2}, \eqref{TTe} give
\be
W[J] = \tfrac{1}{2} \, J\cdot \G \cdot J + T[\, \G \cdot J \,] \, ,
\label{WT}
\ee
and using \eqref{Tlim}
\be
W[J_t]   \tosim{t\to \infty} W_*[J] = T_*[ \, \G_0 \cdot J] \, ,
\label{WTlim}
\ee
This implies for the correlation functions defined in \eqref{corr}
\be
G^{(n)}\big (e^t x_1, \dots , e^t x_n \big ) \tosim{t\to \infty} 
e^{-\frac{1}{2}n(d-2+\eta)t}\, G_*^{(n)}( x_1, \dots ,  x_n) \, ,
\label{Glim}
\ee
with, from \eqref{WTlim},
\be
G_*^{(n)} \big ( x_1, \dots ,  x_n \big ) = 
\prod_{r=1}^n \int \d^d x'{\!}_r \; \G_0(y_r) \;
\frac{\de^n}{\de \vphi(x'{\!}_1) \dots \de \vphi(x'{\!}_n)}\, T_*[\vphi] \, 
\bigg |_{\vphi=0}\, ,
\ee
for $y_r = x_r - x'{\!}_r$.

Since \eqref{Glim} requires $G_*^{(n)}\big (e^t x_1, \dots , e^t x_n \big )
= e^{-\frac{1}{2}n(d-2+\eta)t}\, G_*^{(n)}( x_1, \dots ,  x_n)$ then this
is just the expected scaling behaviour at an IR critical point if $\eta$ is
identified with the $\phi$ anomalous dimension.

\subsection{Gaussian Solution}

Although perhaps somewhat trivial it is illustrative to consider a Gaussian
solution which is quadratic in the fields of the form
\be
\S_t[ \vphi] = \half \, \vphi \cdot F_t \cdot \vphi + c_t V =
\frac{1}{2(2\pi)^d} \, \int \d^d p\; \tphi(p) \, \tF_t(p) \, \tphi(-p) 
+ c_t V \, ,
\label{gauss}
\ee
where $F_t = F_t{\!}^T$.  Substituting \eqref{gauss} into \eqref{basic6}, 
or equivalently \eqref{rg5}, gives an evolution equations for $F_t$ and $c_t$,
\begin{subequations}
\begin{align}
\frac{\pr}{\pr t}  \tF_t(p) = {}& ( 2 - \eta - p \cdot \pr_p )  \tF_t(p) 
+ 2 K'(p^2) \,  \tF_t(p) ^2 - \eta \, \frac{p^2}{K(p^2)} \, , \label{gauss2} \\
\bigg ( \frac{\pr}{\pr t} - d \bigg ) c_t \, V  = {}& 
- \frac{1}{2} \, \tr \big ( G \cdot F_t - \eta \, 1 \big ) \, .
\label{cteq}
\end{align}
\end{subequations}
\eqref{gauss2} can be rewritten as
\be
\frac{\d}{\d t}  \bigg ( e^{-(2-\eta)t} \tF_t(e^t p) + \frac{e^{\eta t}}
{k(t)} \bigg ) = e^{-\eta t}\,  {\dot k}(t)\, \bigg ( \big (
 e^{-(2-\eta)t} \tF_t(e^t p) \big )^2 - \frac{e^{2\eta t}} 
{k(t)^2} \bigg )\, ,
\ee
with $k(t)$ by \eqref{keq}.
This may be solved for any arbitrary initial $\tF_0(p)$ giving
\begin{align}
\frac{p^2/K(p^2)}{\tF_t(p) + p^2/K(p^2)} - \sigma_t(p^2)
= e^{\eta t} \frac{K(p^2)}{K(e^{-2t}p^2)} \;
\frac{p^2/K(e^{-2t} p^2)}{e^{2t} \tF_0(e^{-t} p) + p^2/K(e^{-2t} p^2)} \, ,
\label{solF}
\end{align}
for $\sigma_t$ given by \eqref{sigeq}. The corresponding solution of 
\eqref{cteq} is trivial
\be
c_t \, V = e^{dt} \bigg ( c_0 \, V - \frac{1}{2} \int_0^t \! \d t' \; 
e^{-dt'} \,
 \tr \big ( G \cdot F_{t'} - \eta \, 1 \big )  \bigg ) \, .
\ee
For $\eta=0$ \eqref{solF} may be simplified to
\be
\frac{1}{\tF_t(p)} = \frac{e^{-2t}}{\tF_0(e^{-t} p)} + \frac{1}{p^2}
\Big ( K(e^{-2t} p^2) - K(p^2) \Big ) \, .
\label{solF1}
\ee
For locality it is crucial to assume  $\tF_0(p)$ is
analytic in $p$ for $p \approx 0$. 

If $t\to \infty$ then $\S_t \to \S_*$ where
\be
\S_*[\vphi] = \half \, \vphi \cdot F_*  \cdot \vphi + c_* V  \, ,
\label{Sgauss}
\ee
and, for the limit $c_t \to c_*$ to exist, it is necessary that $c_0$ is 
fine tuned to cancel any $e^{dt}$ terms giving
\be
c_* \, V =   \frac{1}{2d} \, \tr ( G \cdot F_* - \eta 1 ) \, .
\label{cstar}
\ee
  
For $\tF_0(0) \ne 0$, and assuming $\eta<2$, 
from \eqref{solF}
\be
\tF_*(p) = \lim_{t\to \infty} \tF_t(p) = \frac{p^2}{K(p^2)} 
\bigg ( \frac{1}{\sigma_*(p^2)} - 1 \bigg ) \, ,
\label{Fss}
\ee
which is independent of the initial $\tF_0(p)$. Since, from \eqref{s0}, 
$\sigma_*(p^2) \propto p^2$, as $p^2\to 0$, $\tF_*(0)>0$ so this fixed point 
does not lead to any IR long range order. It corresponds to the trivial high 
temperature fixed  point described in the introduction.
If $\tF_0(0)=0$, which defines the critical surface for Gaussian theories, 
so that 
\be
\tF_0(p) = \frac{1}{z} \, p^2 + {\rm O}\big ( (p^2)^2 \big ) \, ,
\ee
and $\eta=0$ also then \eqref{solF1} gives the limit, depending only on $z$,
\be
\tF_*(p) = \frac{p^2}{1+z - K(p^2)} \, .
\ee
This gives rise to an IR fixed point with long range order, the need for 
$\eta=0$ was also made clear by Comellas \cite{Com}.
For $\tF_t(p)$ to be non singular it is necessary that $z> 0$. 
The solutions for $F_*$ in \eqref{Sgauss} are then
\be
F_* = \begin{cases}\ (1- h \cdot \G )^{-1} \cdot h \, , & \eta \ne 0 \, , \\
\ \big ( (1+z) \G_0 - \G \big )^{-1} \, , & \eta =0 \, .
\end{cases}
\label{Fstar}
\ee
Hence from \eqref{philoca} and \eqref{Phi0}
\be
\Phi_\delta =  \begin{cases}\  0 \, , & \eta \ne 0 \, , \\
\ z\, \G_0 \cdot \big (  (1+z) \G_0 - \G \big )^{-1}\cdot \vphi \, , & 
\eta=0 \, , \ \delta= \delta_0 \, .
\end{cases}
\label{PhiG}
\ee
For  $\eta \ne 0$ therefore $\Z=0$, corresponding to the
lack of any condition determining $\eta$. For $\eta=0$ using \eqref{Fstar}
in \eqref{Sgauss} is identical with \eqref{gauss0} for $Z= z/(1+z)$.

The result \eqref{PhiG} can be used to verify the previous formula
for zero modes since from \eqref{zpsi}
\be
\psi_\Z = H \cdot \vphi \, ,  \qquad H = z \, \G \cdot F_*  \, .
\ee
and then \eqref{red} gives
\be
\Z[\vphi] = \vphi \cdot H^T \cdot (F_* + \G^{-1} ) \cdot \vphi - \tr (H) \, ,
\label{ZM}
\ee
where
\be
H^T \cdot (F_* + \G^{-1} ) = z(1+z) \, F_* \cdot \G_0 \cdot F_* 
= - z(1+z)\frac{\pr}{\pr z} \, F_*  \, .
\ee
Since ${\tilde H}(p) = zK(p^2)/ ( 1+z - K(p^2) )$ then
inserting $1=\pr_p \cdot p/d$ and integrating by parts gives
\be
\tr (H) = - z(1+z) \, \frac{2}{d} \, \frac{V}{(2\pi)^d} \int \d^d p \;
\frac{p^2\, K'(p^2) }{(1+z - K(p^2))^2} =  2z(1+z) \frac{\pr}{\pr z}
\, c_* V \, ,
\ee
since, with $\eta=0$,
\be
c_* \, V =  \frac{1}{2d} \, \tr ( G \cdot F_* ) =
\frac{1}{d} \, \frac{V}{(2\pi)^d} \int \d^d p \; \frac{p^2 \, K'(p^2)}
{1+z - K(p^2)} \, .
\label{cstar1}
\ee
Hence
\be
\Z[\phi] = - 2z(1+z) \, \frac{\pr}{\pr z} \S_*[\vphi] \, ,
\ee
in accord with $z$ being a redundant parameter.

Clearly $\S_*$ depends on the cut off function $K$, on the other hand $T_*$
obtained from $\S_*$ according to \eqref{Tstar} is independent of $K$. To 
determine $T_*$ and demonstrate this in the Gaussian case it is sufficient to 
use
\be
e^\Y \, e^{ - \frac{1}{2} \, \vphi \cdot F_* \cdot \vphi} = 
e^{ - \frac{1}{2} \, \vphi \cdot ( \G + F_*{\!}^{-1})^{-1} \cdot  \vphi
 - \frac{1}{2} \, \tr \ln ( 1 + \G \cdot F_* )} \, .
\label{Yact}
\ee
From \eqref{Fstar}
 \be
(\G + F_*{\!}^{-1})^{-1}  = \begin{cases}\  h \, , & \eta \ne 0 \, , \\
\ \frac{1}{1+z}\, \G_0{\!}^{-1}   \, , & \eta =0 \, ,
\label{GF}
\end{cases}
\ee
and assuming
\be
\half \, \tr \ln ( 1 + \G \cdot F_* ) + c_* V = 0 \, ,
\label{GFC}
\ee
then, since for $\eta=0$, $h= \G_0{\!}^{-1}$,  \eqref{Tstar} gives
\be
T_*[\vphi]   = \begin{cases}\  0 \, , & \eta \ne 0 \, , \\
\ \frac{z}{1+z}\, \frac{1}{2} \, \vphi \cdot \G_0{\!}^{-1}  \cdot \vphi  \, , 
& \eta =0 \, .
\end{cases}
\ee
For $\eta =0$ this result shows that
\be
2z(1+z) \, \frac{\pr}{\pr z}  \, T_*[\vphi] = 
\vphi \cdot \frac{\pr}{\pr \vphi} T_*[\vphi] = P_\Z [\vphi] \, , 
\ee
in accord with the expression  \eqref{PZ} for the zero mode. To verify 
\eqref{GFC}, with $c_*V$ given by \eqref{cstar}, we note that
\be
\half \, \tr \ln ( 1 + \G \cdot F_* )  = 
\begin{cases}\ -  \frac{V}{2(2\pi)^d} \int \d^d p \; 
\ln \big ( 1 -  {\tilde h}(p) K(p^2)/p^2  \big ) \, , & \eta \ne 0 \, , \\
\  \frac{V}{2(2\pi)^d} \int \d^d p \;  \ln \big ( 1 + K(p^2)/(1+z - K(p^2) )\big ) \, , & \eta = 0 \, ,
\end{cases}
\ee
and then insert $1= \pr_p \cdot p/d$ and integrate by parts. For $\eta\ne 0$ 
it is necessary to  use
$p \cdot \pr_p \big ( {\tilde h}(p) K(p^2)/p^2 \big ) 
= - \eta \big ( 1 -  {\tilde h}(p) K(p^2)/p^2  \big )
+ {\tilde h}(p) \, 2 K'(p^2)$.

Of course for the Gaussian fixed point all critical exponents can be obtained.
To obtain the scaling dimensions $\Delta$ for local scalar operators
$\Phi[\vphi;x]$ we consider the transformation \eqref{defPP} for $\S_*$ given
by \eqref{Sgauss}. Extending \eqref{Yact}, and using the results already
obtained for $T_*$, ensures that \eqref{defPP} becomes in this case
\be
P_\Phi[\vphi] = e^{\,\vphi\, \cdot \, \ln \G^{-1} \cdot \, \G_*  \,
\cdot \, \frac{\delta}{\delta \vphi}}
\, e^{ \frac{1}{2} \,  \frac{\delta}{\delta \vphi} \, \cdot \, \G_* \cdot
\frac{\delta}{\delta \vphi}} \, \Phi[\vphi] 
= e^{ \frac{1}{2} \,  \frac{\delta}{\delta \vphi} \, \cdot \, \G \cdot \, \G_*{\!}^{-1} \cdot \,
\G \, \cdot \frac{\delta}{\delta \vphi}} \, \Phi[\, \G_* \cdot \G^{-1} \cdot \vphi] \, ,
\label{PPhi}
\ee
for
\be
\G_* = \G - \G \cdot ( \G +  F_*{\!}^{-1} )^{-1} \cdot \G \, , \qquad
\G_*{\!}^{-1} = F_* + \G^{-1} \, .
\label{Gstar}
\ee

For $\eta=0$ 
\be
\G_* = \G - \frac{1}{1+z} \, \G \cdot \G_0{\!}^{-1} \cdot \G \, .
\ee
In this case \eqref{DPP} becomes
\be
D^{(\de_0)} \vphi \cdot \frac{\delta}{\delta \vphi} \, P_{\Phi_\Delta}[\vphi] =
D^{(\Delta)} P_{ \Phi_\Delta}[\vphi] \, .
\label{DPP2}
\ee
With the basis of local operators for
$\M'_{\rm{local}}$ provided by \eqref{localphi}
\be
D^{(\de_0)} \vphi \cdot \frac{\delta}{\delta \vphi} \, P_{n,s}[\vphi;x]
= D^{(\Delta_{n,s})}  P_{n,s}[\vphi;x] \, , \qquad
\Delta_{n,s} = n \, \de_0 + s \, , \ \ n,s = 0,1,2, \dots \, .
\ee
Identifying $P_{\Phi_{n,s}} =  P_{n,s}$
the corresponding operators $\Phi_{n,s}$ are obtained by inverting 
\eqref{PPhi}
\begin{align} 
\Phi_{n,s}[\vphi] = {}&
e^{- \frac{1}{2} \,  \frac{\delta}{\delta \vphi} \, \cdot \, 
\G_* \cdot \frac{\delta}{\delta \vphi}} \, 
e^{\,- \vphi\, \cdot \, \ln \G^{-1} \cdot \, \G_*  \,
\cdot \frac{\delta}{\delta \vphi}} \, P_{n,s}[\vphi] 
=  e^{- \frac{1}{2} \,  \frac{\delta}{\delta \vphi} \, \cdot \,
\G_* \cdot \frac{\delta}{\delta \vphi}} \,  
P_{n,s}[\, \G \cdot \G_*{\!}^{-1} \cdot\vphi] \nn \\
={}& \N_{\G_*} \big [ P_{n,s}[\, \G \cdot \G_*{\!}^{-1} \cdot\vphi]  \big ] \, .
\end{align}
The inversion is well defined acting on monomials of finite order in $\vphi$.
As  special cases
\begin{align}
P_{1,0}[\vphi] = {}& \vphi \qquad \ \Rightarrow \qquad
\Phi_{1,0} = \G \cdot \G_*{\!}^{-1} \cdot \vphi = \frac{1+z}{z} \, 
\Phi_{\delta_0} \, , \nn \\
P_{1,2}[\vphi] ={}& \G_0{\!}^{-1} \cdot \vphi \quad \Rightarrow \quad
\Phi_{1,2} =  \G_0{\!}^{-1} \cdot \G \cdot \G_*{\!}^{-1} \cdot \vphi
= \Phi_{d- \delta_0} \, ,
\label{specP}
\end{align}
where $\Phi_{\delta_0}, \, \Phi_{d- \delta_0}$  are given by \eqref{Phi0},
\eqref{philoc} with \eqref{Sgauss} and \eqref{Fstar}.
In this case we may take $\Z= 1 \cdot \Phi_{2,2}$.

A basis of redundant operators may also be obtained from \eqref{redPP}
by taking
\begin{align}
\O_{\psi_{n,s}} [\vphi] =  {}&
e^{- \frac{1}{2} \,  \frac{\delta}{\delta \vphi} \, \cdot \,
\G_* \cdot \frac{\delta}{\delta \vphi}} \,
e^{\,- \vphi\, \cdot \, \ln \G^{-1} \cdot \, \G_*  \,
\cdot \frac{\delta}{\delta \vphi}} \, \big (\, \vphi \cdot \G_0{\!}^{-1}
\cdot P_{n,s}[\vphi] \, \big ) \nn \\
& \quad \mbox{for} \quad \psi_{n,s} =
\G \cdot \G_0{\!}^{-1} \cdot  P_{n,s} \, .
\label{redG}
\end{align}
In this basis $\Z= \O_{\psi_{1,0}}$ for $P_{1,0}$ as in \eqref{specP}.

For $\eta \ne 0$  \eqref{Gstar} with \eqref{GF} give
\be
\G(p)^{-1} {\tilde \G}_*(p) = \sigma_*(p^2) = {\rm O}(p^2) \quad \mbox{as}
\quad p^2 \to 0 \, .
\ee
In \eqref{PPhi} $ e^{\,\vphi\, \cdot \, \ln \G^{-1} \cdot \, \G_*  \,
\cdot \frac{\delta}{\delta \vphi}}$ then generates contributions which
are singular as $p\to 0$ and so are non local. To construct a local basis
we now write
\be
P_{\Phi_{n,s}} =   e^{\,\vphi\, \cdot \, \ln \G_0{\!}^{-1} \,
\cdot \, \frac{\delta}{\delta \vphi}} P_{n,s} \, .
\ee
This modification generated additional terms in the eigenvalue equation
\be
D^{(\de - \eta)} \vphi \cdot \frac{\delta}{\delta \vphi} \, 
P_{\Phi_{n,s}} = e^{\,\vphi\, \cdot \, \ln \G_0{\!}^{-1} \,
\cdot \, \frac{\delta}{\delta \vphi}} \;
D^{(\de - \eta + 2)} \vphi \cdot \frac{\delta}{\delta \vphi} P_{n,s} 
= D^{(\Delta_{n,s})} P_{\Phi_{n,s}} \, ,
\ee
where now
\be
\Delta_{n,s} = \half ( d+2 -\eta ) n + s \,  .
\label{htl}
\ee
Manifestly there is no zero mode in this case, and for $\eta <2$ there
are no relevant operators even in $\vphi$ except for the identity.

Instead of \eqref{redG} a basis of redundant operators is obtained
for the trivial high temperature fixed point by taking
\be
\O_{\psi_{n,s}} [\vphi] = 
e^{- \frac{1}{2} \,  \frac{\delta}{\delta \vphi} \, \cdot \,
\G_* \cdot \frac{\delta}{\delta \vphi}} \,
e^{\,- \vphi\, \cdot \, \ln \G^{-1} \cdot \, \G_*  \,
\cdot \frac{\delta}{\delta \vphi}} \, \big (\, \vphi 
\cdot P_{\Phi_{n,s}} [\vphi] \, \big ) \, .
\ee
This shows that all scaling operators, with eigenvalues given by 
$\Delta_{n,s}-d$ with $\Delta_{n,s}$ as in \eqref{htl},
are redundant except when $n=s=0$, when $\O \propto V$ and $\lambda = -d$, 
reflecting the triviality of this fixed point (similar results were 
described in \cite{Phase}). The lack of any condition 
constraining $\eta$ also relates to the absence of a zero mode.

\subsection{Legendre Transform}

An alternative simple form for an exact RG equation, valid outside any
perturbation theory, was introduced by Wetterich 
\cite{Wett2}
by considering the RG flow of the one particle generating functional $\Gamma$.
This was shown to be equivalent  to  the standard Polchinski equation by
Morris \cite{Morris} where $\Gamma$ is related to the action $\S$ appearing
in the Polchinski equation by a Legendre transform. Here we show how
this extends to the case when the Polchinski equation is modified by 
allowing for the freedom to introduce the free parameter $\eta$ which
plays the role of an anomalous dimension as in \eqref{basic6}. A related
discussion was recently given by Rosten \cite{Rosten3}.

Before introducing a Legendre transformation the Polchinski equation
\eqref{basic6} is first rewritten in terms of the functional trace \eqref{Itr}
so that
\begin{align}
\bigg ( \frac{\pr}{\pr t}&  +  D^{(\de)} \vphi \cdot \frac{\delta}{\delta \vphi} 
- d \, V \frac{\pr}{\pr V}\bigg ) \, \S_t[\vphi] \nn \\
= {}&  \frac{1}{2}\,  \frac{\delta}{\delta \vphi} \, \S_t[\vphi]\cdot G
\cdot \frac{\delta}{\delta \vphi} \, \S_t[\vphi]
-  \frac{1}{2}\,  \tr \big ( G \cdot \S_t^{(2)}[\vphi]  \big )
- \frac{1}{2} \, \eta \; \big ( \vphi \cdot \G^{-1} \cdot \vphi - \tr(1) \big ) \, ,
\label{basic7}
\end{align}
where now
\be
\S_t^{(2)}[\vphi;x,y]  = 
\frac{\delta^2}{\delta \vphi(x) \, \delta \vphi(y)} \, \S_t[\vphi]  \, ,
\qquad \S_t^{(2)}[\vphi]^T = \S_t^{(2)}[\vphi] \, .
\ee

The Legendre transform determining  $\Gamma$ is then
\be
\S_t[\vphi] = \Gamma_t [ \Phi] + \half \, \Phi \cdot \R \cdot \Phi
+ \half \, \vphi \cdot \Q \cdot \vphi - \vphi \cdot \I \cdot \Phi \, , \quad
\R= \R^T \, , \ \Q = \Q^T \, ,
\label{SGt}
\ee
where $\Phi$ is defined by
\be
\frac{\delta}{\delta \vphi} \S_t[\vphi] - \Q \cdot \vphi = - \I \cdot \Phi \, .
\label{defPhi}
\ee
The Legendre transform leads directly to
\be
\frac{\delta}{\delta \Phi} \Gamma_t[\Phi]  + \Phi  \cdot \R= 
\vphi \cdot \I  \, , \qquad \frac{\pr}{\pr t} \Gamma_t[\Phi] =
\frac{\pr}{\pr t} \S_t[\vphi] \, .
\label{pph}
\ee
It is easy to further obtain
\be
\S_t^{(2)}[\vphi]  - \Q = 
- \I \cdot \big ( \R + \Gamma_t^{(2)} [\Phi]\,  \big )^{-1} \cdot \I^T \, ,
\ee
where
\be
\Gamma_t^{(2)} [\Phi;x,y]  = 
\frac{\delta^2}{\delta \Phi(x) \, \delta \Phi(y)} \, \Gamma_t[\Phi]  \, .
\label{Gam2}
\ee

If we omit a $\Phi$-independent term 
$\half\, \tr( G \cdot \Q - \eta \, 1) \propto V$, 
which is reconsidered later but may be consistently neglected in obtaining 
an equation for $\Gamma_t$, the RG flow equation \eqref{basic7} then becomes
\be
\bigg ( \frac{\pr}{\pr t} +  D^{(\de)} \Phi  \cdot \frac{\delta}{\delta \Phi}
- d \, V \frac{\pr}{\pr V} \bigg )\Gamma_t[\Phi]  = \frac{1}{2} \, \tr \Big (\, 
\I^T \cdot G \cdot \I \cdot 
\big ( \R + \Gamma_t^{(2)} [\Phi]\, \big )^{-1} \Big ) 
\label{wetteq} 
\ee
so long as
\begin{align}
&  - D^{(\de)} \bigg ( \frac{\delta}{\delta \Phi}\Gamma_t [\Phi] \cdot \I^{-1} + 
\Phi \cdot \R\cdot  \I^{-1} \bigg ) \cdot 
( \I \cdot \Phi - \H \cdot \vphi ) \nn \\
 &{}=  
\half \big ( \Phi  \cdot \I^T - \vphi \cdot \Q \big ) \cdot G \cdot \big (
\I \cdot \Phi - \Q \cdot \vphi \big )  - \half \, \eta \; 
\vphi \cdot \G^{-1} \cdot \vphi   +  \frac{\delta}{\delta \Phi}
 \Gamma_t[\Phi]  \cdot D \Phi \, ,
 \end{align}
 using \eqref{defPhi} and \eqref{pph}. Eliminating $\vphi$ through \eqref{pph}
 leads to equations for $\Q, \I$ and $\R$ which can be reduced to
 \begin{subequations}
 \begin{align}
- {D}^{(d-\de)}\, \Q - \Q {\overleftarrow D}{}^{(d-\de)} = {}& 
\Q \cdot G \cdot \Q  - \eta \, \G^{-1} \,  ,  \\
- {D}^{(d-\de)}\,  \I - \I {\overleftarrow D}{}^{(d-\de)} = {}& 
\Q \cdot G \cdot \I \, ,   \label{Ieq} \\
- {D}^{(d-\de)}\,  \R - \R {\overleftarrow D}{}^{(d-\de)} = {}& 
\I^T \! \cdot G \cdot \I   \, ,
 \label{Req}
 \end{align}
 \end{subequations}
 where $\bar D$ is related to $D$ as in \eqref{DDbar}. With \eqref{GK}
 and \eqref{GKp} these become the differential equations
 \begin{subequations}
 \begin{align}
 \big ( p \cdot \pr_p - 2 +\eta \big ) {\tilde \Q}(p) ={}& 
2 K'(p^2) \, {\tilde \Q}(p)^2 - \eta \, \frac{p^2}{K(p^2)} \, , \label{Heqp} \\
 \big ( p \cdot \pr_p - 2 +\eta \big )\, {\tilde \I}(p) 
={}& 2 K'(p^2) \, {\tilde \Q}(p)\,  {\tilde \I}(p)   \, , \label{Ieqp} \\
\noalign{\vskip  4pt}
\big ( p \cdot \pr_p - 2 +\eta \big ) {\tilde \R}(p) = {}& 2 K'(p^2) \, 
{\tilde \I}(-p) \, {\tilde \I}(p) \, . \label{Reqp}
\end{align}
\end{subequations}
\eqref{Heqp} is similar to \eqref{gauss2} and can be solved in an analogous 
fashion
 \be
f(p^2) = (p^2)^{-1+\frac{1}{2}\eta}\, {\tilde \Q}(p) + 
\frac {(p^2)^{\frac{1}{2}\eta}}
{K(p^2)} \ \Rightarrow \  f'(x) = x^{-\frac{1}{2}\eta}
K'(x) \, f(x) \bigg ( f(x) - \frac{2 x^{\frac{1}{2} \eta}}{K(x)} \bigg )\, ,
 \ee
or
 \be
\frac{\rmd}{\rmd x}\bigg ( \frac{1}{f(x)K(x)^2}\bigg ) = x^{-\frac{1}{2}\eta} 
\frac{\rmd}{\rmd x} \frac{1} {K(x)} \, .
\ee
Requiring analyticity for $p \approx 0$ determines a unique solution
\be
{\tilde \Q}(p) = \frac{p^2}{K(p^2)} \bigg ( \frac{1}{\sigma_*(p^2)} - 1
\bigg ) \, ,
\label{Hres}
\ee
where $\sigma_*$ is given by \eqref{hsol}. \eqref{Ieqp} is a linear
homogeneous first order equation for ${\tilde \I}(p)$ which can be easily 
solved by integration. With an arbitrary choice for the overall scale
we have
\be
{\tilde \I}(p) = \frac{p^2}{\sigma_*(p^2)} \, .
\label{Ires}
\ee
With this solution \eqref{Reqp} becomes
\be
{\tilde \R}(p) = (p^2)^{1-\frac{1}{2}\eta} \, r(p^2) \, , \quad
r'(x)  = x^{\frac{1}{2}\eta} \, \frac{K'(x)}{\sigma_*(x)^2} = 
\frac{\rmd}{\rmd x}\bigg ( x^{\frac{1}{2}\eta} \, \frac{K(x)}{\sigma_*(x)}
\bigg ) \, ,
\label{Rreq}
\ee
so that, assuming analyticity for $p\approx 0$ again,
\be
{\tilde \R}(p) =  \frac{p^2K(p^2)}{\sigma_*(p^2)} \, .
\label{Rres}
\ee

From the definition \eqref{hsol}, for general $\eta<2$, the asymptotic behaviour
of $\sigma_*(p^2)$ is given by \eqref{s0} and 
\be
\sigma_*(p^2) \to 1 \quad \mbox{as} \quad p^2\to \infty \, ,
\ee
assuming that $K(p^2) \to 0$ as $p^2\to \infty$ faster than any inverse power
and $K'(p^2) <0$ for all $p^2$.
In consequence ${\tilde \R}(p)$, as defined by \eqref{Rres}, is a finite positive 
constant for $p=0$ and falls off rapidly for large $p^2$. 
Reinstating the cut off scale $\Lambda$ in \eqref{Rres} by dimensional
considerations in the form
\be
{\tilde \R}_\Lambda(p) = \frac{p^2 \, K(p^2/\Lambda^2)}{\sigma_*(p^2/\Lambda^2)} \, ,
\ee
then we may define
\be
{\dot \R} \equiv  \frac{\pr}{\pr t} \R_\Lambda \, \Big |_{\Lambda=1} 
= - {D}^{(d-\de)}\,  \R - \R {\overleftarrow D}{}^{(d-\de)} - \eta \, \R
= \I^T \!\cdot G \cdot \I - \eta\, \R \, ,
\label{Rdot}
\ee
as a consequence of \eqref{Req} or \eqref{Reqp}. 

Using the above results the omitted term in \eqref{wetteq} becomes
\be
\half\, \tr( G \cdot \Q - \eta \, 1) = \half \, 
\tr \big ( {\dot \R} \cdot \R^{-1} -  G \cdot \G^{-1} \big ) \, ,
\ee
which are removed by the natural redefinitions $\Gamma_t[\Phi] - 
\tr ( {\dot \R} \cdot \R^{-1} )/2d \to \Gamma_t[\Phi]$ and also
$\S_t[\vphi] - \tr (G \cdot \G^{-1})/2d \to \S_t[\vphi]$. In terms
of the full action $S_t$, related to $\S_t$ as in \eqref{pol1}, \eqref{SGt}
now becomes
\be
S_t[\vphi] = \Gamma_t [ \Phi] + \half \, 
\big ( \Phi - \G_0 \cdot \G^{-1} \cdot \vphi \big ) \cdot \R \cdot 
\big ( \Phi - \G_0 \cdot \G^{-1} \cdot \vphi \big ) -
\frac{1}{2d} \, \tr ( {\dot \R} \cdot \R^{-1} ) \, ,
\ee
where $\G_0$ is given in \eqref{Gz}. Applying \eqref{Rdot} in \eqref{wetteq}
\be
\bigg ( \frac{\pr}{\pr t} + D^{(\de)} \Phi  \cdot \frac{\delta}{\delta \Phi}
- d \, V \frac{\pr}{\pr V} \bigg )\Gamma_t[\Phi]  = \frac{1}{2} \, \tr \Big (
\big ( {\dot \R } + \eta \, \R \big )  \cdot
\big ( \R + \Gamma_t^{(2)} [\Phi]\, \big )^{-1} \Big ) \, , 
\label{wetteq2}
\ee
for
\be
{\tilde{\dot \R}}(p) = ( p \cdot \pr_p - 2 ){\tilde  \R}(p) \, .
\label{Rdot2}
\ee
\eqref{wetteq2}  is just the standard form of the Wetterich
RG equation \cite{Wett2} extended to include the parameter $\eta$. Assuming
$\R$ is an independent cut off function all dependence on $\eta$ is explicit.

If there is a fixed point as $t\to \infty$ for a suitable choice
of $\eta$ then at the fixed point $\Gamma_*$ must satisfy, by virtue
of \eqref{wetteq} and \eqref{Rdot},
\be
E[\Phi] \equiv \bigg (  D^{(\de)} \Phi  \cdot \frac{\delta}{\delta \Phi}
-  d\, V \frac{\pr}{\pr V} \bigg ) \Gamma_*[\Phi] 
-  \frac{1}{2} \, \tr \Big (\, \big ( {\dot \R} + \eta \R \big ) \cdot
\big ( \R + \Gamma_*^{(2)} [\Phi]\, \big )^{-1} \Big ) = 0 \, .
\label{fxG}
\ee
Corresponding to \eqref{asymS} we then have
\be
\Gamma_t[\Phi] \sim \Gamma_*[\Phi] - \sum_{n\ge 0}
e^{-\lambda_n t}\, \P_n[\Phi] \quad \mbox{as} \quad t \to \infty \, .
\label{asymG}
\ee
Although the relation between $\Phi$ and $\vphi$ in general depends
on $t$ we have to first order in an expansion about the fixed point
\be
\P_n[\Phi] = \O_n[\vphi] \qquad \mbox{for} \qquad
\frac{\delta}{\delta \vphi} \S_*[\vphi] - \Q \cdot \vphi = - 
\I \cdot \Phi \, .
\label{PO}
\ee
The eigenvalue equation, which is equivalent to \eqref{eig} with \eqref{DelS}, 
becomes
\begin{align}
\lambda \, \P[\Phi] = \Delta_{\Gamma_*}  \P[\Phi]  ={}&  
\bigg ( D^{(\de)} \Phi  \cdot \frac{\delta}{\delta \Phi} - 
d\, V\frac{\pr}{\pr V} \bigg ) \P[\Phi] \nn \\
&{} +  \frac{1}{2} \, \tr \Big (\,
\big ( {\dot \R} + \eta \R \big ) \cdot 
\big ( \R + \Gamma_*^{(2)} [\Phi]\, \big )^{-1}\cdot \P^{(2)}[\Phi] 
\cdot \big ( \R + \Gamma_*^{(2)} [\Phi]\, \big )^{-1} \Big ) \, ,
\label{Geig}
\end{align}
with $\P^{(2)}[\Phi;x,y]$ defined as in \eqref{Gam2}.

It is easy to check from \eqref{Geig} and \eqref{fxG} that
\begin{subequations}
\begin{align}
\P[\Phi] = 1 \cdot \Phi \qquad \Rightarrow \qquad \lambda = - \half(d+2-\eta) \,,\\
\P[\Phi] = 1 \cdot  \frac{\delta}{\delta \Phi}  \Gamma_*[\Phi] 
\qquad \Rightarrow \qquad \lambda = - \half(d-2+\eta) \, , \label{redPp}
\end{align}
\end{subequations}
in agreement with \eqref{exacta} and \eqref{exact}.

If we consider redundant operators of the form  \eqref{red} then using \eqref{PO}
to define an equivalent $\P_\Psi[\Phi]$ gives, with the results in \eqref{Hres},
\eqref{Ires} and \eqref{Rres},
\begin{align}
\P_\Psi[\Phi]={}& \Psi \cdot \frac{\delta}{\delta \Phi}  \Gamma_*[\Phi] -
 \tr \Big ( \big ( \R + \Gamma_*^{(2)} [\Phi]\, \big )^{-1} \cdot \R \cdot
\Psi^{(1)} \Big ) \, , 
\label{redP}
\end{align}
for
\be
\Psi^{(1)}[\Phi;x,y] = \frac{\delta}{\delta \Phi(y)} \Psi[\Phi;x] \, , \qquad
{\tilde \Psi}[\Phi;p] = {\tilde \psi}[\vphi;p]\, \frac{1}{K(p^2)}   \, .
\ee
Manifestly $\P$ in \eqref{redPp} is the redundant operator $\P_1$.

For the zero mode operator given by \eqref{ZO} and \eqref{zpsi} the
corresponding operator here is also a redundant operator of the form \eqref{redP}
since
\be
\Z[\Phi] = \P_\Phi[\Phi] = \Phi \cdot \frac{\delta}{\delta \Phi}  \Gamma_*[\Phi]
- \tr \Big ( \big ( \R + \Gamma_*^{(2)} [\Phi]\, \big )^{-1} \cdot \R \Big ) \, .
\label{ZGa}
\ee
To verify that this is a zero mode it is necessary to use
\be
\Delta_{\Gamma_*}  \Z[\Phi]  =  \Phi \cdot \frac{\delta}{\delta \Phi} E[\Phi]
- \tr \Big ( \big ( \R + \Gamma_*^{(2)} [\Phi]\, \big )^{-1} \cdot 
E^{(2)}[\Phi] \Big ) \, ,
\ee
where $E$ is defined in \eqref{fxG} and we require the identity
\begin{align}
- \tr \Big (  & \big ( \R + \Gamma_*^{(2)} [\Phi]\, \big )^{-1} \cdot \R 
\cdot \big ( \R + \Gamma_*^{(2)} [\Phi]\, \big )^{-1} \! \cdot \big ( {D}^{(d-\de)}\,
\Gamma_*^{(2)} [\Phi] +  \Gamma_*^{(2)} [\Phi]  {\overleftarrow D}{}^{(d-\de)} \big )
\Big ) \nn \\
& =  \tr \Big (  \big ( \R + \Gamma_*^{(2)} [\Phi]\, \big )^{-1} \cdot \big (
{\dot \R} + \eta \R \big ) 
\cdot \big ( \R + \Gamma_*^{(2)} [\Phi]\, \big )^{-1} \! \cdot  \Gamma_*^{(2)} [\Phi]
\Big ) \, ,
\end{align}
which depends on \eqref{Rdot}.

If $\Gamma_t[\Phi]$ is restricted to Gaussian form so that
\be
\Gamma_t[\Phi] = \half \, \Phi \cdot \chi_t \cdot \Phi + X_t \, V \, ,
\ee
then the solution of \eqref{wetteq} gives simply
\be
{\tilde \chi}_t(p) = e^{(2-\eta)t} \, {\tilde \chi}_0(e^{-t} p) \, ,
\ee
independent of $\R$. For there to be a limit as $t\to \infty$ with
${\tilde \chi}_0(p)$ analytic in $p$ then it is necessary that $\eta=0$,
${\tilde \chi}_0(p) = r \, p^2 + \dots$ and also $X_0$ to be chosen precisely
to remove any $e^{dt}$ terms, so that from \eqref{fxG}
\be
\Gamma_*[\Phi] = \frac{1}{2} \, \Phi \cdot \chi_* \cdot \Phi -
\frac{1}{2d} \, \tr \big (
( \R + \chi_* )^{-1} \cdot {\dot \R}\big ) \, , \qquad {\tilde \chi}_*(p)= r\, p^2 \, ,
\ee
where, with $\dot \R$ given by \eqref{Rdot2},
\be
\tr \big (
( \R + \chi_* )^{-1}\cdot {\dot \R} \big ) = \frac{V}{(2\pi)^d} \int \d^d p \;
\frac{{\tilde{\dot \R}}(p)}{\R(p) + r \, p^2} \, .
\label{RRdot}
\ee

Applying the formula \eqref{ZGa} to the Gaussian  case gives
\begin{align}
\Z[\Phi] = {}& \Phi \cdot \chi_* \cdot \Phi - \tr \big (
( \R + \chi_* )^{-1}\cdot \R  \big ) \nn \\
 = {}& \Phi \cdot \chi_* \cdot \Phi + \frac{1}{d} \, \frac{V}{(2\pi)^d} 
 \int \d^d p \; \frac{{\tilde{\dot \R}}(p)\; r \, p^2}
 {\big ({\tilde \R}(p) + r \, p^2\big )^2}
 = 2r \frac{\pr}{\pr r} \Gamma_*[\Phi] \, ,
\end{align}
using $1 = \pr_p \cdot p /d$ and integrating by parts with the result
\eqref{RRdot} for $\dot \R$. In terms of the original Polchinski cut off
function $K(p^2)$, and with $\eta=0$, ${\tilde \R}(p) =  p^2 K(p^2)/(1-K(p^2))$,
 ${\tilde {\dot\R}}(p) = 2( p^2)^2 K'(p^2)/(1-K(p^2))^2$.

\subsection{Alternate RG Equations}

By considering a transform akin to that in \eqref{FT0}
the Polchinski RG equation generates a new RG equation 
with some additional desirable features. This equation is equivalent
to considering expansions of the original equation 
in terms of a normal ordered basis,
and is connected with the approach based on using  scaling fields to 
reduce the RG equations to a tractable set of finite equations \cite{Gol}.

To handle $\eta\ne 0$ it is necessary to introduce a new Green function
${\hat \G}$ which is defined by a modification of \eqref{DY} by requiring
\be
\Big [ D^{(\de)} \vphi
\cdot \frac{\delta}{\delta \vphi}  , \, {\hat \Y} \Big ] = 
\frac{1}{2} \,  \frac{\delta}{\delta \vphi}\cdot {G} \cdot
\frac{\delta}{\delta \vphi}  \, , \qquad
{\hat \Y} = \frac{1}{2}  \, 
\frac{\de}{\de \vphi} \cdot {\hG} \cdot \frac{\de}{\de \vphi} \, .
\label{DY2}
\ee
The corresponding equation for $\hG$ is then
\be
\big ( y \cdot \pr_y + d-2 + \eta \big ) \hG(y) = - {G}(y) \qquad \mbox{or}
\qquad
\big ( p \cdot \pr_p  +2 - \eta \big ) {\tilde \hG}(p) = 2K'(p^2) \, ,
\label{Geq4}
\ee
using \eqref{GKp}. This has a solution
\be
{\tilde \hG}(p) = \frac{{\hat K}(p^2)}{p^2} \, , \qquad
{\hat K}(p^2) = - (p^2)^{\frac{1}{2}\eta} \, 
\int_{p^2}^\infty \!\! \d x \; K'(x) \, x^{-\frac{1}{2}\eta} \, ,
\label{Khat}
\ee
where ${\hat K}(p^2)$ has been required to vanish for large $p^2$.
Clearly
\be
{\hat K}(p^2) \tosim{p\to 0} {\hat C}_\eta (p^2)^{\frac{1}{2}\eta}
\, , \qquad {\hat C}_\eta =
- \int_0^\infty \!\! \d x \; K'(x) \, x^{-\frac{1}{2}\eta}
\, ,
\ee
or equivalently
\be
\hG(y)\tosim{y\to \infty}   \frac{k}{(y^2)^{\frac{1}{2}(d-2+\eta)}} \, , 
\label{Gy}
\ee
where
\be
C_\eta = (d-2+\eta)S_d \, k = \frac{2^\eta \; \Gamma(\half d + \half \eta)}
{\Gamma(\half d)\, \Gamma(1-\half \eta)} \, {\hat C}_\eta \, , \qquad
S_d = \frac{2 \pi^{\frac{1}{2}d}}{\Gamma(\half d)} \, .
\label{Ceta}
\ee
For $\eta=0$ it is easy to see that $\hG(y)=\G(y)$ and ${\hat \Y}=\Y$ with
$C_0 = {\hat C}_0 =1$ as a consequence of $K(0)=1$, and the asymptotic
form is the same as given by \eqref{Gas} with \eqref{Gz}. For general $\eta$
if $K(p^2)$ falls off faster than any power then from \eqref{Khat}
\be
{\hat K}(p^2) \sim K(p^2) \quad \mbox{as} \quad p^2 \to \infty \, .
\ee

Using \eqref{DY2}  \eqref{basic6} can be recast
in the form of the functional differential equation, 
\begin{align}
\bigg ( \frac{\pr}{\pr t}&  \, + \, D^{(\de)} \vphi
\cdot \frac{\delta}{\delta \vphi}  - d \, V \frac{\pr}{\pr V} \bigg ) \, 
e^{\hat \Y} \, \S_t[\vphi]  \nn \\   
&{} = 
e^{\hat \Y} \,  \frac{1}{2} \bigg ( \frac{\delta}{\delta \vphi}\S_t[\vphi] \cdot
{G} \cdot \frac{\delta}{\delta \vphi} \S_t[\vphi] 
- \eta \; \big ( \vphi \cdot \G^{-1} \cdot \vphi - \tr(1) \big ) \bigg )  \, . 
\label{De1}
\end{align}
Defining
\be
{\topcirc \S}_t[\vphi] = e^{\hat \Y} \S_t[\vphi] \, ,
\label{SSY}
\ee
then, using $e^{a \frac{\d^2}{\d z^2}} \big ( f(z) \, g(z) \big ) =
e^{2a \frac{\pr^2}{\pr z \pr z'}} \big ( e^{a \frac{\d^2}{\d z^2}}f(z) \,
e^{a \frac{\d^2}{\d z'^2}}g(z') \big ) \big |_{z'=z}$,
\begin{align}
\bigg ( \frac{\pr}{\pr t} \, + \,  &  D^{(\de)}
\vphi \cdot \frac{\delta}{\delta \vphi} - d \, V \frac{\pr}{\pr V} \bigg )  \, 
{\topcirc \S}_t[\vphi]
+ \half \eta \; \big ( \vphi \cdot \G^{-1} \cdot \vphi + 
\tr ( {\hat \G}  \cdot \G^{-1} - 1 ) \big )  \nn \\
=  {}& \exp\bigg (\frac{\de}{\de \vphi} \cdot \hG \cdot \frac{\de} 
{\de \vphi'} \bigg ) \frac{1}{2} \, \frac{\de  {\topcirc \S}_t[\vphi]}{\de \vphi}
\cdot G \cdot \frac{\de  {\topcirc \S}_t[\vphi']}{\de \vphi'} 
\bigg |_{\vphi'=\vphi} \nn \\
= {}&  \frac{1}{2} \sum_{n=0}^\infty \frac{1}{n!}
 \int \d^d x \, \d^d x' \; {G}(y) \;
 \prod_{r=1}^n \d^d x_r \, \d^d x' {\!}_r \, \hG(y_r) \, \nn \\
 \noalign{\vskip -3pt}
& \hskip 2cm {}\times \frac{\de^{n+1} {\topcirc \S}_t[\vphi]}
{\de \vphi_i(x)\de \vphi_{i_1}(x_1) \dots \vphi_{i_n}(x_n) }\;
\frac{\de^{n+1} {\topcirc \S}_t[\vphi]}
{\de \vphi_i(x')\de \vphi_{i_1}(x'{\!}_1) \dots \de \vphi_{i_n}(x'{\!}_n) }\, ,
\label{exp1}
\end{align}
where $y_r = x_r - x'{\!}_r$.
The transformation $\S_t \to {\topcirc \S}_t $ is tantamount to expressing 
$\S_t$ in a normal ordered basis.
If ${\topcirc \S}_t[\vphi]$ is expanded in a basis of monomials
$\P_n[\vphi]$ then $\S_t[\vphi] = e^{-{\hat \Y}} {\topcirc \S}_t[\vphi]$ 
has a corresponding expansion in terms of
the  normal ordered monomials 
$ \N_{\hat \G}(\P_n[\vphi]) = e^{-{\hat \Y}} \P_n[\vphi] $, as in the
definition \eqref{normal}, for the two point function ${\hat \G}$.

For $\eta=0$ and ${\hat \G} = \G$ an equation essentially identical 
with \eqref{exp1} was used as a starting point by Wieczerkowski and Salmhofer 
\cite{Wiec, Salm} in proofs of perturbative renormalisation  for scalar field
theories.
 
\section{Derivative Expansion}

Since the RG flow equations for $\S_t$ ensure that it remains essentially
local it is natural to consider an expansion where $\S_t[\vphi]$ is an 
entirely local functional 
$\int \d^d x \; \cL(\vphi, \pr \vphi, \pr \pr \vphi, \dots)$,  
with the expansion parameter  the total number of derivatives.
Although such a derivative expansion has often been used as an approximation 
to the exact RG functional equations 
\cite{Morris1,Morris2,Com,Morris3,Bervillier,Canet}
the resulting differential equations, at least beyond lowest order, depend on
the detailed form of the cut off function and reliability of any results,
unless some optimisation strategy is used, can be uncertain. It is
also not obvious how to maintain consistency with perturbative
results beyond lowest order.

It is also useful to consider an extension to a $N$-component scalar field 
$\phi_i \to \vphi_i$, assuming now $\cdot $ includes contraction of
indices where appropriate. The quadratic term in \eqref{pol1}
$\frac{1}{2} \, \vphi \cdot \G^{-1} \cdot \vphi$,
which includes the cut off function $K$, is invariant under
$O(N)$ symmetry but this may be reduced depending on the form of
the initial $\S_0[\vphi]$ in the RG flow equations.
For the present discussion we start from the Polchinski equation 
\eqref{basic6} which includes $\eta$. In the multi-component
case this becomes in general a matrix but for simplicity we assume
$\eta_{ij} = \eta \, \delta_{ij}$, 
as would be required by $O(N)$ symmetry.

Usually the derivative expansion is applied directly to $\S_t[\vphi]$.
Here we discuss an alternative form of derivative expansion, different 
from the standard approach, which is in terms of ${\topcirc \S}_t$, defined
in \eqref{SSY}, by writing
\be
{\topcirc \S}_t [\vphi ]  = \int \d^d x \; \big ( {\topcirc V}(\vphi ) + 
\half \pr^\mu \! \vphi_j \pr_\mu \vphi_k \,
{\topcirc Z}_{jk}(\vphi) + \dots \big ) \, .
\label{deS}
\ee
It is easy to obtain
\begin{align}
\bigg ( D^{(\de)}  \vphi
\cdot & \frac{\delta}{\delta \vphi} - d \, V \frac{\pr}{\pr V}\bigg ) 
 {\topcirc \S}_t [\vphi ]  \nn \\
= \int \d^d x \; \bigg ( & - d \, 
 {\topcirc V}(\vphi ) + \de \, \vphi_i\frac{\pr}{\pr \vphi_i}
{\topcirc V}(\phi) \nn \\
\noalign{\vskip -6pt}
&{} + \half \pr^\mu \! \vphi_j \pr_\mu \vphi_k 
\Big ( \eta \, {\topcirc Z}_{jk}(\vphi) 
+  \de \, \vphi_i
\frac{\pr}{\pr \vphi_i}  {\topcirc Z}_{jk}(\vphi)\Big )  + \dots \bigg ) \, . 
\label{DSz}
\end{align}
If \eqref{deS} is inserted on the right hand side of \eqref{exp1} 
the functional derivatives generate $\delta$-functions ensuring that
the integrals over $x_r,x'{\!}_r$ all become trivial to evaluate so
there remain integrals just over $x,x'$. The resulting expression
corresponds to contributions from all two point Feynman graphs.
The leading
terms, with up to two derivatives, are then, after judicious integrations 
by parts,
\begin{align}
\bigg ( \frac{\pr}{\pr t} & \, + D^{(\de)} \vphi
\cdot \frac{\delta}{\delta \vphi}  - d \, V \frac{\pr}{\pr V}  \bigg )  \, 
{\topcirc \S}_t [\vphi] + \half \eta \;\big (\vphi \cdot \G^{-1} \cdot \vphi 
+ \tr ( {\hat \G}  \cdot \G^{-1} - 1 ) \big ) \nn \\
\sim {}& \frac{1}{2} \sum_{n=0}^\infty \frac{1}{n!}
 \int \d^d x \, \d^d x' \;  \bigg \{ 
{\topcirc V}_{i\,i_1\dots i_n}(\vphi) \,
{\topcirc V}_{i\,i_1\dots i_n}(\vphi')\; 
{G}(y) \, \hG(y)^n\nn \\
 \noalign{\vskip -5pt}
& \hskip 1.5cm {}+ \pr^\mu \! \vphi_j \pr_\mu \vphi_k \;
{\topcirc Z}_{jk,i\,i_1\dots i_n}(\vphi) \,
{\topcirc V}_{i\, i_1\dots i_n}(\vphi') \;
{G}(y) \, \hG(y)^n \nn \\
\noalign{\vskip 4pt}
& \hskip 1.5cm {} +
 \pr^\mu \! \vphi_j \pr_\mu \vphi'_k \; \big (
2 {\topcirc Z}_{ji,i_1\dots i_n}(\vphi) -  
{\topcirc Z}_{ii_1,j\, i_2\dots i_{n}}(\vphi) \big )  \,
{\topcirc V}_{k i\,i_1\dots i_n}(\vphi')\; {G}(y) \, \hG(y)^n   \nn \\
& \hskip 1.5cm {}- {\topcirc Z}_{jk,i_1 \dots i_{n-1}}(\vphi) \,
{\topcirc V}_{jk\,i_1 \dots i_{n-1}}(\vphi') \, 
\Big ( \pr^2 {G}(y)\;
\hG(y)^n + n \, {G}(y) \, \pr^2 g(y) \; \hG(y)^{n-1} \Big ) \bigg \} \, ,
\label{exp2}
\end{align}
where in the integral $\vphi'= \vphi(x')$ while $\vphi=\vphi(x)$, as
before $y=x-x'$, and we adopt the notation ${\topcirc V}_{i\, i_1\dots i_n}
(\vphi)  = \pr_i \pr_{i_1} \dots \pr_{i_n} {\topcirc V}(\vphi)$, and
similarly for derivatives of ${\topcirc Z}_{jk}(\vphi)$.

To obtain tractable closed equations which may be solved it is necessary to 
project the right hand side of \eqref{exp2} onto local expressions of the 
same form as feature
in the derivative expansion. This is achieved by assuming that the
products of ${G}(y)$ and $\hG(y)$ that feature in \eqref{exp2}
can be expanded in the form
\begin{align}
{G}(y) \, \hG(y)^n \sim  - c_n \, \de^d(y) - c'{\!}_n \, \pr^2 \de^d(y)
+ \dots \, , \nn \\
\pr^2 {G}(y)\;
\hG(y)^n + n \,  {G}(y) \, \pr^2 \hG(y) \; \hG(y)^{n-1} \sim  
d_n \,  \de^d(y)  + \dots \, .
\label{ggg}
\end{align}
In general $c_n,c'{\!}_n,
d_n$, for arbitrary $n$ and dimension $d$, depend on the detailed form 
of the cut off function. However some results are independent of the 
precise form of the cut off and depend only on the form of $\hG(y)$
for large $y$ as exhibited in \eqref{Gy}. To demonstrate this we 
consider the integrals 
\begin{align}
 {}&  - \int \d^d y \, (y^2)^r {G}(y) \, \hG(y)^n = 
\frac{1}{n+1} \int \d^d y \, (y^2)^r \, 
\big ( y \cdot \pr_y + (n+1)( d-2 + \eta)  \big ) \hG(y)^{n+1} \nn \\
&{} =  \frac{1}{n+1} \int \d^d y \, \pr_\mu \bigg ( y^\mu 
\frac{k^{n+1}}{(y^2)^{\frac{1}{2}d}}\bigg ) = \frac{C_\eta}{d+2r}\,   
k^n \quad \mbox{if} \quad (n+1)(d-2+\eta) - 2r  = d \, ,
\label{int}
\end{align}
using \eqref{Geq4} and where $C_\eta$ is given by \eqref{Ceta}. For $\eta=0$, 
and $C_0=1$, these results are just the coefficients of the logarithmic 
divergencies in the associated two point Feynman graphs in appropriate 
dimensions depending on $n$. Using \eqref{int} it is then easy to see that
for $n=1,2,\dots$,
\be
c_n \big |_{d=(2-\eta)(n+1)/n} = \frac{C_\eta}{d} \, k^n \, , \qquad
c'{\!}_n \big |_{d=((2-\eta)(n+1)+2)/n} =  \frac{C_\eta}{2d(d+2)}\, k^n  \, .
\label{spec}
\ee
There are no such comparable results for $d_n$ and furthermore it is evident 
that $d_0=0$. In consequence we assume the simplest expressions for 
$c_n,c'{\!}_n,d_n$  consistent with this, and which interpolate \eqref{spec} 
for all $n,d$ including $n=0$, and take henceforth
\be
c_n = \frac{C_\eta}{d} \, k^n \, , \qquad 
c'{\!}_n = \frac{C_\eta}{2d(d+2)}\, k^n \, , \qquad d_n = 0  \, .
\label{resgg}
\ee

If \eqref{ggg}, with \eqref{resgg}, is inserted in \eqref{exp2} then,
along with \eqref{DSz}, we obtain equations for the function ${\topcirc V}$,
\begin{align}
\bigg ( \frac{\pr}{\pr t} - d + \de \, \vphi_i \frac{\pr}{\pr\vphi_i}
\bigg ) {\topcirc V}(\vphi) = {}& - \frac{C_\eta}{2d} \, \sum_{n=0}^\infty \;
\frac{k^n}{n!} \; {\topcirc V}_{i_1 \cdots i_{n+1}}(\vphi) \;
{\topcirc V}_{i_1 \cdots i_{n+1}}(\vphi) \nn \\
={}& - \frac{C_\eta}{2d}  \; 
e^{k \, \frac{\pr^2}{\pr \vphi_j \pr \vphi'{\!}_j }} \;
\frac{\pr}{\pr \vphi_i}  {\topcirc V}(\vphi) \;
\frac{\pr}{\pr \vphi'{\!}_i}{\topcirc V}(\vphi') \, \bigg |_{\vphi'=\vphi} \, ,
\label{Veq1}
\end{align}
and also ${\topcirc Z}_{jk}$,
\begin{align}
\bigg ( \frac{\pr}{\pr t}&{} +\eta  + \de \, \vphi_i 
\frac{\pr}{\pr \vphi_i}
\bigg ) {\topcirc Z}_{jk}(\vphi) + \eta \, \de_{jk} \nn \\
= {}& - \frac{C_\eta}{d} \, \sum_{n=0}^\infty \; \frac{k^n}{n!}  \, 
\Big ( {\topcirc V}_{i_1\dots i_{n+1}}(\vphi) \;
{\topcirc Z}_{jk,i_1\dots i_{n+1}}(\vphi) \nn \\
 \noalign{\vskip -6pt} 
&\hskip 2.6cm {} +  
{\topcirc V}_{i_1\dots i_{n+1}(j}(\vphi) \, \big (
2 {\topcirc Z}_{k)\,i_1,i_2\dots i_{n+1}}(\vphi) -
{\topcirc Z}_{i_1i_2,k)\, i_3\dots i_{n+1}}(\vphi) \big )  \, \Big ) \nn \\
&{} + \frac{C_\eta}{2d(d+2)} \, \sum_{n=0}^\infty \;
\frac{k^n}{n!} \; {\topcirc V}_{j\,  i_1 \dots i_{n+1}}(\vphi) \;
{\topcirc V}_{k\, i_1 \dots i_{n+1}}(\vphi)  \, .
\label{Zeq1}
\end{align}
For a single component $\vphi$, $N=1$, this can written in the form
\begin{align}
\bigg ( \frac{\pr}{\pr t}&{} +\eta  + \de \, \vphi
\frac{\pr}{\pr \vphi}
\bigg ) {\topcirc Z}(\vphi) + \eta \,  \nn \\
={}& -e^{k \, \frac{\pr^2}{\pr \vphi \, \pr \vphi'}} \, \frac{C_\eta}{d}
\bigg ( \frac{\pr}{\pr \vphi}  {\topcirc V}(\vphi) \;
\frac{\pr}{\pr \vphi'}  {\topcirc Z}(\vphi') 
+  \frac{\pr^2}{\pr \vphi^2} {\topcirc V}(\vphi) \;
{\topcirc Z}(\vphi') \nn\\
\noalign{\vskip -4pt}
& \hskip 3cm{} -  \frac{1}{2(d+2)} \, \frac{\pr^{2}}{\pr \vphi^{2}}  
{\topcirc V}(\vphi) \;
\frac{\pr^2}{\pr \vphi'^2}  {\topcirc V}(\vphi')\bigg ) \, 
\bigg |_{\vphi'=\vphi} \, ,
\label{Zeq2}
\end{align}

Assuming now the relevant solutions of \eqref{Veq1} satisfy
\be
{\topcirc V}(\vphi) = \frac{2d \omega^2}{C_\eta} \, 
e^{\frac{1}{2}k \, \frac{\pr^2}
{\pr \vphi_i \pr \vphi_i }} \, V  ( \vphi  / \omega ) \, , \qquad
\omega^2 = \half (d-2+\eta) k \, ,
\label{VVZZ}
\ee
then \eqref{Veq1} reduces to
\be
\bigg ( \frac{\pr}{\pr t} - d + \de \, \vphi_i \frac{\pr}{\pr \vphi_i}
- \frac{\pr^2}{\pr \vphi_i \pr \vphi_i }\bigg ) \, V(\vphi) = 
- \bigg ( \frac{\pr}{\pr \vphi_i} V(\vphi) \bigg )^{\! 2} \, .
\label{Veq2}
\ee
Moreover for a single component field if
\be
{\topcirc Z}(\vphi) = e^{\frac{1}{2}k \, \frac{\pr^2}
{\pr \vphi^2}} \, Z  ( \vphi  / \omega ) \, ,
\ee
then \eqref{Zeq2} becomes
\begin{align}
\Big ( \frac{\pr}{\pr t}  +\eta +  \Delta_V + 2 V''(\vphi) \Big ) Z(\vphi) =
- \eta+ \frac{2d}{(d+2)C_\eta} \,  V''(\vphi)^2 \, .
\label{Zeq3}
\end {align}
where
\be
\Delta_V = - \frac{\pr^2}{\pr \vphi^2} +
\de\, \vphi \frac{\pr}{\pr \vphi}
+ 2  V'(\vphi) \frac{\pr}{\pr \vphi} \, .
\ee
Solutions of \eqref{Veq2} and \eqref{Zeq2} clearly generate solutions of
\eqref{Veq1} and \eqref{Zeq2} using \eqref{VVZZ} (in general \eqref{VVZZ}
cannot be inverted but $V,Z$ are defined by the requirement of satisfying
\eqref{Veq2} and \eqref{Zeq3}). It is not clear how to reduce the
more general equation \eqref{Zeq1} to a similar form as in \eqref{Zeq3}
since it is not possible in general to extract a factor 
$e^{k \, \frac{\pr^2}{\pr \vphi_j \pr \vphi'{\!}_j }}$ on the right hand side
of \eqref{Zeq1}, as was done in \eqref{Veq1}.

The procedure adopted here in obtaining a derivative expansion in analogous
to that used in the scaling field approach but there 
${\topcirc V},{\topcirc Z}$
are expanded in a basis of monomials in $\vphi$ and it is then possible to use
the more general form \eqref{ggg} for the products of propagators appearing
in the expansion \eqref{exp2}.

\subsection{Extension of the Local Potential Approximation}

For a single component field at a fixed point \eqref{Veq2} becomes
\be
- d \,V_* (\vphi) + \de \, \vphi V_*{\!}'(\vphi) - V_*{\!}''(\vphi)  =
- V_*{\!}'(\vphi)^{2}  \, .
\label{Veq3}
\ee
It is furthermore consistent to require $\eta = {\rm O}(V_*{\!}^2)$ 
in which case \eqref{Zeq3} may be simplified at the fixed point determined 
by \eqref{Veq3} by restricting, in a expansion in powers of $V_*$,  only 
to terms up to ${\rm O}(Z_*V_*)$ to the form
\be
\big ( \Delta_{V_*} + 2 V_*{\!}''(\phi) \big ) Z_*(\vphi)  = 
-\eta + \frac{2d}{d+2} \, V_*{\!}''(\vphi)^2 \, ,
\label{Zeq4}
\ee
taking also $C_\eta \to 1$.
Combining \eqref{Zeq4} and \eqref{Veq3} 
is a minimal extension of the local
potential approximation to include a non zero anomalous dimension $\eta$
which was by discussed Osborn and Twigg \cite{DT}. The associated eigenvalue 
equation for critical exponents becomes
\be
\begin{pmatrix}\Delta_{V_*} -  d & 0 \\
\noalign{\vskip 2pt}
2 \frac{\rmd}{\rmd \phi} Z_* (\phi) \frac{\rmd}{\rmd \phi}  - 
\frac{4d}{d+2}V_*{\!} ''(\phi)  \frac{\rmd^2}{\rmd \phi^2}
& \Delta_{V_*} + 2 V_*{\!}''(\phi)\end{pmatrix} \begin{pmatrix}f(\phi) \\ 
g(\phi)\end{pmatrix}  =
\lambda \begin{pmatrix}f(\phi) \\ g(\phi)\end{pmatrix} \, ,
\label{fgeq}
\ee
which is easy to analyse numerically \cite{DT} since it can be reduced to
solving $(\Delta_{V_*} -  d) f = \lambda f$ and $(\Delta_{V_*} + 2 V_*{\!}'') g
= \lambda \, g$.

The virtue of \eqref{Veq3} and \eqref{Zeq4} is that the eigenvalue
problem \eqref{fgeq} has exact eigenfunctions and eigenvalues which
match those of the RG equations as discussed in subsections \ref{eigen} and 
\ref{zerom}. Corresponding to \eqref{exacta}, \eqref{exact} in \eqref{fgeq}
\begin{align}
f_\vphi(\vphi) = {}& \vphi - \frac{2}{2-\eta}\, V_*{\!}'(\vphi) \, , \ \
g_\vphi(\vphi) = - \frac{2}{2-\eta}\, Z_*{\!}'(\vphi)  \, , \quad 
\lambda_\vphi = - \half ( d + 2 - \eta ) \, ,  \nn \\
f_r(\vphi)  = {}& V_*{\!}'(\vphi) \, , \ \ g_r(\vphi)  = Z_*{\!}'(\vphi)
\, , \hskip 3.5cm \lambda_r =- \half ( d - 2 + \eta ) \, .
\label{exact2}
\end{align}
Corresponding to the zero mode
\be
f_\Z(\vphi) = 0 \, , \quad g_\Z(\vphi) = 1 - \frac{2}{2-\eta}\, V_*{\!}''(\vphi)\, .
\ee

As a  consequence of this zero mode solution
there exist non trivial solutions of the homogeneous
equation \eqref{Zeq4} so that $ \Delta_{V_*} + 2 V_*{\!}''$ is not invertible.
The existence of a solution for $Z_*$ imposes an eigenvalue condition on $\eta$,
since the right side of \eqref{Zeq4} must be orthogonal to $g_\Z(\vphi)$
with respect to a suitable scalar product, constructed in \cite{DT}, 
for which $\Delta_{V_*}$ is hermitian. For $\eta$ so determined there
is then a line of equivalent fixed points $Z_*(\vphi) \sim Z_*(\vphi)
+ c \, g_\Z(\vphi)$ for any $c$.

These results may illustrated by an epsilon expansion close to
the multi-critical points arising when $\vphi^{2(n+1)}$,  $n=1,2,\dots$, 
becomes a marginal operator which occurs, when $\eta=0$, for $(n+1)\delta_0 = d$ or
$d = d_n = {2(n+1)}/{n}$. Using 
$e^{\frac{1}{2\delta_0}\frac{\d^2}{\d \vphi^2}} \big ( - \frac{\d^2}{\d \vphi^2}
+ \delta_0 \, \vphi \frac{\d}{\d \vphi} \big )  =
\delta_0 \, \vphi \frac{\d}{\d \vphi}\, 
e^{\frac{1}{2\delta_0}\frac{\d^2}{\d \vphi^2}}$ then approximating \eqref{Veq3} gives
\be
d = d_n - \epsilon \, , \qquad e^{\frac{1}{2\delta_0}\frac{\d^2}{\d \vphi^2}}
V_*(\vphi) = g_n \epsilon \; \vphi^{2(n+1)} + 
{\rm O}\big (\epsilon^2 \vphi^{2p}, p \ne n+1 \big ) \, ,  
\ee
where, for $\epsilon>0$,
\be
g_n = n \, n! \delta_0{\!}^{n} \, \bigg ( \frac{n!}{2(2n+1)!} \bigg )^2 \, .
\ee
The corresponding solutions for $V_*(\vphi)$ are expressible in terms of 
Hermite polynomials using the identity
\be
e^{-\frac{1}{4}\, \frac{\d^2}{\d x^2}} (2x)^r = H_r(x) \, .
\ee
At any fixed order in the $\epsilon$-expansion there is a finite sum of $H_r$.

The condition for \eqref{Zeq4} then to have solutions to leading order
becomes
\be
 e^{\frac{1}{2\delta_0}\frac{\d^2}{\d \vphi^2}} \Big ( 
-\eta + \frac{2d_n}{d_n+2} \, V_*{\!}''(\vphi)^2 \Big ) = 
{\rm O} \big ( \vphi^{2p}, p \ne 0 \big ) \, ,
\ee
giving \cite{Phase}
\be
\eta = 8 (n+1)^3 \frac{(2n+1)!}{\delta_0{\!}^{2n}} \, g_n{\!}^2 \, \epsilon^2
= 4n^2 \bigg (\frac{(n+1)!^2}{(2n+2)!} \bigg )^3 \, \vep^2 \, .
\ee
With this result for $\eta$ \eqref{Zeq4} has the solution
$Z_*(\vphi) = c( 1 - V_*{\!}''(\vphi) ) + {\rm O}(\vep^2)$ for arbitrary $c$.

\section{Supersymmetric Example}

As a small example of the derivative expansion we consider a three
dimensional field theory with $\N=2$ supersymmetry. Applications of 
RG flow equations to supersymmetric theories have been described in
\cite{Vian,Arnone,Rosten2,Gies}.
For three dimensional $\N=2$ theories
there are chiral superfields  which may have a holomorphic superpotential,
like $\N=1$ Wess Zumino theories in four dimensions from which they may be
obtained by reduction. Unlike the four dimensional theories \cite{Rosten2}
non trivial IR fixed points in the absence of any gauge fields are not excluded.

As usual with supersymmetry it is convenient to adopt a spinorial notation
using the result that
in three dimensions the gamma matrices may be realised in terms of symmetric
real $2\times 2$ matrices
\be
(\si_a)_{\alpha\beta} = (\si_a)_{\beta\alpha}  \, , \qquad
({\tilde \si}_a)^{\alpha\beta} =  \vep^{\alpha\gamma}\vep^{\beta\delta}
(\si_a)_{\gamma\delta} \, ,
\label{gam}
\ee
with $\alpha,\beta=1,2$ and
\be
\si_a \, {\tilde \si}_b + \si_b \, {\tilde \si}_a = - 2\eta_{ab} \, I \, ,
\ee
with $\eta_{ab}$ the 3-dimensional Minkowski metric with signature $(-1,1,1)$
and $I$ the identity matrix. Any 3-vector $x^a$ is then equivalent
to a symmetric $2\times 2$ matrix
using the $\sigma$-matrices in \eqref{gam},
\be
x^a \to \x_{\alpha\beta}= (x^a \si_a)_{\alpha\beta} \, , \qquad
\tx^{\alpha\beta} = \vep^{\alpha\gamma}\vep^{\beta\delta} \x_{\gamma\delta} \, ,
\ee
so that $\x \tx = - x^2 \, I$. We also define
\be
\pr_{\alpha\beta} = ({\si}^a \pr_a)_{\alpha\beta} \, , \qquad
{\tilde \pr}^{\alpha\beta} = \vep^{\alpha\gamma}\vep^{\beta\delta}\pr_{\gamma\delta} \, ,
\ee
so that
\be
\pr_{\alpha\beta} \, \tx^{\gamma\delta} = - \delta_\alpha{\!}^\gamma
\delta_\beta{\!}^\delta -  \delta_\alpha{\!}^\delta\delta_\beta{\!}^\gamma\, .
\ee

For $\N=2$ superfields there are additional anti-commuting 
Grassmannian coordinates $\theta^\alpha, \bth^\alpha$. 
The associated covariant derivatives are $D_\alpha,\bD_\alpha$, where
$D_\alpha \theta^\beta = \delta_\alpha{\!}^\beta$ and
$\bD_\alpha \bth^\beta = \delta_\alpha{\!}^\beta$,
satisfy
\be
\big \{ D_\alpha , \bD_\beta \big \} = - 2i \, \pr_{\alpha\beta} \, .
\ee
$D_\alpha,\bD_\alpha$ anti-commute with the generators of super-translations. 
The full $\N=2$ superspace ${\mathbb M}^{3|4}$ with coordinates $(x,\theta,\bth)$ 
contains the invariant chiral superspaces ${\mathbb M}^{3|2}$ and 
${\bar {\mathbb M}}^{3|2}$, defined in terms  the chiral coordinates $(x_+, \theta)$
and $(x_-,\bth)$, where
$\x_\pm^{\alpha\beta}$ are required to satisfy
\be
\bD_\gamma \, \x_+^{\,\alpha\beta} = 0 \, , \qquad 
D_\gamma \, \x_-^{\, \alpha\beta} = 0 \, .
\ee 
Chiral superfields are defined on ${\mathbb M}^{3|2}$ and their
anti-chiral conjugates on ${\bar {\mathbb M}}^{3|2}$ so that
\be
\bD_\alpha \phi = 0 \ \ \Rightarrow  \ \ \phi(x_+, \theta) \, , \qquad
D_\alpha \bphi = 0 \ \ \Rightarrow  \ \ \bphi(x_-, \bth)  \, .
\ee
For two points labelled by $(x,\theta,\bth)$ and $(x',\theta',\bth')$ there
is a supertranslation invariant generalisation $y$ of the interval $x-x'$
given by
\begin{align}
{\rm y}^{\alpha\beta} = {}& \tx_+^{\, \alpha\beta} - \tx'{\!}_-^{ \ \alpha\beta}
- 4i \, \theta^{(\alpha} \, \bth'{}^{\beta)} \nn \\
=  {}& \tx^{\, \alpha\beta} - \tx'{}^{\,\alpha\beta} + 2i \, \theta^{(\alpha} \big (
\bth^{\beta)} - \bth'{}^{\beta)}  \big )
- 2i \, \big ( \theta^{(\alpha} - \theta'{}^{(\alpha}\big )
\bth'{}^{\, \beta)}  \, .
\label{defy}
\end{align}
It is easy to see that this satisfies $\bD_\gamma \, {\rm y}^{\alpha\beta}
= D'{\!}_\gamma {\rm y}^{\alpha\beta} = 0$ as $y$ is a function
on ${\mathbb M}^{3|2} \times {\bar {\mathbb M}}'{}^{3|2}$.

For chiral or anti-chiral fields we may extend \eqref{XY} to
\be
\phi \cdot \psi =  \int \d^3 x\, \d^2 \theta  \; \phi(x,\theta) \, \psi(x,\theta) \, , \qquad
\bphi \cdot {\bar \psi} =  \int \d^3 x\, \d^2 \bth  \; \bphi(x,\bth)\, {\bar \psi}(x,\bth) \, ,
\label{ppp}
\ee
and the associated functional derivatives
are correspondingly  defined so that
\begin{align}
\frac{\de}{\de \phi(x,\theta)}  \, \phi (x', \theta') ={}& \delta^3(x-x') \, 
(\theta- \theta')^ 2 \, , \qquad  \theta^2 = 
\vep_{\alpha\beta}\theta^\alpha \theta^\beta\, , \nn \\
\frac{\de}{\de \bphi(x,\bth)}  \, \bphi (x', \bth') = {}& \delta^3(x-x') \, 
(\bth - \bth')^ 2 \, ,  \qquad  \bth^2 = 
\vep_{\alpha\beta}\bth^\alpha \bth^\beta\, ,
\end{align}
where $(\theta- \theta')^ 2 , (\bth - \bth')^ 2 $ play the role of
Grassmannian delta functions, the integrations over $\theta,\bth$ being 
normalised so that
$\int \d^2 \theta \; \theta^2 = \int \d^2 \bth \; \bth^2 = 1$. 
In order to write RG equations analogous to those for simple scalar fields
we define bilinear expressions involving a chiral field $\phi$ and an
anti-chiral field $\bphi$, extending \eqref{XYG}, by
\be
\phi \cdot G \cdot \bphi 
= \int \! \d^3 x_+ \, \d^2\theta  \; \d^3 x'{\!}_-  \,\d^2 \bth' \; 
\phi(x_+,\theta) \, G(y) \, \bphi(x'{\!}_- , \bth') \, ,
\label{ppG}
\ee
for $y$ defined by \eqref{defy}.
Using, with this definition,
\be
{G}(y) = \tfrac{1}{16} \, {\bD}^2 D'{}^2 \big ( {G}(x-x') (\theta-\theta')^2
(\bth - \bth')^2 \big ) \, .
\label{Gss}
\ee
the expression \eqref{ppG}  may be elevated from integrals over 
${\mathbb M}^{3|2}$ and ${\bar {\mathbb M}}'{}^{3|2}$ to integrals over 
the full superspace ${\mathbb M}^{3|4}$ and ${\bar {\mathbb M}}'{}^{3|4}$ by 
letting
\be
\d^3 x_+ \, \d^2 \theta \, (-\tfrac{1}{4} \bD^2) \to \d^3 x \, \d^2 \theta
\, \d^2 \bth \, , \qquad
\d^3 x'{\!}_- \, \d^2 \bth' \, (-\tfrac{1}{4} D'^2) \to 
\d^3 x' \, \d^2 \theta' \, \d^2 \bth' \, ,
\label{intS}
\ee
and hence
\be
\phi \cdot G \cdot \bphi = \int \! \d^3 x \, \d^2 \theta
\, \d^2 \bth  \; \d^3 x' \, \d^2 \theta' \, \d^2 \bth'  \; 
\phi(x_+,\theta) \, G(x-x') (\theta-\theta')^2
(\bth - \bth')^2  \, \bphi(x'{\!}_- , \bth') \, .
\label{ppG2}
\ee
For the identity $I$, where $I(y)=\delta^3(y)$, then
\be
S_0[\phi,\bphi] = \phi \cdot I \cdot \bphi = \int \! \d^3 x \, \d^2 \theta
\, \d^2 \bth  \; \phi(x_+,\theta)  \, \bphi(x_-,\bth) \, .
\ee
is just the standard free kinetic term.

As before in \eqref{scale} it is convenient to rescale by using the cut off 
$\Lambda$ to  dimensionless variables so that
\be
\phi(x,\theta) = \Lambda^{\frac{1}{2}}\, \vphi(x \Lambda, \theta 
\Lambda^{\frac{1}{2}}) \, , \qquad \bphi(x,\bth) = \Lambda^{\frac{1}{2}}\, 
\bvphi(x \Lambda, \bth \Lambda^{\frac{1}{2}}) \, .
\label{rescale}
\ee

Just as for scalar field theories RG flow equations analogous
to \eqref{basic6} may be written in a similar form for $\S_t[\vphi,\bvphi]$
where
\begin{align}
\bigg ( \frac{\pr}{\pr t}&  + D^{(\de)}
\vphi \cdot \frac {\delta}{\delta \vphi} +D^{(\de)} \bvphi \cdot 
\frac{\delta}{\delta \bvphi} \bigg )  \S_t [\vphi,\bvphi] \nn \\
= {}& \frac{\delta} {\delta \vphi} \S_t[\vphi,\bvphi] \cdot G \cdot
\frac{\delta} {\delta \bvphi} \S_t[\vphi,\bvphi] -
\frac{\delta} {\delta \vphi} \cdot G \cdot 
\frac{\delta} {\delta \bvphi}\, \S_t[\vphi,\bvphi] 
- \eta \, \vphi \cdot \G^{-1} \cdot \bvphi \, ,
\label{srg2}
\end{align}
for, arising from the rescaling in \eqref{rescale},
\begin{align}
{D}^{(\de)} \vphi(x,\theta) = {}&  \big ( x \cdot \pr_x + \half \, 
\theta^\alpha \pr_\alpha + \de  \big )  \vphi(x,\theta) \, , \nn \\
{D}^{(\de)} \bvphi(x,\bth) = {}&  \big ( x \cdot \pr_x + \half \,
\bth^\alpha{\bar \pr}_\alpha + \de \big ) \bvphi(x,\bth) \, ,
\end{align}
with $\pr_\alpha
= \pr/\pr \theta^\alpha, \, {\bar \pr}_\alpha = \pr/\pr \bth^\alpha$.
In this case 
\be
\de=\half + \half \eta \, , 
\ee
with $\eta$ the anomalous dimension.
From the definition \eqref{defy}
\be
D^{(\de)}F(y) + F(y) {\overleftarrow D}{}^{(\de)} = 
(y \cdot \pr_y + 2\de ) F(y) \, .
\label{Feq}
\ee
In this case there are no $V$ terms as the superspace volume vanishes.
In \eqref{srg2} we have introduced $\eta$ representing a one parameter
arbitrariness in the choice of RG flow equations. As in the earlier discussion
this is expected to be constrained for $\S_t[\vphi,\bvphi]$ to have a well
defined limit, realising a non trivial IR fixed point, as $t\to \infty$.
We also require that $G, \G^{-1}$ satisfy the corresponding equation to
\eqref{Geq} which, as a consequence of \eqref{Feq}, leads to the same relation
as in \eqref{Geq2} for $\delta_0=\half, \, d=3$.

The previous discussion of a derivative expansion can be easily adapted to
the present case.  Instead of \eqref{DY2} we require
\be
\bigg [ D^{(\de)} \vphi \cdot 
\frac {\delta}{\delta \vphi} + D^{(\de)} \bvphi \cdot 
\frac{\delta}{\delta \bvphi}\,   , \, {\hat \Y} \bigg ] 
= \frac{\delta} {\delta \vphi} \cdot G \cdot
\frac{\delta} {\delta \bvphi} \, ,
\ee
which gives now
\be
{\hat \Y} =  \frac{\delta}{\delta \vphi} \cdot {\hat \G} \cdot
 \frac{\delta}{\delta \bvphi} \, .
\label{DY3}
\ee
with ${\hat \G}(y)$ satisfying \eqref{Geq2} as before. 
Following \eqref{SSY} so that 
${\topcirc S}_t[\vphi,\bvphi] = e^{\hat \Y} {S}_t[\vphi,\bvphi]$  then 
\eqref{exp1} becomes in this case
\begin{align}
& \bigg ( \frac{\pr}{\pr t}  + 
D^{(\de)} \vphi \cdot \frac {\delta}{\delta \vphi} +
D^{(\de)}\bvphi \cdot \frac{\delta}{\delta \bvphi} 
\bigg )  \, {\topcirc \S}_t[\vphi,\bvphi]
+  \eta \, \vphi \cdot \G^{-1} \cdot \bvphi \nn \\
&{} = 
\exp\bigg (\frac{\de}{\de \vphi} \cdot {\hat \G} \cdot \frac{\de} {\de \bvphi'}
+ \frac{\de}{\de \vphi'} \cdot {\hat \G} \cdot \frac{\de} {\de \bvphi} \bigg ) 
\frac{\delta} {\delta \vphi} {\topcirc \S}_t[\vphi,\bvphi] \cdot G \cdot
\frac{\delta} {\delta \bvphi'} {\topcirc\S}_t[\vphi',\bvphi']
\bigg |_{\vphi'=\vphi,\bvphi'=\bvphi} \, .
\label{sSS}
\end{align}
For a derivative expansion we may then write
\begin{align}
{\topcirc \S}_t[\vphi,\bvphi] = {}& \int  \d^3 x\, \d^2 \theta \; 
{\topcirc W}(\vphi)
+  \int \d^3 x\, \d^2 \bth \; {\topcirc {\bar W}}(\bvphi) \nn \\
&{}+  \int \d^3 x\, \d^2 \theta \, \d^2\bth \; \Big [
{\topcirc K(\vphi,\bvphi)} + D^2 \vphi \, {\topcirc L}(\vphi,\bvphi) + 
\bD^2 \bvphi  \, {\topcirc {\bar L}}(\vphi,\bvphi) + \dots \Big ] \, , 
\label{WKL}
\end{align}
where only the superpotential terms ${\topcirc W},{\topcirc{\bar W}}$ 
involve just integrals over the chiral superspaces $\S_\pm$ and
in the second line $\vphi(x_+,\theta),\bvphi(x_-,\bth)$. In general
${\topcirc K(\vphi,\bvphi)}$ is a K\"ahler potential such that
\be
{\topcirc K(\vphi,\bvphi)} \sim {\topcirc K(\vphi,\bvphi)} + 
f(\vphi) + {\bar f}(\bvphi) \, .
\label{arbK}
\ee
The separation in \eqref{WKL} between local superpotential terms
involving integrals over ${\mathbb M}^{3|2}$, ${\bar {\mathbb M}}^{3|2}$ 
and the remainder which contains
only full superspace integrations is valid beyond any derivative
expansion and is at the root of the non renormalisation theorems
for supersymmetric quantum field theories.

Inserting \eqref{WKL} on 
the left hand side of \eqref{sSS} we may use
\begin{align}
\! \bigg ( D^{(\de)} &  \vphi \cdot \frac {\delta}{\delta \vphi} +
{\D}^{(\de)} \bvphi \cdot \frac{\delta}{\delta \bvphi}\ \bigg ) 
{\topcirc \S}_t[\vphi,\bvphi]    \nn \\
= {}& \!\! \int  \! \d^3 x\, \d^2 \theta \, \Big ( -
2 {\topcirc W}(\vphi) + \de \, \vphi \frac{\pr}{\pr \vphi} 
{\topcirc W}(\vphi) \Big ) 
+  \int \! \d^3 x\, \d^2 \bth \, \Big ( - 2{{\bar W}}(\bvphi) 
+ \de \, \bvphi \frac{\pr}{\pr \bvphi} 
{\topcirc {\bar W}}(\bvphi) \Big )\nn\\
&{}+  \int \!  \d^3 x\, \d^2 \theta \, \d^2\bth \, \bigg [
- {\topcirc K}(\vphi,\bvphi) 
+ \de \, \Big ( \vphi \frac{\pr}{\pr \vphi}
+  \bvphi \frac{\pr}{\pr \bvphi} \Big ) {\topcirc K}(\vphi,\bvphi) \nn \\
&\hskip 2.7cm {} 
+ \de \, D^2 \vphi \bigg ({\topcirc L}(\vphi,\bvphi) 
+  \Big ( \vphi \frac{\pr}{\pr \vphi} +  
\bvphi \frac{\pr}{\pr \bvphi} \Big ){\topcirc L} (\vphi,\bvphi) \bigg ) 
\nn \\
&\hskip 2.7cm {}+ \de \,\bD^2 \bvphi  \bigg (
{\topcirc {\bar L}}(\vphi,\bvphi) +  \Big ( \vphi \frac{\pr}{\pr \vphi}
+  \bvphi \frac{\pr}{\pr \bvphi} \Big ) 
{\topcirc {\bar L}}(\vphi,\bvphi)  \bigg )  + \dots \bigg ] \, .
\label{Sder}
\end{align}
On the right hand side of \eqref{sSS} we may expand the exponential.
As a result of \eqref{Gss}, and similarly for ${\dot g}(y)$, all
integrations are over the full superspace $\S$. Consequently there are
no contributions involving integrals just over 
${\mathbb M}^{3|2}$, ${\bar {\mathbb M}}^{3|2}$ which would
correspond to terms in the RG flow of the superpotentials. Hence we have, 
without approximation, the linear equations
\be
\bigg ( \frac{\pr} {\pr t} + \de \, \vphi \frac{\pr}{\pr \vphi}
- 2 \bigg ) {\topcirc W}(\vphi) = 0 \, , \qquad
\bigg ( \frac{\pr} {\pr t} + \de \, \bvphi \frac{\pr}{\pr \bvphi}
- 2 \bigg ){\topcirc {\bar W}}(\bvphi) = 0 \, .
\ee
The solution for any $t$, and $\eta$, is simple
\be
{\topcirc W}_t(\vphi) = e^{2t}\, {\topcirc W}_0
\big ( e^{-\de \,t} \vphi \big ) \, ,
\qquad {\topcirc {\bar W}}_t(\bvphi) = 
e^{2t}\, {\topcirc {\bar W}}_0\big (e^{-\de \, t} \bvphi \big ) \, .
\ee

As discussed earlier $\eta$ is constrained by the requirement of a limit
as $t\to \infty$. ${\topcirc W}(\vphi)$ and ${\topcirc {\bar W}}(\bvphi)$ 
are required to
be holomorphic functions for any finite $\vphi,\bvphi$ so the possible
limits can only be of the form
\be
{\topcirc W}_t(\vphi)  \to {\topcirc W}_*(\vphi) = g_* \vphi^n \, ,  \qquad
{\topcirc {\bar W}}_t(\vphi)  \to {\topcirc {\bar W}}_*(\vphi) = 
{\bar g}_* \bvphi^n \, ,
\qquad n=2,3 , \dots \, .
\label{Wlim}
\ee
This requires
\be
\eta = \frac{4}{n} - 1 \,,
\ee
and also, to achieve the limit for any particular $n$, that the initial 
${\topcirc W}_0(\vphi)$ and ${\topcirc {\bar W}}_0(\bvphi)$ are
constrained to contain no terms ${\rm O}(\vphi^p, \bvphi^p)$ with $p<n$.
Any terms with $p>n$ are irrelevant and vanish in the limit \eqref{Wlim}.
For unitary theories $\eta\ge 0$, or $n\le 4$,  with $\eta=0$, when $n=4$,
corresponding to a free massless theory. The $n=2$ case is also a
massive free theory. However in three dimensions, unlike the four
dimensional case, there is the possibility of a non trivial IR fixed
point when $n=3$ and $\eta= \frac{1}{3}$. Given that $\eta$ is not then close
to zero this fixed point is clearly non perturbative.

Higher order equations in the derivative approximation may be obtained in 
the same fashion as for simple scalar field theories. Using just
\be
{G}(y) \, \hG(y)^n \sim \frac{C_\eta}{3} \, k^n \de^3(y)
=  \frac{C_\eta}{3} \, k^n \,
\tfrac{1}{16} \, {\bD}^2 D'{}^2 \big ( \de^3(x-x') (\theta-\theta')^2
(\bth - \bth')^2 \big ) 
\ee
we may obtain from \eqref{sSS} and \eqref{Sder}, using \eqref{WKL} with
\eqref{intS},
\begin{align}
\bigg ( \frac{\pr}{\pr t} -1 + \de \, 
\Big ( \vphi \frac{\pr}{\pr \vphi}&{} + \bvphi \frac{\pr}{\pr \bvphi} \Big ) 
\bigg ) {\topcirc K}(\vphi,\bvphi) \nn \\
= {}& - \eta \, \vphi \bvphi 
+  \frac{2C_\eta}{3} \sum_{n=0}^\infty \frac{k^n}{n!} \;
\frac{\pr^{n+1}}{\pr \vphi^{n+1}} {\topcirc W}(\vphi) \; 
\frac{\pr^{n+1}}{\pr \bvphi^{n+1}}  {\topcirc{\bar W}}(\bvphi) \, ,
\label{KWW}
\end{align}
up to contributions reflecting the freedom in \eqref{arbK}.
Defining
\begin{align}
{\topcirc K}(\vphi,\bvphi) = {}& \omega^2\,
e^{k \frac{\pr^2}{\pr \vphi \pr \bvphi}}
K(\vphi/\omega,\bvphi/\omega) \, , \qquad \omega^2 = (1+\eta) k \nn \\ 
{\topcirc W}(\vphi) = {}& A \omega^2 \, W(\vphi/\omega) \, , 
\ \ {\topcirc {\bar W}}(\vphi) = A \omega^2\, {\bar W}(\vphi/\omega) \, ,
\qquad A^2 = \frac{3}{2C_\eta} \, ,
\end{align}
then \eqref{KWW} becomes
\be
\Big ( \frac{\pr}{\pr t} -1 + \Delta \Big ) {K}(\vphi,\bvphi) 
=   - \eta \, \vphi \bvphi  +  {W}'(\vphi) \; {{\bar W}}'(\bvphi) \, , 
\label{KWW2}
\ee
for
\be
\Delta =  - \frac{\pr^2}{\pr\vphi\,\pr\bvphi} + \de \,
\Big ( \vphi \frac{\pr}{\pr \vphi} + \bvphi \frac{\pr}{\pr \bvphi} \Big )\, .
\label{defD}
\ee
To eliminate the freedom in \eqref{arbK} we may define
\be
Z(\vphi,\bvphi) =  \frac{\pr^2}{\pr\vphi\,\pr\bvphi} {K}(\vphi,\bvphi) \, ,
\ee
and  then \eqref{KWW2} becomes
\begin{align}
\Big ( \frac{\pr}{\pr t} +\eta + \Delta \Big ) {Z}(\vphi,\bvphi) 
=  - \eta + {W}''(\vphi) \; {{\bar W}}''(\bvphi) \, .
\label{KWW3}
\end{align}
In contrast to scalar theories, $\Delta$ in \eqref{KWW2} and \eqref{defD}
is just  a straightforward derivative operator.

\section{Discussion}

The notion of the Wilsonian effective action plays a crucial role in 
the conceptual understanding of quantum field theories. Expressed in
terms of local fields describing the relevant degrees of freedom
it may be used to calculate physical amplitudes at energy scales $\epsilon$
for $\epsilon < \Lambda$ for $\Lambda$ an energy cut off which is implicit
in the effective action. The Wilson effective action is required to be
quasi-local in the sense that expanding in terms of monomials in the
fields there should be no singularities in the coefficient functions
for momenta less than $\Lambda$. For application to a description
of low energies $\epsilon$ the action may then be identified in terms
of an expansion in terms of a basis of local operators, formed the fields
and derivatives, consistent with the assumed symmetries
of the theory, so long as there are no anomalies, with the expansion
essentially a series in powers of $\epsilon/\Lambda$.

Nevertheless a precise construction of the  Wilsonian effective action
is in general more elusive. For the restricted world of scalar field
theories the Wilson/Polchinski exact RG equations offer a prescription
for the effective action for an arbitrary continuously variable RG
scale $\Lambda$ in a form which lends itself to analytic treatment. 
It is also possible  to formulate various approximation schemes although
in general these lack systematic control. Many of these features can 
be extended to theories with fermion fields but gauge theories present
major problems. A cut off restricted to the quadratic part of the initial
action does not respect gauge invariance. Although various attempts have
been made to extend exact RG equations to gauge theories they lack
the essential simplicity of the scalar field equations and are also
unable to demonstrate the presence of IR fixed points which are
know to exist in very many quantum field theories in three and four
dimensions. Moreover such RG equations do not manifestly  describe
the flow of a Wilsonian effective action without any long distance
singularities.

At any IR fixed point with scale invariance there is expected to be
also conformal symmetry. For conformal field theories the additional
symmetry constrains the correlation function such that two and
three point functions of conformal primary fields are fully determined
up to an overall constant. It is natural to consider how conformal
symmetry is realised in an exact RG framework, if only for scalar
field theories. An extension of the Wilson exact RG equation to
include conformal transformations was considered long ago \cite{Schafer}.
In terms of the fixed point action $S_*$ it is possible to define
functional generators which satisfy the algebra of the conformal 
group \cite{HOR}. For any conformal field theory there is a natural
scalar product defined by the two point function as a consequence
of the state/operator correspondence. A precise  definition of a scalar
product such that the generator of scale transformations 
$\Delta_{\S_*,{\rm loc}}$ is a hermitian operator would enable
significant additional results for critical exponents to be 
obtained.

\bigskip
\leftline{\bf Acknowledgements}
\medskip
This paper has had a lengthy gestation. In large part it has developed
as part of a dialogue with Oliver Rosten,  to whom we are very
grateful.

\newpage

\appendix\section{Redundant Operators and RG Flow}

The results obtained in the text derive from the particular 
Polchinski RG equation. We show here how some results are also valid
for more general RG equations in which $\Psi_t$ in \eqref{basic3} is
allowed to be essentially arbitrary.

For any RG equation there are redundant operators which correspond
to redefinitions of the basic fields. We extend here the previous discussion
to show that under RG flow these form a closed subspace. 
The flow equations are therefore applied to  actions $\{S[\vphi]\}$ depending
on a field $\vphi$ and containing a cut off, which are invariant under
the appropriate symmetry group and which are at most smooth deformations
of local actions formed by integrals over an invariant action density 
${\cal L}$ constructed from a sum of monomials in the fields and derivatives 
at the same point. Such actions form a manifold $\M$ which is in general 
infinite dimensional. Coordinates on $\M$ may be identified with the couplings 
$\{g\}$.
The basic flow equation, defining trajectory $S_t[\vphi]\in \M$, has  the form
\be
\frac{\pr}{\pr t} S_t = - D\vphi \cdot \frac{\de}{\de \vphi} S_t
+ \Psi_t \cdot  \frac{\de}{\de \vphi} S_t -  
\frac{\de}{\de \vphi} \cdot \Psi_t \, ,
\label{RGP}
\ee
where  we assume $\vphi\in V_\vphi$, the vector space formed by local
fields, with a dual $V_\vphi{\!}^*$ and $\cdot$ then denotes the associated 
product on $V_\vphi\times V_\vphi{\!}^* \to \V = T\M_S$, the tangent space 
to $\M$ at $S$.   With $D : V_\vphi \to V_\vphi$ we require
\be
{\bar \vphi} \cdot \vphi = \vphi \cdot {\bar \vphi} \, , \quad 
{\bar \vphi} \cdot D \vphi = {\vphi} {\overleftarrow D} \cdot {\bar \vphi} = 
\vphi \cdot \bD {\bar \vphi} \, , \qquad 
\vphi \in V_\vphi  , \, {\bar \vphi} \in V_\vphi{\!}^* \, .
\ee
Also in \eqref{RGP} $\Psi\in V_\vphi$ is here arbitrary except that we assume it 
has the functional form
\be
\Psi\Big (S,  \frac{\de}{\de \vphi} \Big )  \, ,
\label{choice}
\ee
so that $\Psi_t = \Psi (S_t,  \frac{\de}{\de \vphi} )$.
In \eqref{RGP} $S_t[\vphi]$ is arbitrary up to the 
addition of terms that do not depend on $\vphi$. This freedom may be used
to absorb any $\vphi$-independent terms that may arise in any rearrangements
of  \eqref{RGP} giving alternative forms.

At a fixed point of the flow equation \eqref{RGP}
\be
 D\vphi \cdot \frac{\de}{\de \vphi} S_* - \Psi_* \cdot  \frac{\de}{\de \vphi} 
S_*  +  \frac{\de}{\de \vphi} \cdot \Psi_* =0 \, ,
\label{fixS}
\ee
with $\Psi_*= \Psi(S_*, \frac{\de}{\de \vphi} )$. In the neighbourhood
of the fixed point we may write
\be
S = S_* + \epsilon \, \O \, ,
\label{Svar}
\ee
for $\O\in \V_* = T\M_{S_*}$ the space of operators for the critical point 
theory. The variation \eqref{Svar} induces a change in $\Psi$ of the form
\be
\Psi\Big (S, \frac{\de}{\de \vphi} \Big )= \Psi_* + \epsilon \, \D \, \O \, ,
\label{Pvar}
\ee
defining the linear operator $\D:\V_*\to V_\vphi$. Hence the operator 
$\Delta_{S_*}:\V_*\to \V_*$ given by
\be
\Delta_{S_*} =  D\vphi \cdot \frac{\de}{\de \vphi} -
\Psi_* \cdot  \frac{\de}{\de \vphi} 
-  \frac{\de}{\de \vphi} S_* \cdot \D +  \frac{\de}{\de \vphi} \cdot \D \,.
\label{Dop}
\ee
determines the critical exponents through the eigenvalue equation 
\be
\Delta_{\S_*}  \O = \lambda \, \O \, ,
\label{eig2}
\ee
as in \eqref{eig} but for the more general $\Delta_{S_*}$ given 
by \eqref{Dop}.  Note that from  \eqref{fixS}
\be
\Delta_{S_*} \frac{\delta}{\delta \vphi} S_*  =  
-  \bD \frac{\delta}{\delta \vphi} S_*  \, .
\label{eigS}
\ee

Redundant operators $\{\O_\psi\}$ here have the form 
\be
\O_\psi = \psi \cdot  \frac{\de}{\de \vphi} S_* -  \frac{\de}{\de \vphi}
\cdot \psi \, ,
\label{red1}
\ee
for some $\psi\in V$ and defining a subspace $\V_R \subset \V_*$.
In \eqref{red1} the additional term present in \eqref{red} is omitted due to
\eqref{pol1}. The aim here is to show, as is necessary for such operators 
to form a closed space under RG flow, that $\Delta_{S_*}:\V_R\to \V_R$ or
\be
\Delta_{S_*} \O_\psi = \O_{\smash{\tilde \psi}} \, , \qquad
{\tilde \psi} = {\tilde \Delta}_{S_*} \, \psi \, ,
\label{red2}
\ee
for some linear operator ${\tilde \Delta}_{S_*}$. To demonstrate \eqref{red2} 
we first note that
\be
\Delta_{S_*} \O_\psi = \Big (\,D\vphi \cdot \frac{\de}{\de \vphi} - \Psi_* 
\cdot \frac{\de}{\de \vphi} \Big )  \O_\psi  - \O_{\D \O_\psi} \, .
\ee
It is then sufficient to show, using \eqref{fixS},
\begin{align}
& \Big (  D\vphi \cdot \frac{\de}{\de \vphi} -  \Psi_* \cdot  
\frac{\de}{\de \vphi} \Big )\, \psi \cdot \frac{\de}{\de \vphi} S_* \nn \\
&{} = \bigg ( \Big ( D\vphi \cdot \frac{\de}{\de \vphi}  -
\Psi_* \cdot  \frac{\de}{\de \vphi} \Big ) \psi + 
\psi \cdot  \frac{\de}{\de \vphi} \Psi_*  - D \psi \bigg ) 
\cdot \frac{\de}{\de \vphi}  S_*
- \psi \cdot \frac{\de}{\de \vphi} \Big (
\frac{\de}{\de \vphi} \cdot \Psi_* \Big ) \, ,
\end{align}
and also
\be
\Big (   D\vphi \cdot \frac{\de}{\de \vphi} -
\Psi_* \cdot  \frac{\de}{\de \vphi} \Big )\, \frac{\de}{\de \vphi}\cdot \psi
= \frac{\de}{\de \vphi}\cdot \bigg (
\Big ( D\vphi \cdot \frac{\de}{\de \vphi}  -
\Psi_* \cdot  \frac{\de}{\de \vphi} \Big ) \psi - D \psi \bigg ) 
+  \Big ( \frac{\de}{\de \vphi} 
\Psi_* \cdot \frac{\de}{\de \vphi} \Big ) \cdot \psi \, ,
\ee
Since
\be
\psi \cdot \frac{\de}{\de \vphi} 
\Big ( \frac{\de}{\de \vphi} \cdot \Psi_* \Big ) 
+ \Big ( \frac{\de}{\de \vphi} \Psi_*
\cdot \frac{\de}{\de \vphi} \Big ) \cdot \psi =
\frac{\de}{\de \vphi} \cdot \bigg (
\Big ( \psi \cdot \frac{\de}{\de \vphi} \Big ) \Psi_* \bigg ) \, ,
\ee
we then have in \eqref{red2}
\be
{\tilde \psi} = 
\Big (   D\vphi \cdot \frac{\de}{\de \vphi} 
- \Psi_* \cdot  \frac{\de}{\de \vphi} \Big )\, \psi  
+ \psi \cdot \frac{\de}{\de \vphi} \Psi_* - \D \, \O_\psi - D \psi\, .
\label{tpsi}
\ee
determining ${\tilde \Delta}_{S_*}$ in \eqref{red2}. 

For the particular case of Wilson/Polchinski equations it is sufficient to 
require
\be
\Psi = \frac{1}{2} \, {G} \cdot \frac{\de}{\de \vphi} S \, , \qquad
\D =   \frac{1}{2} \, {G} \cdot \frac{\de}{\de \vphi} \, ,
\label{pol}
\ee
which leads to equations of the form \eqref{basic4}.
With this choice \eqref{tpsi} then ensures that in \eqref{red2}
\be
{\tilde \Delta}_{S_*}  = \Delta_{S_*}  -  D  \, .
\label{polD}
\ee

For a zero mode corresponding to a redundant marginal operator
we must have
\be
\Delta_{S_*}\, \Z = 0 \, , \qquad \Z = \O_{\psi_\Z} \, , \quad
{\tilde \Delta}_{S_*} \, \psi_\Z = 0 \, .
\label{zero}
\ee
For the equations obtained by requiring \eqref{pol} and \eqref{polD},
it is easy to construct such a $\psi_\Z$. Assuming
\be
\psi_\Z = \vphi + \GH \cdot \frac{\delta}{\delta\vphi}\, S_* \, ,
\ee
then using \eqref{eigS}
\be
{\tilde \Delta}_{S_*} \psi_\Z =  - \big ( {G} + D \, \GH \big )\cdot 
\frac{\delta}{\delta \vphi}\, S_*  
- \GH \cdot \bD  \frac{\delta}{\delta\vphi}\, S_*\, ,
\ee
and therefore \eqref{zero} is satisfied if
\be
D \, \GH + \GH {\overleftarrow D} = - {G} \, .
\label{Heq}
\ee
In solving \eqref{Heq} for $\GH$ it is necessary to impose the boundary 
conditions so that $\psi_\Z$ depends essentially locally on $\vphi$.
In the context of the earlier discussion in \ref{eigen} $D$ takes the form 
$D \vphi = D^{(\de)} \vphi + G \cdot \G^{-1} \cdot \vphi$, 
for $\delta$ as in \eqref{deeq}, and then \eqref{Heq} becomes explicitly 
$( p \cdot \pr + 2 - 4 p^2 K'(p^2)/K(p^2) -\eta ){\tilde \GH}(p)= 2 K'(p^2)$, 
which has a solution identical with that in \eqref{zpsi}. 

A crucial issue in the discussion of exact RG equations such as \eqref{RGP}
is the extent to which $\Psi$ is restricted while maintaining an 
equation for the RG flow equation which allows non trivial IR fixed
points to be realised as $t\to \infty$, even assuming \eqref{choice}.
The trivial choice $\Psi=0$ cannot lead to any non trivial fixed 
points, unlike the non linear equations obtained by taking \eqref{pol}. 

In order to clarify such issues we consider 
infinitesimal variations in the  functional form for $\Psi$
in \eqref{choice}, $\delta \Psi(S,\frac{\de}{\de \vphi})$, leading to 
\be
S_* \to S_* + \delta S_*  \quad \Rightarrow \quad 
\Psi_* \to \Psi_* + \D \delta S_* + \delta \Psi_* \, ,
\label{pert1}
\ee
with $\de \Psi_* = \delta \Psi(S_*,\frac{\de}{\de \vphi})$.
The variation of the fixed point equation \eqref{fixS} then gives
\be
\Delta_{S_*} \delta S_* - \delta \Psi_* \cdot  \frac{\de}{\de \vphi} 
S_*  +  \frac{\de}{\de \vphi} \cdot  \delta \Psi_* =0  \, , \quad \mbox{or}
\quad \Delta_{S_*} \delta S_*  =  \O_{\delta \Psi_*} \, .
\label{varP}
\ee
Special cases of \eqref{varP} are given by \eqref{var1} and \eqref{var2}.
As a consequence of \eqref{red2}, \eqref{varP} requires
\be
\delta S_*  = \O_\psi \, , \qquad 
{\tilde \Delta}_{S_*}  \psi = \delta \Psi_* \, .
\label{pert2}
\ee

We here verify how perturbations such as \eqref{pert1}, with
$\delta S_*$ a redundant operator given by \eqref{pert2}, 
lead to equivalent results, as far as the eigenvalues determined by
\eqref{eig2} are concerned, for apparently arbitrary choices for $\de\Psi$.  
To show invariance of the spectrum corresponding to non redundant 
operators $\O$ in \eqref{eigO2} it is sufficient to demonstrate 
that the operator $\O$ can be modified to $\O + \delta \O$ such that
\be
\Delta_{S_*} \delta \O + \delta \Delta_{S_*} \O = \lambda\; \delta \O
+ \O_{\chi} \, ,
\label{veig}
\ee
for some suitable $\chi$ and where from \eqref{pert1}
\be
\delta \Delta_{S_*} =-  \big ( \D \de S_* + \de \Psi_* \big ) \cdot  
\frac{\delta}{\delta \vphi}
- \Big ( \frac{\delta}{\delta \vphi} \de S_*  \Big ) \cdot \D
- \frac{\delta}{\delta \vphi} S_* \cdot \delta \D + 
\frac{\delta}{\delta \vphi} \cdot \delta \D \, .
\label{varD}
\ee
The precise form for $\delta \D$, which is determined by extending
\eqref{Svar} and \eqref{Pvar}, is subsequently irrelevant. The result
\eqref{veig} ensures that, to linear order, the perturbed eigenvalue
equation is of the form \eqref{eigO2} for the same eigenvalue $\lambda$ as
in the unperturbed case.

To achieve compatibility with \eqref{veig} it is sufficient to take
\be
\delta \O = \psi  \cdot \frac{\delta}{\delta \vphi} \, \O \, ,
\ee
where $\psi$ is determined by \eqref{pert2}. In general, using \eqref{Dop} 
for $\Delta_{S_*}$,
\be
\Delta_{S_*} \delta \O = \Big (  D\vphi \cdot \frac{\de}{\de \vphi} -
\Psi_* \cdot  \frac{\de}{\de \vphi} \Big ) \, \de \O - \O_{\D \de \O} \, ,
\label{DOZ0}
\ee
and the basic eigenvalue equation \eqref{eig2} with \eqref{tpsi} ensures that
\begin{align}
\Big (  D\vphi \cdot \frac{\de}{\de \vphi} -
\Psi_* \cdot  \frac{\de}{\de \vphi} \Big ) 
\Big ( \psi  \cdot \frac{\delta}{\delta \vphi} \, \O \Big ) 
= {}& \lambda \; \psi  \cdot \frac{\delta}{\delta \vphi} \, \O 
+ \psi  \cdot \frac{\delta}{\delta \vphi}
\bigg ( \Big ( \frac{\delta}{\delta \vphi}S_* - \frac{\delta}{\delta \vphi}
\Big ) \cdot \D \O \bigg ) \nn \\
&{} + \big ( {\tilde \Delta}_{S_*}\psi + \D \O_\psi \big )
\cdot \frac{\delta}{\delta \vphi} \O \, .
\label{DOZ1}
\end{align}
Furthermore from \eqref{varD}
\be
\delta \Delta_{S_*} \O =
- \big ( \D \O_\psi + \de \Psi_* \big ) \cdot  \frac{\delta}{\delta \vphi}\O
- \Big ( \frac{\delta}{\delta \vphi} \O_\psi \Big ) \cdot \D \O
- \O_{\de \D \, \O}  \, . 
\label{DOZ2}
\ee
Hence combining \eqref{DOZ0}, \eqref{DOZ1} and \eqref{DOZ2}, and using \eqref{pert2},
\begin{align}
\Delta_{S_*} \delta \O + \delta \Delta_{S_*} \O - \lambda\; \delta \O 
= &{} - \O_{\de \D \, \O} - \O_{\D \de \O} \nn \\
&{} -  \frac{\delta}{\delta \vphi} \O_\psi \cdot \D \O
+  \psi  \cdot \frac{\delta}{\delta \vphi}\bigg ( 
\Big ( \frac{\delta}{\delta \vphi}S_* - \frac{\delta}{\delta \vphi}
\Big ) \cdot \D \O \bigg ) \, .
\end{align}
This may be simplified by virtue of
\be
\psi \cdot \frac{\delta}{\delta \vphi}
\bigg ( \Big ( \frac{\delta}{\delta \vphi}S_* - \frac{\delta}{\delta \vphi}
\Big ) \cdot \D \O \bigg ) - \frac{\delta}{\delta \vphi} \Big (
\psi \cdot \frac{\delta}{\delta \vphi} S_* - \frac{\delta}{\delta \vphi}
\psi \Big ) \cdot \D \O = \xi \cdot \frac{\delta}{\delta \vphi} S_*
- \frac{\delta}{\delta \vphi} \cdot \xi \, ,
\ee
where
\be
\xi =  \psi  \cdot \frac{\delta}{\delta \vphi} \, \D \O
- \D O \cdot  \frac{\delta}{\delta \vphi} \, \psi  \, .
\ee
This then ensures that  \eqref{veig} is satisfied if we take
\be
\chi = \xi - \delta \D \, \O - \D \, \delta \O \, .
\ee

If arbitrary variations $\de \Psi$ were allowed then it would be possible
to continuously transform $\Psi_*$ to zero, contradicting the
expected triviality of the RG equations for $\Psi =0$.
However there are potential obstructions to unconstrained variations
$\delta \Psi_*$ at non trivial fixed points
due to the presence of zero modes satisfying \eqref{zero}. 
To show this we assume there is a scalar product with respect to which
$\Delta_{S_*}$ is hermitian so that
\be
\big \langle \O' , \Delta_{S_*} \O \big \rangle
= \big \langle \Delta_{S_*} \O' , \O \big \rangle \, .
\label{herm}
\ee
From this it follows that the critical exponents defined \eqref{eig2}
must be real. As a direct consequence of \eqref{herm},
\eqref{varP} also requires that $\delta \Psi$ must satisfy
\be
\big \langle \Z , \O_{\delta \Psi_*} \big \rangle = 0 \, .
\ee
This requires that $\delta \Psi_*$ is restricted to a surface of
codimension one in the space of all possible variations. 
Such a condition should ensure that the anomalous dimension $\eta$,
which was initially introduced in terms of a contribution to $\Psi$,
cannot be varied at the critical point when $\Psi\to \Psi_*$.
This quantisation of $\eta$ is then directly associated with the presence
of the zero mode $\Z$.

\section{Perturbative Considerations}

Any exact RG equation should of course generate the usual perturbative results
in a weak coupling expansion. This requires that at a fixed point,
if this is accessible perturbatively as in the $\vep$-expansion, the
critical exponents found by using an exact RG equation should match those
found in an $\vep$-expansion. Of course with approximations this may no
longer hold but may perhaps be used as a form of boundary condition to
constrain any free parameters which appear for example in a derivative
expansion. That lowest order perturbative results can  be matched with those
obtained in a derivative expansion is demonstrated in this section
although extending this beyond lowest order is more problematic.

For scalar theories we consider therefore the simple Lagrangian
\be
 \cL_V = \half \, \pr^\mu \phi \cdot \pr_\mu \phi + V(\phi)  \, .
\label{lag}
\ee
Formally for dimensions $d_n=2(n+1)/n$, $n=1,2,\dots$, this has a 
renormalisable perturbative expansion if $V(\phi)$ is a polynomial of 
degree $2(n+1)$ which may then be expressed in terms of a finite linear 
sum $\sum_I g^I \O_I(\phi)$ over all linearly independent monomials 
$\O_I(\phi)$, with coefficients the couplings $g^I$
parameterising the renormalisable theory. Renormalisability ensures
that the  $\beta$-functions determining the perturbative RG flow
are also expressible in terms of a similar expansion 
$\beta_V(\phi) = \sum_I \beta^I(g) \O_I(\phi)$.
Using dimensional regularisation and including the 
canonical dimension then the RG flow becomes
\be
\frac{\d}{\d t} V(\phi) = -B_V(\phi) \, ,
\ee
where
\be
B_V(\phi) =  (\Gamma_\phi \phi) 
\cdot \pr \, V(\phi) - d \, V(\phi) + \tbeta_V(\phi) \, , \qquad
\Gamma_{\phi,ij} =  \half (d-2)\delta_{ij} + \gamma_{\phi,ij} \
\label{BV}
\ee
In \eqref{BV} the contribution to the $\beta$-functions involving the 
anomalous dimension matrix, $\gamma_{\phi,ij} $, for $\phi$, has been isolated,
the usual $\beta$-function is then $\beta_V(\phi)= \tbeta_V(\phi)+ 
(\gamma_\phi \phi) \cdot \pr V(\phi)$.
Using dimensional regularisation by letting $d=d_n-\vep$, and assuming minimal 
subtraction of poles in $\vep$, $\tbeta_V(\phi)$ does not depend on $d$
explicitly and in each order of a loop expansion $\tbeta_V(\phi)$ is a 
scalar polynomial formed from contractions of
$V_{i_1 i_2 \dots i_k}(\phi)= \pr_{i_1} \dots \pr_{i_k} V(\phi)$ with 
$k\ge 2$. Fixed points arise when $V=V_*$ if
\be
B_{V_*}(\phi) = 0 \, .
\label{fixp}
\ee

For discussion of the critical exponents which are associated with 
mixing of scalar non derivative and derivative operators the fundamental 
Lagrangian is extended to 
\be
\cL = \cL_V + \cL_{F,G} \, ,
\quad \cL_{F,G} =
F(\phi) + \half G_{ij}(\phi) \pr^\mu \phi_i \pr_\mu \phi_j \, ,
\label{LFG}
\ee
with $\cL_V$ as in \eqref{lag} and, as discussed above, $V(\phi)$ a polynomial
of degree $2(n+1)$. The theory defined by \eqref{LFG} is renormalisable, 
in the sense that all counterterms linear in  $F,G_{ij}$, may be absorbed into 
a bare Lagrangian $\cL_0$ which has the same functional form 
as \eqref{LFG} if,  in dimensions $d_n$, $F(\phi)$ is a polynomial in $\phi$ of
degree $4n-1$  and $G_{ij}(\phi)$ also a polynomial of degree $2n-1$. 
For higher degree four derivative terms are also necessary in \eqref{LFG}.

In a similar fashion to the definition of the usual $\beta$-functions we may 
define linear operators acting on $F,G_{ij}$,
\begin{align}
& \Delta_V \! \begin{pmatrix} F(\phi) \\ G_{ij}(\phi) \end{pmatrix} \nn \\
&{} =
\begin{pmatrix}  (\Gamma_\phi \phi) \cdot \pr F(\phi) 
- d F(\phi) + \gamma_{FF} F(\phi) + \gamma_{FG,ij} G_{ij}(\phi) 
\\    (\Gamma_\phi \phi) 
\cdot \pr G_{ij}(\phi) + \gamma_{\phi,ik}  G_{kj}(\phi)
+ \gamma_{\phi,jk}  G_{ik}(\phi)  + 
\gamma_{GF,ij} F(\phi) + \gamma_{GG,ijkl} G_{kl}(\phi)   \end{pmatrix} \, ,
\end{align}
with $\Gamma_\phi$ defined as in \eqref{BV} and $\gamma_{FF}, \gamma_{FG,ij}, 
\gamma_{GF,ij}, \gamma_{FG,ijkl}$ are
differential operators depending on the couplings $g^I$ or $V$.
If $F(\phi)$ is restricted to be a polynomial of degree $2(n+1)$,  then
\be
\tbeta_{V+F}(\phi) = \beta_V(\phi) 
+ \gamma_{FF} F(\phi) + {\rm O}(F^2) \, .
\ee

At a fixed
point the exponents are defined by the coupled linear equations
\be
\Delta_{V_*} \! \begin{pmatrix} F(\phi) \\ G_{ij}(\phi) \end{pmatrix}
= -\lambda \begin{pmatrix} F(\phi) \\ G_{ij}(\phi) \end{pmatrix} \, .
\label{exFG}
\ee

The action defined by the lagrangian in \eqref{LFG} is invariant under
\be
\delta \cL_V = \delta_{F,G} \cL_{F,G} \quad \mbox{for} \quad
\delta \phi_i =  v_i(\phi) \, , \quad \delta \pr_\mu \phi_i = v_{i,j}(\phi)
\pr_\mu \phi_j \, ,
\ee
if
\be
\delta_{F,G}  F(\phi) = v(\phi)\cdot \pr V(\phi)
\, , \quad  \delta_{F,G} G_{ij}(\phi) = \pr_i v_j(\phi) + \pr_j v_i(\phi) \, .
\label{diffeo}
\ee
Assuming $F(\phi),G_{ij}(\phi)$ are restricted to ensure that no mixing with 
four  derivative operators arises then it is necessary to require $v_i(\phi)$ 
is a  polynomial in $\phi$ of degree $2n$. In \cite{DT} it was shown how this
leads to identities for $\Delta_V$ such that
\begin{align}
\Delta_V \! \begin{pmatrix} v(\phi)\cdot \pr V(\phi) \\ \pr_i v_j 
(\phi)+ \pr_j v_i(\phi)  \end{pmatrix} =
\begin{pmatrix} U(\phi) \cdot \pr V(\phi)  + v(\phi) \cdot \pr B_V(\phi) \\
 \pr_i U_j(\phi) + \pr_j  U_i(\phi) \end{pmatrix} \, ,
\label{DDeq} 
\end{align}
for
\be
U_i(\phi)=  (\Gamma_\phi\phi) \cdot \pr  v_i(\phi) + \gamma_{FF} v_i(\phi)
- \Gamma_{\phi,ij} v_j(\phi )  \, .
\label{defU}
\ee

The result \eqref{DDeq} ensures that at a critical point, where 
\eqref{fixp} holds, there are solutions of \eqref{exFG} if
\be
U_i(\phi) = - \lambda \, v_i(\phi) \, .
\ee
In particular since $\gamma_{FF}$ involves at least second order derivatives 
with respect to $\phi$ there are exact zero modes, with $\lambda=0$, obtained 
by taking $v_i(\phi)\to \phi_i$,
\be
F_0 (\phi) = \phi \cdot \pr \, V(\phi) \, , \qquad
G_{0,ij} = 2\, \delta_{ij} \, .
\ee

In perturbative calculations  \eqref{DDeq} is equivalent to
\be
\gamma_{FF}  \big ( v(\phi) \cdot \pr \, V(\phi) \big )
+ \gamma_{FG,ij} \big ( \pr_i v_j(\phi) + \pr_j v(\phi) \big )
= \big ( \gamma_{FF} v(\phi) \big ) \cdot \pr\, V(\phi)
+ v(\phi) \cdot \pr \tbeta_V(\phi) \, ,
\label{geq1}
\ee
and
\begin{align} 
\gamma_{GF,ij}&  \big ( v(\phi)\cdot \pr \, V(\phi) \big ) +
\gamma_{GG,ijkl} \big ( \pr_k v_l(\phi) + \pr_l v_k(\phi) \big ) \nn \\
&{} = \pr_i \big ( \gamma_{FF} v_j(\phi) - 2\, \gamma_{\phi,jk} v_k(\phi) \big )
+ \pr_j \big ( \gamma_{FF} v_i(\phi) 
- 2\, \gamma_{\phi,ik} v_k(\phi) \big ) \, .
\label{geq2}
\end{align}

For $d=d_n$ there are logarithmic divergencies at $pn$ loops for $p=1,2,\dots$
which give rise to contributions to the $\beta$-functions.
At lowest order for the $\beta$-function associated with $V$
\be
\tbeta^{(n)}_V(\phi) = a_n \, V_{i_1 \dots i_{n+1}}(\phi)
 V_{i_1 \dots i_{n+1}}(\phi) \, ,
\ee
from which it is easy to see that
\be
\gamma^{(n)}_{FF} F(\phi) = 2a_n \, V_{i_1 \dots i_{n+1}}(\phi)
 F_{i_1 \dots i_{n+1}}(\phi) \, .
\ee
We also have \cite{DO,DT}
\begin{align}
\gamma^{(n)}_{FG,ij} G_{ij}(\phi) 
= &{}- a_n \!\!\!\!\!
\sum_{\genfrac{}{}{0pt}{}{r,s,t\ge 1}{r+s+t=n+2}} \!\!\!\!
\frac{(n+1)!}{r!\,s!\,t!} \, {\hat K}_{rst} \,
V_{i_1 \dots i_r k_1 \dots k_t} (\phi) V_{j_1 \dots j_s k_1 \dots k_t} (\phi)\,
G_{i_1j_1,i_2\dots i_r j_2 \dots j_s} (\phi) \nn \\
&{}+ a_n \!\!\!\!\!
\sum_{\genfrac{}{}{0pt}{}{r\ge 2,s,t\ge 1}{r+s+t=n+2}} \!\!\!\!
\frac{(n+1)!}{r!\,s!\,t!} \, {\hat K}_{rst} \,
V_{i_1 \dots i_r k_1 \dots k_t} (\phi) V_{j_1 \dots j_s k_1 \dots k_t} (\phi)\,
G_{i_1i_2,i_3\dots i_r j_1 \dots j_s} (\phi) \, ,
\label{bnF}
\end{align}
and
\begin{subequations}
\begin{align}
\gamma^{(n)}_{GG,ijkl} G_{kl}(\phi) =2a_n \big ( &
V_{i_1 \dots i_{n+1}} (\phi) \, G_{ij,i_1 \dots i_{n+1}} (\phi)
+ 2 V_{i_1 \dots i_{n+1}(i} (\phi) \, G_{j)\,i_1,i_2 \dots i_{n+1}} (\phi) \nn \\
&{} - V_{i_1 \dots i_{n+1}(i} (\phi) \, 
G_{i_1i_2,j)\,i_3 \dots i_{n+1}} (\phi) \big ) \, , \\
\gamma^{(n)}_{FG ,kl} G_{kl}(\phi) = {}& 0 \, .
\label{bnG}
\end{align}
\end{subequations}
The precise expressions for $a_n$ and $ {\hat K}_{rst}$ are given in 
\cite{DT} but are unimportant here, it is only necessary to note that
$ {\hat K}_{rst}$ is symmetric and satisfies
\be
{\hat K}_{1st} = 1 \, , \quad s+t = n+1 \, .
\ee
With these results it is straightforward to verify that
\eqref{geq1} and \eqref{geq2} are satisfied
since $\gamma^{(n)}_{\phi,ij}=0$ at this order.

At the next order
\begin{align}
\tbeta^{(2n)}_V(\phi) = {}& -\frac{b_n}{3n} \!\! 
\sum_{\genfrac{}{}{0pt}{}{r,s,t\ge 1}{r+s+t=2n+2}} \!\!\!\!
\frac{(2n+2)!}{r!\,s!\,t!} \; {\hat K}_{rst} \,
V_{i_1 \dots i_r j_1 \dots j_s} (\phi) \,
V_{i_1 \dots i_r k_1 \dots k_t} (\phi)\,
V_{j_1 \dots  \dots j_s  k_1 \dots k_t} (\phi) \, , \nn \\
\gamma^{(2n)}_{\phi,ij} = {}& 2b_n \, g_{i\, i_1 \dots i_{2n+1}} \, 
g_{j\, i_1 \dots i_{2n+1}}
\end{align}
where $g_{i_1 \dots i_{2n+2}} = V_{i_1 \dots i_{2n+2}}(\phi)$ is the
dimensionless coupling and $b_n = ((n+1)!)^2 a_n{\!}^2/(2n+2)!$.
${\hat K}_{rst}$ includes the effect of subdivergencies when any $r,s,t$
are equal to $n+1$.

\renewcommand{\baselinestretch}{0.6}

\end{document}